\documentstyle[12pt,amssymb,epsfig,epsf]{article}

\renewcommand{\theequation}{\arabic{section}.\arabic{equation}}

\newcommand{\eps}{\epsilon}

\newcommand{\be}{\begin{equation}}
\newcommand{\ee}{\end{equation}}
\newcommand{\ba}{\begin{eqnarray}}
\newcommand{\ea}{\end{eqnarray}}
\newcommand{\baa}{\begin{eqnarray*}}
\newcommand{\btab}{\begin{tabular}}
\newcommand{\etab}{\end{tabular}}
\newcommand{\eaa}{\end{eqnarray*}}

\newcommand{\derleft}{\stackrel{\leftarrow}{D}}
\newcommand{\derright}{\stackrel{\rightarrow}{D}}

\def \labeltest #1 {\label{#1}}
\newcommand\re[1]{(\ref{#1})}

\newcommand\lr[1]{{\left({#1}\right)}}

\newcommand \rvev [1] {\langle{#1}\rangle}

\def \e {\mbox{e}}
\def \CO {{\cal O}}

\begin{document}

\begin{titlepage}
\begin{flushright}
\begin{tabular}{l}
LPT--Orsay--98--83\\
NORDITA--99--11--HE\\
SPbU--IP--99--04\\
hep-ph/9902375
\end{tabular}
\end{flushright}

\vskip1cm
\begin{center}
  {\large \bf
     Baryon Distribution Amplitudes in QCD
  \\}

\vspace{1cm}
{\sc V.M.~Braun}${}^{1}$,
{\sc S.\'{E}. Derkachov}${}^2$,
{\sc G.P.~Korchemsky}${}^3$
          and {\sc A.N.~Manashov}${}^4$
\\[0.5cm]

\vspace*{0.1cm} ${}^1${\it
          NORDITA, Blegdamsvej 17, DK-2100 Copenhagen, Denmark
                       } \\[0.2cm]
\vspace*{0.1cm} ${}^2$ {\it Department of Mathematics,
St.-Petersburg Technology Institute,\\ St.-Petersburg, Russia
                       } \\[0.2cm]
\vspace*{0.1cm} ${}^3$ {\it
Laboratoire de Physique Th\'eorique%
\def\thefootnote{\fnsymbol{footnote}}%
\footnote{Unite Mixte de Recherche du CNRS (UMR 8627)},
Universit\'e de Paris XI, \\
91405 Orsay C\'edex, France
                       } \\[0.2cm]
\vspace*{0.1cm} ${}^4$ {\it
Department of Theoretical Physics,  Sankt-Petersburg State
University, \\ St.-Petersburg, Russia
                       } \\[1.0cm]


\def\thefootnote{\arabic{footnote}}
\setcounter{footnote} 0

\vskip0.8cm
{\bf Abstract:\\[10pt]} \parbox[t]{\textwidth}{
We develop a new theoretical framework for the description
of leading twist
light-cone baryon distribution amplitudes which is based on
integrability of the helicity $\lambda=3/2$  evolution equation to
leading logarithmic accuracy. A physical interpretation is that
one can identify a new `hidden' quantum number which distinguishes
components in the $\lambda=3/2$ distribution amplitudes with different
scale dependence. The solution of the corresponding evolution equation
is reduced to a simple three-term recurrence relation. The exact analytic
solution is found for the component with the lowest anomalous dimension
for all moments $N$, and the WKB-type expansion is constructed
for other levels, which becomes asymptotically exact at large $N$.
Evolution equations for the $\lambda=1/2$ distribution amplitudes
(e.g. for the nucleon) are studied as well. We find that the
two lowest anomalous dimensions for the  $\lambda=1/2$ operators
(one for each parity) are separated from the rest of the spectrum
by a finite  `mass gap'. These special states can be interpreted
as scalar diquarks.
}
\vskip1cm

\end{center}

\end{titlepage}

{\small \tableofcontents}

\section{Introduction}
\setcounter{equation}{0}

There exists a general consensus that exclusive processes involving
large momentum transfers are dominated by
`valence' components in hadron wave functions
with the minimum number of Fock constituents \cite{BLreport,CZreport}.
It is equally generally accepted that the asymptotic behavior of exclusive
amplitudes is in most cases determined by the so-called `hard-rescattering'
mechanism involving configurations of partons with almost zero
transverse separations, although the theoretical status of the
dominance of small transverse distances is somewhat weaker for baryons
\cite{earlybaryon} than for pions \cite{ER,earlypion}.

As always in a field theory, extraction of the asymptotic behavior
introduces divergences. In the present context, infrared divergences
in perturbative diagrams describing the hard rescattering are removed
by renormalization of nonperturbative scale-dependent
{\em distribution amplitudes} which are defined in terms of the
 Bethe-Salpeter wave functions integrating out transverse
degrees of freedom
\begin{equation}
 \phi(x_1,\ldots,x_n;\mu) \sim \int\limits^{|k_{i,\perp}|<\mu}
  \!\!\! d^2k_{1,\perp}\ldots d^2k_{n,\perp}
  \,\Phi_{BS}(x_1,k_{1,\perp};\ldots;x_n,k_{n,\perp})
\end{equation}
with $x_i$ being the longitudinal momentum fractions carried by partons.
The  concept of distribution amplitudes is central for the theory of
hard exclusive processes where their r\^ole is analogous to that of
more familiar parton distributions in the description of inclusive
processes.

The theoretical basis for studies of distribution amplitudes \cite{ER} is
provided by their definition in terms of hadron-to-vacuum
transition matrix elements of non-local gauge-invariant
light-cone operators of the type (in a suitably chosen gauge)
\begin{eqnarray}
  &&\bar q(z_1n) q(z_2n),
\nonumber\\
  &&\varepsilon_{ijk}q^i(z_1n)q^j(z_2n) q^k(z_3n)
\end{eqnarray}
for meson and baryon distributions, respectively. Here $q^i$ is a generic
quark field with the color $i$, $n$ is an auxiliary light-like vector $n^2=0$ and $z_i$
are real numbers that specify quark (antiquark) separations.
More specific definitions will be given below.
The nonlocal operators as above are understood as generating functionals
for the series of local operators obtained by their Tailor expansion
at short distances (contraction of the derivatives with the light-like
vector ensures taking the leading twist part) and the precise relation
is such that {\em moments} of distribution amplitudes are
given by matrix elements of the contributing local operators \cite{ER}.
The scale dependence of the moments of distribution
amplitudes corresponds to the renormalization group (RG) evolution of local
operators and can be studied using familiar methods.

The  specific problem for distribution amplitudes is to take into account
the additional mixing with operators containing total derivatives
that cannot be neglected in contrast to inclusive processes where
only forward matrix elements are being considered.
It was noticed that
the RG evolution is driven to leading logarithmic accuracy by tree-level
counterterms which thus have the symmetry of the bare QCD
Lagrangian and, in particular, the conformal symmetry.
As a consequence,
operators belonging to different
irreducible representations of the  conformal group cannot mix
under renormalization in one loop \cite{ER,B+,Makeenko,O82,Mueller}.
This observation
solves the mixing problem for meson (two-particle) distributions
in which case a single independent local conformal operator exists
for each moment.
The corresponding anomalous dimension can be continued analytically
to non-integer (complex) moments, defining the Altarelli-Parisi
evolution kernel: coefficients in the expansion of meson distributions
in the basis of Gegenbauer polynomials are renormalized multiplicatively and
with the same anomalous dimensions as in deep inelastic scattering \cite{ER}.
Consequently, assuming `good' behavior at complex infinities,
the distribution amplitude can be restored by inverse Mellin transform
from analytically continued values of the moments; hence the partonic
interpretation of the distribution amplitude proves to be consistent
with its renormalization properties (scale dependence).

The three-quark baryon distribution amplitudes bring in a complication
of principle. The conformal symmetry allows to resolve the mixing with
total derivatives but is not sufficient to diagonalize the mixing
matrix completely. For fixed operator dimension, alias for fixed total
number $N$ of covariant derivatives, $D_\mu=\partial_\mu-igA_\mu$,
there exist $N+1$ independent local operators (modulo operators with
the total derivatives)
\begin{equation}
 {\cal O}_{N,k} = q\, (\derleft\cdot n)^{k} q \,(n\cdot\!\derright)^{N-k} q,~~~
    k=0,\ldots,N
\end{equation}
corresponding to $N+1$ genuine independent degrees of freedom of the
three-quark system. One is left with a nontrivial $(N+1)\times(N+1)$
mixing matrix that has to be diagonalized explicitly order by order;
see, e.g., \cite{LB79,P79,T82,O82,Nyeo,Stefanis}. The resulting $N+1$
multiplicatively renormalizable operators have different (in general)
anomalous dimensions whose analytic expressions
are not known. Apart from mathematical incompleteness, absence
of analytic results means that the general structure of the spectrum
is unknown and, in particular, analytic continuation of the anomalous
dimensions to complex moments $N$ is not possible. This, in turn,
implies that partonic interpretation of different `components'
in baryons is not understood beyond the tree level.

This problem was well known but considered as a relatively
minor one and did not attract due attention in the past.
One reason was that the scale dependence of distribution amplitudes
turned out to be  rather mild in a perturbative domain and it seemed premature
to elaborate on the evolution before gross features of nonperturbative
distributions were understood at low scales.
We think that this logic is flawed and
the general experience with hard processes in QCD
rather suggests that `intrinsic' parton distributions
at scales of order 1 GeV cannot be viewed as purely
nonperturbative and disconnected from perturbative evolution.
Despite an obvious fact that perturbative calculations
cannot be made quantitative at low scales, there is increasing
evidence that general patterns of the perturbative gluon emission
are continued to very low momenta. For example, the  shape of deep inelastic
structure functions at 1 GeV appears to be largely determined by perturbative
soft gluon radiation. Small differences in the perturbative
evolution of different components in nucleon distribution amplitudes are
strongly amplified in the nonperturbative domain and one may think
that such differences build up  gross features of distribution
amplitudes at scales of order 1 GeV, where from the
perturbation theory becomes quantitative.
Viewed from this perspective, a detailed study of the evolution of baryon
distribution amplitudes becomes mandatory.

In this paper we suggest a new approach to the construction of baryon
distribution amplitudes which is based on the recent finding \cite{BDM}
that the evolution equation for the baryon distribution amplitudes with
maximum helicity $\lambda=3/2$ is completely integrable. That is
it possesses a nontrivial integral of motion which we identify as a new
`hidden' quantum number that distinguishes components in the $\lambda=3/2$
distribution amplitudes with different scale dependence.
It is interesting to note that the $\lambda=3/2$ evolution equation is
equivalent to the quantum mechanical problem that has already been
encountered in QCD in the studies of the Regge asymptotics of the
scattering amplitudes \cite{FK,Lip} and in the theory of integrable models
as the so-called Heisenberg XXX${}_{s=-1}$ spin magnet \cite{FK,TTF}. This
problem has
been studied in some detail using nontrivial mathematical methods and
the results can be adapted to the present context.

Our approach is advantageous compared to the standard formulation
in at least two aspects. First,  from practical point of view an important
simplification is that diagonalization of a $(N+1)\times (N+1)$ matrix
is replaced by solution of a simple three-term recurrence relation, which
reduces the computer time significantly. Second, we obtain explicit analytic
solutions to the evolution equations in all important limits. In particular,
we will be able to identify trajectories of the anomalous dimensions
and calculate them (and the corresponding eigenfunctions)
using a WKB type expansion for large values of $N$.

These results apply in full to the $\lambda=3/2$ distribution function
of the $\Delta$-resonance and allow for a fairly complete description.
The evolution equation for the $\lambda=1/2$ distributions is not exactly
solvable, but the difference to the $\lambda=3/2$ evolution can be
considered as a small (calculable)
perturbation for the most part of the spectrum. On the other hand,
the structure of the lowest eigenstates is changed drastically.
As we will demonstrate, the two lowest anomalous dimensions for the
$\lambda=1/2$ operators decouple from the rest of the spectrum and
are separated from it by a finite `mass gap'. These two special
states (one for each parity) can be interpreted as bound states
in the corresponding quantum-mechanical model, and, somewhat imprecisely,
can be thought of as corresponding to formation of scalar diquarks.

As a byproduct of our study, we construct  a convenient orthonormal basis
for the expansion of three-particle distribution amplitudes, which is
more suitable compared to  standard Appell polynomials.

The presentation is organized as follows.
Section 2 introduces definitions and the general framework
for the construction of  baryon distribution functions and
their renormalization.
Section 3 is devoted to the conformal symmetry of the evolution
equation and the conformal expansion of distribution amplitudes.
Section 4 presents a detailed study of the
exactly solvable evolution equation for the maximum helicity
$\lambda=3/2$. In Section 5 we consider the evolution
equation for the $\lambda=1/2$ distributions.  A short summary of main
results of phenomenological relevance is given in
Section 6 and the general conclusions in Section 7.
Appendix A presents an explicit construction for the Racah $6j-$coefficients
of the $SL(2)$ group and in Appendix B we consider the effective Hamiltonian
for low-frequency modes of the $\lambda=1/2$ evolution equation.

\section{General framework}
\subsection{Distribution amplitudes}
\setcounter{equation}{0}

Following Refs.~\cite{CZ84} we define the leading twist nucleon
distribution amplitude as the corresponding
matrix element of the gauge-invariant three-quark nonlocal operator
\begin{eqnarray}
\lefteqn{ \langle 0 | u_\alpha^{i'}(z_1)u_\beta^{j'}(z_2)d_\gamma^{k'}(z_3)
\, U_{i'i}(z_1,z_0)U_{j'j}(z_2,z_0)U_{k'k}(z_3,z_0)
\epsilon^{ijk}|P(p,\lambda)\rangle = }
\labeltest{defnuc}
\\
&& =\frac{f_N}{4}\left \{(\not\! pC)_{\alpha\beta}(\gamma_5 N)_\gamma V(z_i p)+
(\not\! p \gamma_5 C)_{\alpha\beta} N_\gamma A(z_i p)
+\left (i\sigma_{\mu\nu}p^\nu C\right)_{\alpha\beta} (\gamma_\mu\gamma_5 N)_\gamma T(z_ip)
\right \},\nonumber
\end{eqnarray}
where $\sigma_{\mu\nu}=\frac{i}{2}[\gamma_\mu,\gamma_\nu]$,
$C$ is the charge conjugation matrix, $|P(p,\lambda)\rangle $
is the proton state with momentum $p$ and helicity $\lambda$,
 and $N$ is the proton spinor.
All the interquark separations are assumed to be light-like,
e.g. $u(z_1)$ denotes the u-quark field at the space point $z_1n$
with $n^2=0$, and $U(z_n,z_0)$ are non-Abelian
phase factors (light-like Wilson lines)
\begin{equation}
   U(z_n,z_0) \equiv {\rm P\ exp\,}
   \left[ig\int_0^1\!dt\, (z_n-z_0)\,n_\mu A^\mu
   (tz_n+(1-t)z_0)\right].
\end{equation}
Because of the light-cone kinematics, the matrix element in fact does not
depend on $z_0$ and the phase factors can be eliminated by choosing
a suitable gauge. To save space we do not show the gauge phase factors
in what follows, but imply that they are always present.

The  invariant functions
$V,A,T$ have the following symmetry properties~\cite{CZ84}
\be
V(1,2,3)=V(2,1,3), \ \ \ \ A(1,2,3)=-A(2,1,3), \  \ \ \ \ T(1,2,3)=T(2,1,3).
\labeltest{12nuc}
\ee
and can be expressed in terms of a single function $\phi_N$  as
\begin{eqnarray}
2T(1,2,3)&=&\phi_N(1,3,2)+\phi_N(2,3,1),\nonumber\\
\phi_N(1,2,3)&=&V(1,2,3)-A(1,2,3).
\end{eqnarray}
Here $\phi_N$ is the leading twist proton distribution amplitude.
If it is presented in the form
\be
\phi_N(z_i p)\equiv
 \int\!{\cal D}x\, \exp \Big[-i\sum x_i z_i (p\cdot n)\Big] \phi_N(x_i),
\labeltest{xp}
\ee
where
\be
 \int\!{\cal D} x \equiv \int_0^1\! dx_1\,dx_2\,dx_3\,\delta
     (1-x_1-x_2-x_3),
\labeltest{Dx}
\ee
then the variables $x_i$ have the meaning of
the longitudinal momentum fractions carried by the three quarks in
the nucleon, $0\le x_i\le 1$ and $\sum x_i=1$.

For what follows,
it is convenient to introduce quark fields with definite chirality
\be
q^{\uparrow (\downarrow)}=\frac12 \left (1\pm \gamma_5\right) q.
\ee
The definition in (\ref{defnuc}) is equivalent to the following
form of the proton state~\cite{CZ84,FZOZ88}
\begin{equation}
|P(p,\lambda=+1/2) \rangle = f_N \int \frac{{\cal D}x\, \phi_N(x)}
                                    {2\sqrt{24x_1x_2x_3}}
\left \{
|u^\uparrow(x_1)u^\downarrow(x_2) d^\uparrow(x_3)\rangle
-
|u^\uparrow(x_1)d^\downarrow(x_2) u^\uparrow(x_3)\rangle \right \},
\end{equation}
where the standard relativistic normalization for the states and
Dirac spinors is implied~\cite{BD65}. The distribution amplitude
$\phi_N$ can be defined in terms of  chiral fields:
\begin{eqnarray}
\lefteqn{
\langle 0|\epsilon^{ijk}(u_i^\uparrow(z_1)\,C\!\!\not\!nu_j^\downarrow(z_2))
       \not\!nd_k^\uparrow(z_3)|P(p)\rangle =}
\nonumber\\&&{}\hspace*{0.5cm}=
    - \frac{1}{2}f_N (pn)\not\!n N^\uparrow(p) \int \!{\cal D}x\,
\exp\big[-ipn(z_1x_1+z_2x_2+z_3x_3)\big]\,\phi_N(x_i,\mu^2)
\labeltest{Nucleon}
\end{eqnarray}
so that moments of $\phi_N$
\be
 \phi_N(k_1,k_2,k_3)=
  \int\!{\cal D}x \,x_1^{k_1} x_2^{k_2} x_3^{k_3}\, \phi_N(x_1,x_2,x_3,\mu^2)
\ee
can be calculated as reduced matrix elements of the renormalized three-quark
leading twist operators
\be
O_{k_1,k_2,k_3}^{\uparrow\downarrow\uparrow}
 = (nD)^{k_1} u^{\uparrow}(0)(C\!\not\! n)
 (nD)^{k_2} u^{\downarrow}(0)
 (nD)^{k_3} \not\!n d^ {\uparrow}(0),
\ee
where $D_\mu=\partial_\mu-igA_\mu$ is a covariant derivative.

The leading twist distribution amplitudes of the $\Delta$ resonance
can be obtained in the similar manner~\cite{FZOZ88}.
For definiteness, we will consider the distribution amplitudes of
$\Delta^{++}$ only, all the other ones can be reconstructed with the help of
the isospin symmetry, see~\cite{FZOZ88}.
For this case one writes
\begin{eqnarray}
&& \langle 0 |
u_\alpha^i(z_1)u_\beta^j(z_2)u_\gamma^k(z_3) | \Delta^{++}(p,\lambda)\rangle
\epsilon^{ijk}= \\
&& =\frac{\lambda^{1/2}_\Delta}{4}\left \{(\gamma_\mu C)_{\alpha\beta}(\Delta)^\mu_\gamma
V_\Delta (z_i p)+
(\gamma_\mu \gamma_5 C)_{\alpha\beta} (\gamma_5 \Delta^\mu)_\gamma A_\Delta(z_i p)
\right.
\nonumber\\
&&-\left.
  \frac12 \left ( i\sigma_{\mu\nu}C\right)_{\alpha\beta} (\gamma_\mu \Delta^\nu)_\gamma
T_\Delta(z_ip)\right \}
-\frac14 f_\Delta^{3/2}
\left ( i\sigma_{\mu\nu}C\right)_{\alpha\beta}
\left (
p_\mu\Delta^\nu-\frac12M_\Delta\gamma_\mu\Delta^\nu\right )_\gamma\phi_\Delta^{3/2}(z_i p).
\nonumber
\labeltest{def2}
\end{eqnarray}
Here $\Delta_\gamma^\mu(p)$ is the $\Delta$ resonance spin$-\frac32$ vector:
\be
(\not\! p -M_\Delta)\Delta^\mu=0, \ \ \ \bar \Delta^\mu\Delta_\mu=-2M_\Delta, \ \ \
\gamma_\mu\Delta^\mu(p)=p_\mu\Delta^\mu(p)=0.
\ee
The dimensionless amplitudes
 $V_\Delta(x),A_\Delta(x),T_\Delta(x)$ determine the distribution
of quarks in the $\Delta(|\lambda|= 1/2)$ state and
 satisfy the following symmetry relations
\begin{eqnarray}
V_\Delta(1,2,3)&=&V_\Delta(2,1,3), \ \ \ \ A_\Delta(1,2,3)=-A_\Delta(2,1,3),
\nonumber \\
T_\Delta(1,2,3)&=&T_\Delta(2,1,3), \\
T_\Delta(1,2,3)&=&V_\Delta(2,3,1)-A_\Delta(2,3,1).
\nonumber
\labeltest{12del}
\end{eqnarray}
Therefore,  only one function is independent;
we choose (cf. \cite{FZOZ88})
\begin{equation}
  \phi_\Delta^{1/2}(x_1,x_2,x_3) \equiv
 V_\Delta(x_1,x_2,x_3)-A_\Delta(x_1,x_2,x_3)
  = T_\Delta(x_3,x_1,x_2)
\end{equation}
as the distribution amplitude of the $\Delta(|\lambda|= 1/2)$ resonance.
The remaining function $\phi_\Delta^{3/2}$ is totally symmetric in all
its  arguments and determines the distribution of quarks in the
$\Delta(|\lambda|= 3/2)$ state.

The structure, again,  becomes more transparent when going over to
chiral quark fields. The definition in Eq.~(\ref{def2})
is equivalent to the following structure of
the $\Delta$ resonance states~\cite{FZOZ88}
\begin{eqnarray}
|\Delta(\lambda=1/2) \rangle &=&f_\Delta^{1/2} \int {\cal D}x
\frac{\phi_\Delta^{1/2}(x_i)}{4\sqrt{24x_1 x_2 x_3}}
|u^\uparrow(x_1)u^\downarrow(x_2) u^\uparrow(x_3)\rangle , \nonumber \\
|\Delta(\lambda=3/2) \rangle &=&f_\Delta^{3/2} \int {\cal D}x
\frac{\phi_\Delta^{3/2}(x_i)}{6\sqrt{24x_1 x_2 x_3}}
|u^\uparrow(x_1)u^\uparrow(x_2) u^\uparrow(x_3)\rangle,
\end{eqnarray}
where $f_\Delta^{3/2}=\sqrt{\frac23}\lambda_\Delta^{1/2}/M_\Delta$,
and the distribution amplitudes can be defined through the
nonlocal matrix elements:
\begin{eqnarray}
\lefteqn{\hspace*{-1cm}
 \langle 0|\epsilon^{ijk}(u_i^\uparrow(z_1)\,C\!\!\not\!nu_j^\downarrow(z_2))
       \not\!nu_k^\uparrow(z_3)|\Delta(p)\rangle =}
\nonumber\\&&{}=
   -\frac{1}{2}\lambda_\Delta^{1/2}n^\mu (\!\not\!n\Delta)^\uparrow_\mu
\int \!{\cal D}x\,
\exp\big[-ipn(z_1x_1+z_2x_2+z_3x_3)\big]\,\phi_\Delta^{1/2}(x_i,\mu^2)
\labeltest{Delta1/2}
\end{eqnarray}
and
\begin{eqnarray}
\lefteqn{\hspace*{-1cm}
 \langle 0|\epsilon^{ijk}
(u_i^\uparrow(z_1)\,C\sigma_{\mu\nu}n_\nu u_j^\uparrow(z_2))
       (\bar\Delta^\mu\!\!\not\!n)\,u_k^\uparrow(z_3)
|\Delta(p)\rangle =}
\nonumber\\&&{}\hspace*{0.0cm}=
   {i}\, f_\Delta^{3/2} (pn)\bar\Delta^\uparrow_\nu \!\not\!n \Delta^{\uparrow,\nu}
\int \!{\cal D}x\,
\exp\big[-ipn(z_1x_1+z_2x_2+z_3x_3)\big]\,\phi_\Delta^{3/2}(x_i,\mu^2).
\labeltest{repr3/2}
\end{eqnarray}

\subsection{Renormalization}

In this paper we will be interested in the scale dependence of the baryon
distribution amplitudes. For each of them,
$\phi \equiv \phi_N,\phi_\Delta^{3/2},\phi_\Delta^{1/2}$,
we anticipate an expansion of the type
\be
\phi(x_i,\mu^2)=x_1 x_2 x_3 \sum_n \phi_n\, P_n(x_i)
\left (\frac{
\alpha_s(\mu)}{\alpha_s(\mu_0)}
\right )^{\gamma_n/b_0},
\labeltest{expansion}
\ee
where $b_0=11/3 N_c-2/3n_f$,
$P_n(x_i)$ are certain polynomials,
$\gamma_n$ are the corresponding anomalous dimensions
and $\phi_n(\mu_0)$ are dimensionless nonperturbative parameters.
The prefactor $x_1 x_2 x_3$ suggests the vanishing of the
distribution amplitude  at the end points $x_k=0$ and as we will show
its presence is closely related to the conformal invariance of
the evolution equations.
Finding $\gamma_n$ and $P_n$ corresponds to explicit diagonalization of
the mixing matrix of the three-quark composite operators and is
fully equivalent to the solution of the corresponding Brodsky-Lepage
equations~\cite{earlybaryon}.

Renormalization properties of the relevant three-quark
operators are most  conveniently presented in terms  of
their generating functionals (nonlocal operators) with three
spinor indices~\cite{BB88}:
\begin{eqnarray}
B^{3/2}_{\alpha\beta\gamma}(z_1,z_2,z_3) &=&
\varepsilon^{ijk}(\!\not\!n q_{i}^\uparrow)_\alpha(z_1n)
(\!\not\!n q_{j}^\uparrow)_\beta(z_2n)
(\!\not\!n q_{k}^\uparrow)_\gamma(z_3n),
\labeltest{B3/2}
\\
B^{1/2}_{\alpha\beta\gamma}(z_1,z_2,z_3) &=&
\varepsilon^{ijk}(\!\not\!n q_{i}^\uparrow)_\alpha(z_1n)
(\!\not\!n q_{j}^\downarrow)_\beta(z_2n)
(\!\not\!n q_{k}^\uparrow)_\gamma(z_3n),
\labeltest{B1/2}
\end{eqnarray}
with $q_i$ being a quark field of color $i$.
$B_{3/2}$ gives rise to the
distribution amplitude $\phi_\Delta^{3/2}$ and $B_{1/2}$ is relevant
both for $\phi_N^{1/2}$ and $\phi_\Delta^{1/2}$.
The nonlocal operators $B_{3/2}$ and $B_{1/2}$ do not mix with each other
since they  belong to different representations of the Lorentz group:
$(3/2,0)$ and $(1,1/2)$, respectively\footnote{
The transformation properties can be made manifest by going over
to the two-component spinors and $\gamma$-matrices in the
Weyl representation \cite{Nyeo}.}.
For most of the discussion
we will assume that all three quarks $q$ have different flavor.
Identity of the quarks does  not influence renormalization but
rather introduces certain selection rules
which pick up the eigenstates with particular symmetry, to be detailed
later.

The renormalization group equation for the nonlocal operators (\ref{B3/2}),
(\ref{B1/2}) can be written as~\cite{BFLK,BB88}
\begin{equation}
 \left\{\mu\,\frac{\partial}{\partial \mu}
         +\beta(g)\,\frac{\partial}{\partial g}\right\}B = {\Bbb H}\cdot B,
\labeltest{RG}
\end{equation}
where ${\Bbb H}$ is some integral operator corresponding, to the one-loop
accuracy, to contributions  of the Feynman diagrams shown in
Fig.~\ref{figure1}.

\setlength{\unitlength}{0.7mm}
\begin{figure}[t]
\vspace{5.3cm}
\begin{picture}(120,200)(0,1)
\mbox{\epsfxsize16.0cm\epsffile{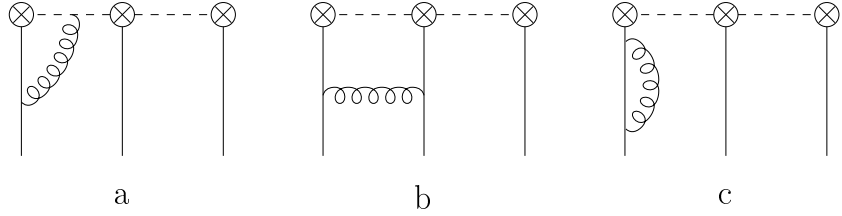}}
\end{picture}
\vspace*{-15.0cm}
\caption{\labeltest{QCD96fig1}
\small Examples of a `vertex' correction (a), 'exchange' diagram (b) and
self-energy insertion (c) contributing to the renormalization of
three-quark operators in Feynman gauge. Path-ordered gauge factors
are shown by the dashed lines. The set of all diagrams includes
possible permutations.
}
\label{figure1}
\end{figure}

To simplify notations, we factor out the color factors and trivial
contributions of the self-energy insertions:
\begin{equation}
  {\Bbb H} = (1+1/N_c){\cal H}+3C_F/2,
\end{equation}
with $C_F=(N_c^2-1)/(2N_c)$.
It is easy to see that the gluon exchange diagram in Fig.~\ref{figure1}b
vanishes unless
the participating quarks have opposite chirality. The renormalization
of the $\lambda=3/2$ operator $B_{3/2}$ (\ref{B3/2})
is therefore determined by the vertex correction
in Fig.~\ref{figure1}a alone (in Feynman gauge).
By explicit calculation one finds~\cite{LB79,BFLK,P79,T82,Nyeo}
\be
{\cal H}_{3/2}={\cal H}_{12}^{v} +{\cal H}_{23}^{v}+{\cal H}_{13}^{v},
\labeltest{H32}
\ee
where ${\cal H}_{ik}^{v}$ are the two-particle kernels involving the
$i$-th and $k$-th quarks, for example,
\begin{eqnarray}
{\cal H}_{12}^{v}\,B(z_i)&=&
-\int_{0}^{1}\frac{d\alpha}{\alpha}\Big \{
{\bar\alpha}
\left [
B(z_{12}^\alpha,z_2,z_3)- B(z_1,z_2,z_3)\right]
  \nonumber\\
&&{}\hspace*{1cm}
+{\bar\alpha}\!\left [
B(z_1,z_{21}^\alpha,z_3)
-B(z_1,z_2,z_3)
\right ] \! \Big \},
\labeltest{H32-part}
\end{eqnarray}
with $\bar \alpha \equiv 1-\alpha$
and $z_{ik}^\alpha\equiv z_{i}\bar\alpha+z_{k}\alpha$.

In the case of $B_{1/2}$ the
vertex correction remains the same, but one has to add
contributions of gluon exchange between the quarks with opposite chirality.
One obtains
\be
{\cal H}_{1/2}={\cal H}_{3/2} - {\cal H}_{12}^e - {\cal H}_{23}^e,
\labeltest{H12}
\ee
where we assume that the first and the third quark have the same chirality,
as in (\ref{B1/2}).
The kernels $H^e_{ik}$ act on $i$-th and $k$-th arguments of
the nonlocal operators only, and can be written in the form
\begin{equation}
{\cal H}_{12}^{e}\,B(z_i)=
\int \!{\cal D}\alpha\,
B(z_{12}^{\alpha_{1}},z_{21}^{\alpha_{2}},z_3)\,,
\labeltest{H12-part}
\end{equation}
with the integration measure ${\cal D}\alpha$ defined in (\ref{Dx}).

Going over to local operators corresponds to the Taylor expansion
of the generating functionals at small distances:
\be
B(z_1,z_2,z_3) = \sum_N\sum_{k_1+k_2+k_3=N}\frac{z_1^{k_1}}{k_1!}
                \frac{z_2^{k_2}}{k_2!}
                \frac{z_3^{k_3}}{k_3!}
 (nD)^{k_1} q(0) (nD)^{k_2} q(0) (nD)^{k_3} q(0).
\labeltest{Tailor}
\ee
The total number of derivatives $N$ is preserved by the evolution
so that the integro-differential  equation (\ref{RG})
takes the matrix form, with the square matrix of size $N(N+1)/2$
for each given $N$ subsector.

A generic local operator with $N$ derivatives can be written as sum of
monomials entering the expansion (\ref{Tailor})
with arbitrary coefficients
\be
{\cal O}=\sum_{k_1+k_2+k_3=N}c_{k_1,k_2,k_3}
(nD)^{k_1} q(0)(nD)^{k_2} q(0) (nD)^{k_3} q(0),
\labeltest{operF}
\ee
and can be represented by a polynomial in three variables
\be
{\Psi(x_1,x_2,x_3)}=\sum_{k_1+k_2+k_3=N}c_{k_1k_2k_3}
x_1^{k_1} x_2^{k_2} x_3^{k_3}.
\labeltest{coefF}
\ee
In what follows we refer to $\Psi(x_i)$ as coefficient function
of a local operator. To justify the name,
note that $\Psi(x_i)$ serves as a projector
separating the contribution of the local operator $O_\Psi$
to the nonlocal operator $B(z_i)$, which can be made explicit
by writing
\begin{equation}
{\cal O}_\Psi=\Psi(\partial_1,\partial_2,\partial_3)B(z_1,z_2,z_3)|_{z_i=0}.
\labeltest{PsiOp}
\end{equation}

Local operators having the same number of derivatives  all
mix together so that the size of the mixing matrix for given $N$ is
$N(N+1)/2$.
Since a local operator is completely determined by its coefficient
function, diagonalization of the mixing matrix for operators
can be reformulated as diagonalization
of the mixing matrix for the coefficient functions.
Requiring that ${\cal O}_\Psi$ (\ref{PsiOp}) is
 multiplicatively renormalized,
one ends up  with  a matrix equation in the space
of homogeneous polynomials of degree $N$ of three variables%
\footnote{Notice that the action of the evolution kernel
${\cal H}$ on the space of the coefficient functions is
different from that on the nonlocal operator $B(z_i)$}
\begin{equation}
    {\cal H}\cdot \Psi_{N,q}  =  {\cal E}_{N,q} \Psi_{N,q}\,,
\labeltest{Sch}
\end{equation}
whose eigenvalues correspond to the anomalous dimensions
\begin{equation}
  \gamma_{N,q} \equiv (1+1/N_c)\,{\cal E}_{N,q} +3/2\, C_F.
\label{anomal}
\end{equation}
Note that the eigenfunctions and the eigenvalues have two indices:
$N$ which refers to the degree of polynomial alias the total number
of derivatives, and $q$ which enumerates the energy levels. In the
case of ${\cal H}_{3/2}$ we will later identify $q$ with a conserved
charge.
The $SL(2)$ symmetry of the equation (\ref{Sch})
(see below) implies that the anomalous dimensions take real
quantized values and
the corresponding
eigenfunctions are mutually orthogonal with the weight function
$x_1x_2x_3$
\begin{equation}
 \int {\cal D}x\, x_1 x_2 x_3 \,\Psi_{N,q}(x_i)\Psi_{N,q'}(x_i)
 \sim \delta_{q,q'}
\,.
\labeltest{aaa}
\end{equation}
The same property allows to identify the eigenfunctions as
the polynomials
entering the expansion of the distribution amplitudes in (\ref{expansion})
\begin{equation}
     P_{n}(x_i)\equiv \Psi_{N,q}(x_i).
\labeltest{Pn}
\end{equation}
To see this, consider the matrix element
$\langle 0| {\cal O}_\Psi |B\rangle=
\Psi(\partial_1,\partial_2,\partial_3)
\langle 0| B(z_1,z_2,z_3)|B\rangle|_{z_i=0}$  and
use the representation similar to (\ref{repr3/2}) to get
\begin{eqnarray}
 \langle 0| {\cal O}_\Psi |B(p)\rangle &\sim &
\int \!{\cal D}x\, \left[\Psi(\partial_1,\partial_2,\partial_3)\,
{\rm e}^{-ipn(z_1x_1+z_2x_2+z_3x_3)}\right]_{z_i=0}\,\phi_B(x_i,\mu^2)
\nonumber\\
&=& (-ipn)^N
\int \!{\cal D}x\, \Psi(x_1,x_2,x_3)\,\phi_B(x_i,\mu^2).
\end{eqnarray}
Substituting the expansion (\ref{expansion}) into this relation
and taking into account that the operator $O_\Psi$ is renormalized
multiplicatively (by construction), one immediately finds that,
first, the polynomials $P_n$ have to coincide with $\Psi_{N,q}$
up to arbitrary normalization and, second, the nonperturbative
coefficient $\phi_n$ is given by the reduced matrix element
of the operator ${\cal O}_\Psi$.

Note that the `Hamiltonians' ${\cal H}_{3/2}$ and  ${\cal H}_{1/2}$
acting in the space of coefficient functions in (\ref{Sch})
are not the same as those acting on nonlocal operators, although
they are related, of course, and the precise
connection can easily be established.
Explicit expressions for ${\cal H}_{3/2}$ and  ${\cal H}_{1/2}$
in the matrix representation
can be found in~\cite{LB79,T82,O82,Nyeo}.

\section{Conformal invariance}
\setcounter{equation}{0}
The Lagrangian of massless QCD is known to be invariant under conformal
transformations. This symmetry survives for evolution equations
at one-loop level since
breaking of the conformal Ward identities induced by the nonzero trace of the
stress-energy tensor is proportional to the QCD $\beta$ function
and is of order $\alpha_s^2$~\cite{O82,Mueller}. One should expect, therefore,
that the evolution operator ${\cal H}$ introduced in the previous section
has the same symmetry and in particular commutes with the generators
of the conformal group. This property imposes strong constraints
on a possible form of the eigenfunctions: In a generic
situation the eigenfunctions of two-particle operators are uniquely determined
by conformal invariance whereas for three-particle operators one is
left with an arbitrary function of one variable.
Aim of this section is to work out the necessary framework.

\subsection{Collinear subgroup ${\it SL}(2,{\Bbb R})$ of the conformal group}

Algebra of the full conformal group contains the generators of dilatations
$\bf{D}$ and special conformal transformations ${\bf K}_\mu$
in addition to the Poincare generators ${\bf P}_\mu$ and ${\bf M}_{\mu\nu}$.
The algebra reads
\begin{eqnarray}
 &&[{\bf D},{\bf K}_\mu] =i{\bf K}_\mu,~~~
   [{\bf K}_\mu,{\bf P}_\nu] =-2i(g_{\mu\nu}{\bf D}+{\bf M}_{\mu\nu}),~~~
 [{\bf D},{\bf M}_{\mu\nu}] = 0,
\nonumber\\
 &&[{\bf D},{\bf P}_\mu] =-i{\bf P}_\mu,~~~
 [{\bf K}_\rho,{\bf M}_{\mu\nu}] =i(g_{\rho\mu}{\bf K}_\nu-
                                    g_{\rho\nu}{\bf K}_\mu),~~~
 [{\bf K}_\mu,{\bf K}_{\nu}] = 0,
\label{comrel}
\end{eqnarray}
plus usual relations for Poincare generators.
Action of these generators
on an arbitrary quantum field $\Phi$ (e.g. quark or gluon) is given
by (see, e.g.~\cite{MS69,O82})
\begin{eqnarray}
&&\left[
{\bf P}_\mu,\Phi(z)
\right]=-i\partial_\mu \Phi(z), \ \ \ \
\left[
{\bf M}_{\mu\nu}\,,\Phi(z)
\right]=\left[
i(z_\nu\partial_\mu-z_\mu \partial_\nu)-{\bf \Sigma}_{\mu\nu}
\right]\Phi(z), \nonumber\\
&&\left[
{\bf D}, \Phi(z)
\right]=-i(z^\nu \partial_\nu+l)\Phi(z)\,,\nonumber\\
&&
\left[
{\bf K}_\mu,\Phi(z)
\right]=
-i\left(
2z_\mu z^\nu \partial_\nu-z^2\partial_\mu+2 l z_\mu -2iz^\nu
{\bf \Sigma}_{\mu\nu}
\right)
\Phi(z)\,.
\labeltest{Cgroup1}
\end{eqnarray}
Here $l$ is the canonical dimension of $\Phi$ ($l=3/2$ for quarks) and
${\bf \Sigma}_{\mu\nu}$ stands for the spin part of the
angular momentum operator.
For a  quark field $\Phi(z)\equiv q(z)$
\begin{equation}
{\bf \Sigma}_{\mu\nu}q\equiv-\frac12 \sigma_{\mu\nu}q.
\end{equation}
In this paper we will be interested in the conformal transformations
for the fields `living' on the light-cone
\begin{equation}
\Phi(z)\equiv\Phi(z n_\mu)\,, \qquad n^2=0\,.
\end{equation}
One can check that the
only remaining nontrivial generators are ${\bf P}_{+},\> {\bf D},\>
{\bf M}_{+-}$ and ${\bf K}_{-}$,
where `$+$' and `$-$' stand for the projection on $n_\mu$
and on the alternative light-like vector $\bar n $, $\bar n n =1$,
respectively.
We will further assume that the field $\Phi$ is chosen to be
an eigenstate of the spin operator $\Sigma_{-+}$, that is it
has fixed spin projection $s$ on the `+' direction\footnote{
This property is automatically satisfied for
leading twist operators which correspond to the maximum spin projection;
in the general case one should use suitable projection operators to separate
different spin components, see e.g. \cite{BF90}.}
\begin{equation}
 \Sigma_{-+}\,\Phi = i\,s\, \Phi.
\end{equation}
For the leading-twist quark operators (\ref{B3/2}), (\ref{B1/2}) $s=+1/2$
for each of the three quarks since $\Sigma_{-+}(\!\not\!n q) =
+i (\!\not\!n q)/2$.

To bring the commutation relations (\ref{comrel}) to the standard form
it is convenient to consider  the following linear combinations:
\begin{eqnarray}
{\bf L}_{+}&=&{\bf L}_1-i{\bf L}_2=\frac{i}{2}{\bf K}_{-}, \ \ \ \ \ \ \
{\bf L}_{-}={\bf L}_1+i{\bf L}_2=-i{\bf P}_{+},\nonumber\\
{\bf L}_0&=&\frac{i}{2}({\bf D}+{\bf M}_{-+}), \ \ \ \ \ \ \
{\bf E}=\frac{i}{2}({\bf D}-{\bf M}_{-+}).
\labeltest{generators}
\end{eqnarray}
The operators ${\bf L}_i$ form the so-called collinear subalgebra $SL(2,{\Bbb R})$
of the conformal algebra:
\be
\left[{\bf L}_{+},{\bf L}_{-} \right]=2{\bf L}_0,\ \ \ \ \
\left[ {\bf L}_{0},{\bf L}_{+} \right]={\bf L}_{+},\ \ \ \ \
\left[{\bf L}_{0},{\bf L}_{-}\right]=-{\bf L}_{-}.
\labeltest{algebra}
\ee
Most importantly, action of the group generators (\ref{generators})
on quantum fields $\Phi$ (which is derived from (\ref{Cgroup1}) by  simple
algebra) can be replaced by differential operators
acting on the field coordinates and satisfying the same $SL(2)$
commutation relations:\footnote{
Note that we use boldface letters for the generators acting on quantum fields
to distinguish from the corresponding differential operators acting on the
field coordinates.}
\ba
{} [ {\bf L}_-,\Phi(z) ]  &=& -\frac{d}{dz} \Phi(z)\equiv L_{-} \Phi(z)\,,
\nonumber
\\
{} [ {\bf L}_+,\Phi(z) ] &=& \left(z^2\frac{d}{dz}+2\,j\, z\right) \Phi(z)
\equiv L_{+} \Phi(z)\,,
\labeltest{SL2-field}
\\
{} [ {\bf L}_0,\Phi(z) ] &=& \left(z\frac{d}{dz}+j\right) \Phi(z)
\equiv L_{0} \Phi(z)\,.
\nonumber
\ea
A one-particle operator $\Phi(z)$ is an eigenstate of the quadratic
Casimir operator
\begin{equation}
  {\bf L}^2 = {\bf L}_0^2+{\bf L}_1^2+{\bf L}_2^2 =
{\bf L}_0^2 -{\bf L}_0 +{\bf L}_+{\bf L}_-,~~~~~[{\bf L}^2,{\bf L}_\alpha]=0,
\labeltest{L1}
\end{equation}
\begin{equation}
 [{\bf L}^2,\Phi(z)] = j(j-1)\Phi(z),
\labeltest{cspin1}
\end{equation}
where
\begin{equation}
  j=\frac{1}{2}(l+s).
\labeltest{cspin2}
\end{equation}
We will refer to $j$ as {\em conformal spin} of $\Phi$ in
what follows.
The remaining generator  ${\bf E}$ counts the twist $t=l-s$ of $\Phi$:
\be
\left[{\bf E},\Phi(z)\right]=\frac12(l-s)\Phi(z).
\ee
It commutes with all ${\bf L}_i$ and is not relevant
for further discussion.

It is helpful to have in mind that the operators $L_\alpha$ generate the
projective (M\"obius) transformations on the line in the `+' direction
on the light-cone:
\begin{eqnarray}
z&\to& z'=\frac{az+b}{cz+d}\,,\qquad ad-bc=1\,,
\nonumber\\
  \Phi(z)&\to& \Phi'(z)=(cz+d)^{-2j}\Phi\left(\frac{az+b}{cz+d}\right)
\labeltest{Moebius}
\end{eqnarray}
with $a,b,c,d$ real.
The collinear conformal transformations of the three-quark  operators
$B(z_1,z_2,z_3)$ defined in (\ref{B3/2}) and (\ref{B1/2})
correspond to independent transformations
(\ref{Moebius}) for each of the fields; the group generators
are given  by the sum of one-particle generators acting
on light-cone coordinates of the quarks:
\begin{equation}
    L_\alpha B(z_1,z_2,z_3) =
     (L_{1,\alpha}+L_{2,\alpha}+L_{3,\alpha})B(z_1,z_2,z_3),
\labeltest{L-alpha}
\end{equation}
where $\alpha =\{0,+,-\}$ and $L_{k,\alpha}$ is the differential
operator (\ref{SL2-field}) acting on the argument of the $k-$th
quark $z_k$, $k=\{1,2,3\}$.
For further use, we introduce two- and three-quark Casimir operators
\begin{eqnarray}
   L^2_{ik} &=& \sum_{\alpha=0,1,2}(L_{i,\alpha}+L_{k,\alpha})^2,
  ~~~~i,k = 1,2,3,
\nonumber\\
   L^2_{} &=& \sum_{\alpha=0,1,2}
                 (L_{1,\alpha}+L_{2,\alpha}+L_{3,\alpha})^2
\nonumber\\
&=& L^2_{12}+L^2_{23}+L^2_{31}-j_1(j_1-1)-j_2(j_2-1)-j_3(j_3-1).
\end{eqnarray}
The last three terms in the last line vanish for quark fields for which
$j=1$,
see~(\ref{cspin2}). The two-particle Casimir operators $L^2_{ik}$
can be written (for quarks) as
\begin{equation}
    L^2_{ik} = - \partial_i\partial_k z_{ik}^2,
\end{equation}
where $\partial_k \equiv \partial/\partial z_k$ and $z_{ik}=z_i-z_k$.
Obviously  $[L_{ik}^2,L^2_{}]=0$.

\subsection{Brodsky-Lepage equations in the $SL(2)$ covariant form}

The expected conformal invariance of the evolution equation for baryonic
operators implies that the two-particle kernels ${\cal H}_{ik}$
commute with the generators of the ${\it SL}(2)$ transformations $L_\alpha$
defined in ({\ref{L-alpha}) and (\ref{SL2-field}).
To show this, consider the following expression
that generalizes both  \re{H32-part} and \re{H12-part}:
\be
{\cal H}_{12} B(z_1,z_2,z_3) =
\int {\cal D}\alpha\, \omega(\alpha_1,\alpha_2)
B(z_1-\alpha_1 z_{12},z_2+\alpha_2 z_{12},z_3),
\labeltest{ansatz}
\ee
where $z_{12}=z_1-z_2$ and the integration measure was defined in (\ref{Dx}).
This operator has a simple meaning --- acting on
the three-particle nonlocal operator $B(z_1,z_2,z_3)$ it displaces
the quarks with the coordinates $z_1$ and $z_2$ on the light-cone in
the direction of each other.

It easy to see that for this ansatz $[{\cal H}_{12},L_-] =
[{\cal H}_{12},L_0]=0$ for an arbitrary function
$\omega(\alpha_1,\alpha_2)$, whereas the condition
$[{\cal H}_{12},L_+]=0$ leads to the following constraint:
\begin{equation}
\left(\frac{\partial}{\partial \alpha_1} \alpha_1\bar\alpha_1
+2\alpha_1\,j_1\right) \omega(\alpha_1,\alpha_2)
=
\left(\frac{\partial}{\partial \alpha_2} \alpha_2\bar\alpha_2
+2\alpha_2\,j_2\right) \omega(\alpha_1,\alpha_2),
\end{equation}
where $\bar\alpha=1-\alpha$. Its general solution has the form
\begin{equation}
\omega(\alpha_1,\alpha_2) = \bar\alpha_1^{2j_1-2}
\bar\alpha_2^{2j_2-2} \varphi\left(\frac{\alpha_1\alpha_2}
{\bar\alpha_1\bar\alpha_2}\right),
\end{equation}
with an arbitrary $\varphi$. However, remembering that the
function $\varphi$ should result from the calculation of the
one-loop diagrams shown in Fig.~\ref{figure1} and must lead to nonsingular
(bounded) operator ${\cal H}_{12}$, one may conclude that the form
of $\varphi$ is almost uniquely fixed.
Note that $j_k=1$ for all the three quark fields entering (\ref{B3/2}) and
(\ref{B1/2}).
Then, notice that the gluon exchange between quarks in Fig.~\ref{figure1}b
amounts to the displacement of the two participating quarks along the light-cone
and the function $\varphi$ must have a smooth behavior
around $\alpha_1=\alpha_2=0$ and $\bar\alpha_1=\bar\alpha_2=0$.
These conditions leave us with the only choice
$\varphi(x)=1$ and its substitution into \re{ansatz} yields indeed
the kernel \re{H12-part}. In a similar way,
the `vertex correction' in
Fig.~\ref{figure1}a obviously corresponds to the displacement of just one of the quark
operators and this leads to the second structure $\varphi(x)=\delta(x)$
which reproduces the two-particle kernel \re{H32-part}.
\footnote{The second possible candidate $\varphi(x)=\delta(1-x)$
is ruled out since for $\alpha_1+\alpha_2\rightarrow 1$
it gives rise to the operator $B(z_1-\alpha_1 z_{12},z_1-(1-\alpha_2)z_{12},z_3)$
which becomes local in two quark fields. Such `contact interaction' terms
possess additional UV singularities
and are not expected to appear.}

Once conformal symmetry of the two-particle kernels is established,
the group theory tells  that ${\cal H}_{ik}$ may only depend on the
corresponding two-particle Casimir operators $L^2_{ik}$. To find the functional
form of this dependence, one has to compare
their action on a suitable
basis of trial functions.
The trick which we use below is general,
and the calculation presents an example of the  use of the `dual basis'
which is elaborated later in Sect.~3.5.

For definiteness, let us find  ${\cal H}_{12}$
as a function of $L^2_{12}$. To this end, it is enough to compare their
action on the homogeneous polynomials of two variables $z_1$ and $z_2$:
$$
B(z_1,z_2,z_3) \longrightarrow b_n(z_1,z_2)
$$
which we choose to be eigenfunctions of the operator
$L_{12}^2=-\partial_1\partial_2z_{12}^2$.

It is easy to see that
the thus defined
polynomials form an (infinite-dimensional) representation of the $SL(2)$
group on which
the operators $L_+\equiv L_{1,+}+L_{2,+}$
and $L_-\equiv L_{1,-}+L_{2,-}$ act as
rising and lowering operators, respectively.
It is thus sufficient
to consider only the functions (polynomials) annihilated by $L_-$, or
equivalently the highest weight of the representation, since
all other eigenfunctions of $L_{12}^2$ can then be obtained by a repeated
application of $L_+$. Since $L_- = -(\partial_1+\partial_2)$ the latter
condition is simply the
translation invariance which leaves one with  
\begin{equation}
 b(z_1,z_2) = (z_1-z_2)^n \equiv z_{12}^n,~~~ n=0,1,2,\ldots
\end{equation}
An explicit calculation gives
\begin{eqnarray}
  L_{12}^2 z_{12}^n &=& (n+2)(n+1)z_{12}^n,
\nonumber\\
  {\cal H}_{12}^{v}z_{12}^n &=& 2[\psi(n+2)-\psi(2)] z_{12}^n,
\nonumber\\
  {\cal H}_{12}^{e}z_{12}^n &=& 1/[(n+2)(n+1)] z_{12}^n,
\labeltest{H-L}
\end{eqnarray}
where $\psi(x)=d\ln\Gamma(x)/{dx}$ is the Euler $\psi$-function.
To cast (\ref{H-L}) in an operator form, define $J_{12}$
as a formal solution of the operator relation\footnote{Since
(\ref{def:J12}) is invariant under the substitution
$J_{12}\to 1-J_{12}$ one has specify which one of the two
formal solutions of (\ref{def:J12}) to choose; the simplest way to fix
the solution is to take the one with larger eigenvalues.}
\begin{equation}
  L_{12}^2 = J_{12}(J_{12}-1).
\labeltest{def:J12}
\end{equation}
The eigenvalues of $J_{12}$ equal $j_{12} = n+2$ and specify
the possible values of the sum of two $j=1$ conformal spins of
quarks in the $(12)$-pair, cf. (\ref{cspin1}).
Then
\begin{eqnarray}
  {\cal H}_{12}^{v} &=& 2[\psi(J_{12})-\psi(2)],
\nonumber\\
  {\cal H}_{12}^{e} &=& 1/[J_{12}(J_{12}-1)] = 1/L^2_{12}.
\labeltest{H-SL2}
\end{eqnarray}
Substituting the representation  \re{H-SL2} into \re{H32} and \re{H12} one
obtains the Schr\"odinger equation \re{Sch} for the three particles
on the (light-cone) line with the coordinates $z_1$, $z_2$ and $z_3$.
The `Hamiltonians' ${\cal H}_{3/2}$ and ${\cal H}_{1/2}$ entering this
equation for different baryon states have a pairwise structure and
are expressed in terms of the corresponding two-particle Casimir
operators \re{H-SL2}.
Furthermore, as we will elaborate in the next section,
the Hamiltonian ${\cal H}_{3/2}$ possesses an additional `hidden' symmetry:
One can construct an integral of motion (conserved charge) that commutes
with ${\cal H}_{3/2}$ and with the $SL(2)$ generators:
\begin{eqnarray}
  Q &=& \frac{i}{2}[L_{12}^2,L_{23}^2]
= i^3 \partial_1\partial_2\partial_3 z_{12} z_{23} z_{31}\,,
\nonumber\\
&&[Q, L_\alpha] = [Q, {\cal H}_{3/2}] =0.
\labeltest{Q3}
\end{eqnarray}
Its presence makes the corresponding Schr\"odinger equation completely
integrable and allows us to calculate the spectrum of
the anomalous dimensions analytically by applying a powerful technique
of integrable models.
The commutativity  $[Q, {\cal H}_{3/2}] =0$ is a consequence of the
commutation relations between $Q$ and two-particle Hamiltonians
\begin{equation}
 [{\cal H}_{12}^v,Q] =i(L_{23}^2-L_{31}^2)\,, \quad
 [{\cal H}_{23}^v,Q] =i(L_{31}^2-L_{12}^2)\,, \quad
 [{\cal H}_{31}^v,Q] =i(L_{12}^2-L_{23}^2)\,.
\labeltest{Q3com}
\end{equation}
The easiest way to prove these operator identities is
to calculate both sides using the conformal basis of functions
introduced below in Sect.~3.4 (see Eqs.~(\ref{threediagonal}) and
(\ref{Q-matrix})).

\subsection{Conformal symmetry of the eigenfunctions}

Equations~(\ref{H-SL2})
define the Brodsky-Lepage evolution kernels in the most
general form, independent on the representation.
The particular choice of the $SL(2)$ generators
(\ref{SL2-field}) corresponds to the evolution of the
nonlocal operator $B(z_1,z_2,z_3)$. As we have argued in Sect.~2.2,
diagonalization of the evolution equation for baryon distribution
amplitudes rather involves solution of the corresponding Schr\"odinger
equation for the local operators, or, equivalently, their
coefficient functions.
This corresponds, formally, to going over to a different
representation, and it is important to realize that
the action of the generators of the collinear conformal group on the
elementary fields and on the coefficient functions
of local operators defined through Eq.~(\ref{PsiOp}) is not the same.
By requiring
\be
[\widehat L_{\pm,0}\Psi(\partial_1,\partial_2,\partial_3)] B(z_1,z_2,z_3)
\bigg|_{z_i=0}
 \equiv
\Psi(\partial_1,\partial_2,\partial_3))[L_{\mp,0} B(z_1,z_2,z_3)]
\bigg|_{z_i=0}
\ee
one finds  the following `adjoint' representation of the generators
acting on the  space of coefficient functions $\Psi(x_1,x_2,x_3)$:
\begin{eqnarray}
\widehat L_{k,0}\Psi(x_i) &=& \left(x_k\partial_k +1\right) \Psi(x_i)\,,\quad
\nonumber\\
\widehat L_{k,+}\Psi(x_i) &=& -x_k \Psi(x_i)\,,\quad
\nonumber\\
\widehat L_{k,-}\Psi(x_i) &=&
\left(x_k\partial_k^2 +2\partial_k\right)\Psi(x_i),
\labeltest{adjoint}
\end{eqnarray}
where, in  order to maintain the same commutation relations (\ref{algebra}),
we have defined $\widehat L_-$ as the adjoint to $L_+$, and vice versa.
To simplify the notations, in what follows we drop the `hat' from
the generators in the adjoint representation, which, hopefully,
will not yield confusion.
Thus, the two-particle Hamiltonians entering the Schr\"odinger equation
for the coefficient functions, (\ref{Sch}), are given by the same operator expressions
(\ref{H-SL2}) but with the $SL(2)$ generators defined by (\ref{adjoint}).

As usual in quantum mechanics, symmetry of the Hamiltonian implies
that the eigenstates are degenerate: applying the $SL(2)$
generators to a particular eigenstate $\Psi_{N,q}$ one arrives at
a yet another eigenstate with the same value of energy ${\cal E}_{N,q}$.
It is then natural to parameterize the eigenstates $\Psi_{N,q}$
by a complete set of mutually commuting conserved charges.
The conformal symmetry allows to identify two such quantum numbers:
the total conformal spin, $L^2$, and its projection, $L_0$, which are
common to both Hamiltonians ${\cal H}_{1/2}$ and ${\cal H}_{3/2}$.

The construction of conformal eigenstates is fully analogous to
the construction of the eigenstates of angular momentum in standard
textbooks on quantum mechanics, with the $O(3)$ symmetry replaced
by $SL(2)$. We require that $\Psi(x_i)$ should
diagonalize simultaneously two integrals of motion
\be
L^2 \Psi(x_i) = h(h-1) \Psi(x_i)\,,\qquad
L_0 \Psi(x_i) = (N+3) \Psi(x_i) \,.
\labeltest{O3}
\ee
Here, the first condition defines the conformal spin of the state, $h$,
and
the second one follows trivially from the fact that $\Psi(x_i)$ is a
homogeneous polynomial of degree $N$ in three variables $x_i$.
Assuming that there
exists a positive definite scalar product on the space of the coefficient
functions (see (Eq.~(\ref{sc-prod}) below) one can easily prove that
eigenvalues of $L_+L_-$ are non-positive. From  the definition
$L^2=L_0(L_0-1)+L_+L_-$ it then follows that $h\le N+3$.
Moreover, the eigenstate with the largest conformal spin $h=N+3$
has to be  annihilated by the lowering
operator $L_-$\footnote{
 $L_+$ and $L_-$ act as the rising and the lowering
operators in the (infinite dimensional) representation labeled
by the spin $h$, respectively,
so that if $L_0\Psi = h_0\Psi$ then $L_0L_\pm\Psi =
(h_0\pm 1)\Psi$.}
\be
L_-\, \Psi^{(0)}(x_i)=0\,,\qquad h=N+3
\labeltest{HW}
\ee
and is, thus, the highest weight of the representation. All other states
can be obtained from the highest weight by a repeated application of the
rising operator $L_+$ which acts trivially on the coefficient
function
\begin{eqnarray}
&&
\Psi^{(n)} (x_i)
=L_+^n \Psi^{(0)}(x_i)
=(-1)^n(x_1+x_2+x_3)^n \Psi^{(0)}(x_i)\,,
\nonumber
\\
&&
L_0\, \Psi^{(n)}(x_i)= (h+n)\Psi^{(n)}(x_i)
\end{eqnarray}
and amounts to `dressing' the corresponding local operator by
the $n-$th power of a total derivative.
These states form an infinite dimensional representation
of the $SL(2)$ group of a positive discrete series labeled by the integer
conformal spin $h=N+3$.
They all have the same energy and, being substituted
into (\ref{expansion}) and (\ref{Pn}),
 lead to identical  contributions to the baryon distribution amplitude
due to the condition $x_1+x_2+x_3=1$. For this reason we can neglect
such states altogether and impose (\ref{HW})
as an additional constraint on the solutions of the Schr\"odinger
equation
(\ref{Sch})\footnote{As familiar from quantum mechanics, the highest weight
states exhibit additional symmetry. In our case,
it is easy to find that the local operator corresponding to
the highest weight transforms under the  $SL(2)$
transformation according to (\ref{Moebius}).}.

Note that the conformal spin $h$ of the three-quark state
which satisfies the highest weight condition (\ref{HW})
is related to the total number of derivatives.
As a consequence, conformal operators with different $N$  do not mix
with each other under renormalization since they belong to different
representations of the collinear conformal group. This condition
is yet not sufficient to diagonalize the evolution equation
since, as will  become clear in the next section,
for fixed $N$ there exist $N+1$
different conformal operators mixing between which is allowed
by conformal symmetry and exists, in general. The size of
the mixing matrix is, however, reduced from $N(N+1)/2$ to $N+1$.
The impact of conformal symmetry is that one can
eliminate all mixing with operators containing total derivatives.

It is straightforward to check that the $SL(2)$ generators $L_\alpha$
as well as the Hamiltonians in~(\ref{H12}) and (\ref{H32})
are Hermitian with respect to the $SL(2)$-invariant scalar product:
\begin{equation}
\langle \Psi_1|\Psi_2\rangle =
120 \int\!{\cal D}x\,
x_1 x_2 x_3\, {\Psi_1^\ast(x_i)}\Psi_2(x_i)\,.
\labeltest{sc-prod}
\end{equation}
Hermiticity implies  that the scalar product of two
eigenfunctions with different eigenvalues vanishes, {\it i.e.}\ coefficient
functions of any two operators that do not mix under renormalization
are orthogonal with this weight function, cf. (\ref{aaa}).

\subsection{The conformal basis}

To find the general  solution
of the evolution equation~(\ref{Sch})
with the Hamiltonian ${\cal H}$ given in (\ref{H32}) or (\ref{H12})
it proves  convenient  to decompose the eigenfunctions $\Psi(x_i)$
over a suitable basis of functions $\Psi^{(12)3}_n(x_i)$
having the same conformal properties as
$\Psi(x_i)$:
\begin{equation}
\Psi(x_1,x_2,x_3) = \sum_{n=0}^{N} i^n {u_n}{f_n^{-1}}
\Psi^{(12)3}_n(x_1,x_2,x_3)\,.
\labeltest{expand1}
\end{equation}
Here, the factor $i^n$ is inserted in order that the coefficients $u_n$ are
real, as will become clear in the next section. The numerical factor
\begin{eqnarray}
 f_n &\equiv& \frac{(n+1)(n+2)}{2(2n+3)}(N-n+1)(N+n+4)
\nonumber\\
&=& \frac{j_{12}(j_{12}-1)}{2(2j_{12}-1)}[h(h-1)-j_{12}(j_{12}-1)]
\labeltest{pn}
\end{eqnarray}
is included for later convenience and
the notations $\Psi^{(12)3}_n$ and $j_{12}\equiv n+2$ will be
explained below.

Aim of this section is to construct such a basis.
To this end, we require that the functions $\Psi^{(12)3}$ have the same
conformal properties as $\Psi$, that is
\begin{eqnarray}
{L_{0}} \, \Psi^{(12)3}_n(x_i) &=& h \Psi^{(12)3}_n(x_i)\,,\qquad
h=N+3,
\nonumber\\
{L_{-}} \, \Psi^{(12)3}_n(x_i) &=& 0,
\nonumber\\
{L^2} \, \Psi^{(12)3}_n(x_i) &=& h(h-1) \Psi^{(12)3}_n(x_i).
\labeltest{basis3}
\end{eqnarray}

The second-order differential equations Eq.~(\ref{basis3})
do not specify the set of polynomials $\Psi_n^{(12)3}$ uniquely,
but rather allow to choose them as a linear combination of $(N+1)$ solutions
with arbitrary coefficients. To fix these coefficients one has to supplement
Eq.~(\ref{basis3}) by some additional condition.
The traditional choice~\cite{earlybaryon}
is  to expand $\Psi(x_i)$
over the set of Appell polynomials \cite{ER53} (for $x_1+x_2+x_3=1$):
\begin{equation}
{\cal A}_{n,N-n} (x_1,x_2)
  \sim  \left[x_1x_2(1-x_1-x_2)\right]^{-1}\! \partial_1^n\partial_2^{N-n}
 x_1^{1+n}x_2^{1+N-n}(1-x_1-x_2)^{1+N}.
\end{equation}
In this way, solving the evolution equation (\ref{Sch}), one is left with
a complicated $(N+1)\times (N+1)$
mixing matrix for the coefficients in front of Appell polynomials
with the same $N$ but different $n$, which does not have any obvious
structure.  This basis is also inconvenient  for calculations
since Appell polynomials with different values of $n$ are not
mutually orthogonal.

The expansion in Appell polynomials is, however, not warranted and
in this paper we suggest a different basis which is orthonormal and
better suited for the solution of the evolution equation.
To this end, we require that in addition to (\ref{basis3}) the
polynomials $\Psi_n^{(12)3}(x_i)$ $(n=0\,,...\,,N)$ should diagonalize
the two-particle Casimir operator
in the channel defined by the $(12)$-quark pair:
\begin{eqnarray}
&
L_{12}^2 \Psi^{(12)3}_n(x_i)
 = j_{12}(j_{12}-1) \Psi^{(12)3}_n(x_i),
&
\nonumber\\
&
j_{12}= n+2\,,\qquad
0 \leq n \leq N\,.
&
\labeltest{basis2}
\end{eqnarray}
The particular choice of a quark pair is of course arbitrary and
we might use, e.g.,  $L_{23}^2$ for the same purpose.
In this way one obtains a different basis of functions $\Psi^{1(23)}_n(x_i)$
that are linear related to $\Psi^{(12)3}_n(x_i)$
through the Racah $6j-$symbols of the $SL(2)$ group (see Appendix~A):
\begin{equation}
\Psi^{1(23)}_k(x_i)=\sum_{n=0}^N \Omega_{kn}
\Psi^{(12)3}_n(x_i)\,.
\labeltest{Racah}
\end{equation}
Here, the superscript indicates the order in which the tensor product
of three $SL(2)$ representations has been decomposed into the irreducible
components.

The solution of the combined Eqs.~(\ref{basis3})
and (\ref{basis2}) can be obtained either solving the corresponding
second-order differential equations explicitly or
making use of the conformal OPE.
The result reads
(in a certain convenient normalization):
\begin{equation}
 \Psi_{N,n}^{(12)3}(x_i)
=(N+n+4)(x_1+x_2)^n(x_1+x_2+x_3)^{N-n}
    \!P^{(2n+3,1)}_{N-n}\!\left(\frac{x_3-x_1-x_2}{x_1+x_2+x_3}\right)
   \!C_n^{3/2}\!\left(\frac{x_1-x_2}{x_1+x_2}\right)
\labeltest{basis}
\end{equation}
where $C^{3/2}_n(x)$ and $P^{(\alpha,\beta)}_k(x)$ are Gegenbauer and
Jacobi polynomials \cite{ER53}, respectively.
Note that each function $\Psi^{(12)3}_{N,n}$ is
specified by a pair of integers $N,n$ which are related in a obvious
way to the total conformal spin of the three-quark operator $h=N+3$
and the
conformal spin of the (12)-pair $j_{12}=n+2$, respectively.
In what follows we often drop the subscript `$N$' if it is clear
from the context.

\subsubsection{Properties of the conformal basis}

The following features of the new basis are especially important.

First, the functions $\Psi^{(12)3}_{N,n}(x_i)$ are mutually
orthogonal with
respect to the $SL(2)$ scalar product (\ref{sc-prod})
\begin{equation}
\langle \Psi^{(12)3}_{N,n} | \Psi^{(12)3}_{M,m}\rangle\equiv
120 \int \!\!{\cal D}{x} \,
x_1 x_2 x_3 \Psi^{(12)3}_{N,n}(x_i)\Psi^{(12)3}_{M,m}(x_i)=
\delta_{MN}\delta_{mn}\frac{60f_{n}}{2N+5}.
\labeltest{norm-basis}
\end{equation}
The integration measure $\int\! {\cal D}x$ is defined in (\ref{Dx}),
$f_{N,n}$ is given in (\ref{pn}) and the factor 120 is introduced in order
 that $\int\! {\cal D}x\cdot 120 x_1 x_2 x_3 =1$.

Second, action of the Casimir operators of the collinear conformal group
in this basis is rather simple. By construction, $\Psi^{(12)3}_{N,n}(x_i)$
diagonalize $L_{}^2$ and $L_{12}^2$ whereas the remaining two
two-particle Casimir operators turn out to be  three-diagonal:
\begin{eqnarray}
L_{23}^2\, \Psi^{(12)3}_n(x_i) &=&
f_n\!\left[
\phantom{-}
\frac{1}{(n+1)}\Psi^{(12)3}_{n-1}(x_i)+
\frac{2n+3}{(n+2)(n+1)}\Psi^{(12)3}_n(x_i)
                                +\frac{1}{(n+2)}\Psi^{(12)3}_{n+1}(x_i)
\right]\!,
\nonumber\\
L_{31}^2\, \Psi^{(12)3}_n(x_i)&= &
f_n\!\left[
-\frac{1}{(n+1)}\Psi^{(12)3}_{n-1}(x_i)+\frac{2n+3}{(n+2)(n+1)}\Psi^{(12)3}_n(x_i)
                                -\frac{1}{(n+2)}\Psi^{(12)3}_{n+1}(x_i)
\right]\!.
\nonumber
\\
\labeltest{threediagonal}
\end{eqnarray}
This property turns out to be crucial for simplification
of the evolution equation.
In particular, using the definition
(\ref{Q3}) one finds that the operator $Q$  can be represented in
the conformal basis by a $(N+1)\times (N+1)$ matrix with only two
subleading diagonals nonzero
\be
Q\, \Psi^{(12)3}_n(x_i) = if_n\left[
\Psi^{(12)3}_{n+1} (x_i)-\Psi^{(12)3}_{n-1} (x_i)
\right].
\labeltest{Q-matrix}
\ee

Finally, the factorized form of $\Psi^{(12)3}_{n}(x_i)$ as a product of
polynomials depending separately on $s=(x_1-x_2)/(x_1+x_2)$ and
$t=(x_1+x_2-x_3)/(x_1+x_2+x_3)$
is convenient for applications. Note that the integration measure is
also factorized:
$\int\! {\cal D}x = \int_{-1}^1 \!dt\, (t+1)/4\int_{-1}^{1}ds$.

\subsubsection{Special cases}

The definition in (\ref{basis}) is valid for arbitrary $x_1,x_2,x_3$.
One important special case is $x_1+x_2+x_3=1$
which corresponds to the expansion of distribution amplitudes
so that  ${x_i}\equiv\{x_1,x_2,x_3\}$ can be identified with
the set of quark momentum fractions. For this
case we obtain a complete  set of polynomials
\begin{equation}
\Psi^{(12)3}_{N,n}(x_i) =
 (N+n+4)(x_1+x_2)^n
    P_{N-n}^{(2n+3,1)}\left({2x_3-1}\right)
   C_n^{3/2}\left(\frac{x_1-x_2}{x_1+x_2}\right)\,,
\label{DAbasis}
\end{equation}
which, as we are going to argue, are much superior
for studies of the three-particle distribution amplitudes as
compared to Appell polynomials.

Another important case is $x_1+x_2+x_3=0$ which corresponds to neglecting
contributions of all operators containing total derivatives. This choice is
relevant if, for example, one considers only forward matrix elements.
It also allows to abstract from unnecessary `kinematical' complications related to
the conformal symmetry and consider the dynamical
mixing problem in the most pure form.
Note that the basis functions (\ref{basis}) become very simple:
\begin{equation}
\Psi_{n}^{(12)3}(x_i)  
\bigg|_{\sum x_i=0} =
w(N,n)\,
(x_1+x_2)^N C_n^{3/2}\left(\frac{x_1-x_2}{x_1+x_2}\right),
\labeltest{basis-dir}
\end{equation}
where
\begin{equation}
  w(N,n) = (-1)^{N-n}\frac{(2N+4)!}{(N+n+3)!(N-n)!} =
      -(-1)^{h-j_{12}}\frac{\Gamma(2h-1)}{\Gamma(h+j_{12}-1)\Gamma(h-j_{12})}.
\end{equation}
Instead of the full coefficient function $\Psi(x_i)$ one can
consider the function of one variable $\widetilde\Psi(x)$ defined
as
\begin{equation}
 x_1 x_2 x_3\, \Psi(x_i) \bigg|_{\sum x_i=0}
 =-
(x_1+x_2)^{N+3}\widetilde\Psi\left(\frac{x_1-x_2}{x_1+x_2}\right),
\labeltest{tildepsi}
\end{equation}
so that if $\Psi$ is expanded in the basis of $\Psi^{(12)3}_{n}(x_i)$
with the coefficients as in Eq.~(\ref{expand1}), then
\begin{equation}
 \widetilde\Psi(x) =\frac{1-x^2}{4}\sum\limits_{n=0}^N i^n u_n
 f_{n}^{-1}\, w(N,n)\, C_n^{3/2}(x).
\labeltest{reduc}
\end{equation}
Note that although $\widetilde\Psi(x)$ was obtained from
the coefficient function $\Psi(x_i)$ by reduction
to the subspace $x_1+x_2+x_3=0$, it
contains all nontrivial dynamics of the
problem. If $\widetilde\Psi(x)$ is known,
then the full function $\Psi(x_i)$ of three
variables can easily be recovered through its expansion (\ref{expand1})
since
\begin{equation}
 u_n = i^n (-1)^N\frac{(N-n+1)!(N+n+4)!}{(2N+4)!}
           \int_{-1}^1\!dx\, \widetilde\Psi(x)\, C_n^{3/2}(x).
\labeltest{geg}
\end{equation}
In physical terms, existence of such a relation is a consequence
of the triangular structure of the mixing matrix with the operators
containing total derivatives, familiar from studies of meson distribution
amplitudes. Similar to the latter case, it is sufficient
for calculation of  the anomalous dimensions
to consider forward matrix elements of three-quark operators for free
quarks. After this is done, the coefficient functions of multiplicatively
renormalizable operators can be obtained from (\ref{geg}),
(\ref{expand1}).

The algebraic structure of this connection is, however, complicated,
which can be traced to the fact that the lowering operator $L_-$
is nontrivial in the `adjoint'  representation (\ref{adjoint}).
As a consequence, there exists  no simple
way to resolve the constraints imposed on the form of the
function $\Psi(x_i)$ by the highest weight condition $L_{-}\Psi=0$.

In what follows we suggest an alternative  basis in which the
eigenfunctions have a much simpler form that is useful in some
applications.

\subsection{The dual conformal  basis}

Once the evolution `Hamiltonians' ${\cal H}$ are
written in terms of the $SL(2)$ generators, one
can abstract from the `physical' Hilbert space
spanned by the coefficient functions $\Psi(x_i)$
and try to find an equivalent representation
of the $SL(2)$ group
with simpler properties of the highest weights.
The calculation made in Sect.~3.2 suggests that the
conformal symmetry properties
of polynomials $\Phi(z_i)$ of the
light-cone coordinates $z_i$ might be
simpler than polynomials $\Psi(x_i)$ of the momentum fractions
since according to (\ref{SL2-field})
the highest weight condition $L_-\Phi(z_i)=0$ translates  to the translation
invariance of $\Phi(z_i)$\footnote{
The basis of the functions $\Phi(z_i)$ is dual to the
conformal basis $\Psi(x_i)$ in the same sense as the light-cone
coordinates $z_i$ are dual to the light-cone momentum fractions $x_i$.
The `physical' coordinate space distribution amplitudes were introduced
in Ref.~\cite{BB88}: these are states which
diagonalize the lowering operator $L_-\Psi = -ip\Psi$ and thus resemble
coherent states in the standard field theory terminology.
Here $p$ has a physical
meaning of  the momentum of the hadronic `wave packet' propagating
along the light-cone.
}.
The translation invariance, combined with the restriction to
homogeneous polynomials $\Phi(z_i)$
of degree $N$ in three light-cone coordinates $z_i$,
implies that $\Phi(z_i)$ essentially reduces to a polynomial
of degree $N$ of a single variable, times a simple
 overall factor:
\be
\left.
\begin{array}{c}
 L_0\,\Phi(z_i)=(N+3)\Phi(z_i)
\\[3mm]
L_-\,\Phi(z_i)=0
\end{array}
\right\}
\quad\Longrightarrow\quad
\Phi(z_i)=(z_1-z_2)^N
\widetilde\Phi\left(
\frac{z_3-z_2}{z_1-z_2}
\right)\,.
\labeltest{dual-cond}
\ee
Note similarity to, and at the same time difference
with Eq.~(\ref{tildepsi}) defining the coefficient
function of one variable $\widetilde\Psi(x)$ for the special choice
of momentum fractions: In both representations the conformal symmetry allows
one to reduce the evolution equation involving three variables
to an equation involving a function of a single variable ---
$\widetilde\Psi(x)$ or $\widetilde\Phi(z)$, respectively.
\footnote{As we will show in Sect.~4.3, these two functions are related
to each other through the duality transformation.}
At the same time, while
the one- and the three-variable descriptions are essentially
equivalent in position space thanks to the translation invariance,
 the relation between $\widetilde\Psi(x)$ and $\Psi(x_i)$
appears to be  much less transparent, see Sect.~3.4.

The easiest way to construct the  basis of polynomials in position space
explicitly is to identify
them with suitable correlation functions in a
certain two-dimensional conformal field theory.
Let ${\cal O}_\Psi$ be a local conformal operator with
spin $h=N+3$ corresponding to the coefficient function $\Psi(x_i)$
so that it is transformed as an elementary field with spin $h$
under the projective transformations (\ref{Moebius}).
In so far as only these transformation properties are important, we can
replace formally the quarks by free scalar fields $\phi(z_k)$
with the same conformal spin $j_k=1$:
${\cal O}_\Psi(\xi) = \Psi(\partial_1,\partial_2,\partial_3)
\phi(\xi_1)\phi(\xi_2)\phi(\xi_3)\big|_{\xi_i=\xi}$.
In the terminology of conformal field theories such operators
are called quasiprimary fields. Correlation functions of them
with elementary fields are known to satisfy the conformal Ward identities
which take the form of the highest weight conditions, (\ref{O3}) and (\ref{HW}),
that we are looking for. This suggests to define the polynomial $\Phi(z_i)$ as
{\em dual} to the coefficient function $\Psi(x_i)$ by the following
correlation function:
\begin{eqnarray}
\Phi(z_1,z_2,z_3)&\equiv& w_1^2 w_2^2 w_3^2
\langle 0|
{\cal O}_\Psi(0) \phi(w_1)\phi(w_2)\phi(w_3)
|0\rangle
\bigg|_{z_i=1/w_i}
\nonumber
\\
&=&
\Psi(\partial_{\xi_1},\partial_{\xi_2},\partial_{\xi_3}) \prod_{k=1}^3
(1-z_k\xi_k)^{-2}\bigg|_{\xi_k=0},
\end{eqnarray}
where we used the expression for a propagator of free field
$\langle 0|\phi(w)\phi(0)|0 \rangle = w^{-2}$.
By construction, the $SL(2)$ generators have the standard representation
(\ref{SL2-field}) on the space of dual polynomials and
it is straightforward to verify 
that
$\Phi(z_i)$ defined in this way satisfies the conditions (\ref{dual-cond}).

The two polynomials  $\Phi(z_1,z_2,z_3)$ and $\Psi(x_1,x_2,x_3)$ have
the same degree $N$ and are related to each other by the Mellin
transformation~\cite{DKM}
\be
\Phi(z_1,z_2,z_3)=\int_0^\infty \prod_k
dt_k\,t_k\,\e^{-t_k}\, \Psi(z_1t_1,z_2t_2,z_3t_3)
\labeltest{trans}
\ee
that amounts to the redefinition of the coefficients
$c_{n_1,n_2,n_3} \to c_{n_1,n_2,n_3}(n_1+1)!(n_2+1)!(n_3+1)!$
in the polynomial $\Psi(z_1,z_2,z_3)=\sum_{n_1,n_2,n_3} c_{n_1,n_2,n_3}
z_1^{n_1}z_2^{n_2}z_3^{n_3}$.

The $SL(2)$ invariant scalar product on the space of
coefficient functions (\ref{sc-prod}) can equivalently be rewritten  as
\begin{eqnarray}
 \langle\Psi_1|\Psi_2\rangle &=&\frac{\Gamma(2N+6)}{\Gamma(6)}
\Psi_1^\ast(\partial_{z_1},\partial_{z_2},\partial_{z_3})
\Phi_2(z_1,z_2,z_3) \bigg|_{z_k=0}
\nonumber\\
&=&\frac{\Gamma(2N+6)}{\Gamma(6)}
\Psi_2(\partial_{z_1},\partial_{z_2},\partial_{z_3})
\Phi_1^\ast(z_1,z_2,z_3) \bigg|_{z_k=0},
\end{eqnarray}
where $\Phi_{1(2)}$ is a dual of $\Psi_{1(2)}$.

The two representations for the eigenfunctions,
$\Phi(z_i)$ and $\Psi(x_i)$, are equivalent
from the point of view of diagonalization of the evolution
equation: they give rise to the same  energy spectrum
and are related to each other through the
transformation (\ref{trans}). However, the use of the dual
representation can be advantageous due to the particular
simple structure (\ref{dual-cond}).

Applying the transformation (\ref{trans}) to the both sides of
(\ref{expand1}) one can construct the dual conformal basis
$\Phi^{(12)3}(z_i)$. The functions $\Phi^{(12)3}_{N,n}$ can be
defined as translation-invariant homogeneous polynomials of three variables
which diagonalize the Casimir operator  $L^2_{12}$
(in the standard representation (\ref{SL2-field})):
\begin{eqnarray}
&&\Phi^{(12)3}_{N,n}(z_i)=(z_1-z_2)^N \varphi_{N,n}(z)\,,\quad
z = \frac{z_3-z_2}{z_1-z_2},
\labeltest{basis-dual}
\\
&&
L_{12}^2 \Phi^{(12)3}_n(z_i) = (n+2)(n+1) \Phi^{(12)3}_n(z_i)\,,
\end{eqnarray}
with $n=0\,,...,N$.
Solving the last condition one gets the explicit expression for
the functions $\varphi_{N,n}(z)$:
\begin{equation}
\varphi_{N,n}(z) =
f_n \frac{(N+n+3)!(n+1)!}{(2n+2)!}
   \,z^{N-n}\,
  {}_2F_1\left({{n-N,n+2}\atop{2n+4}}\bigg|z^{-1}\right)\,,
\end{equation}
which defines $\varphi_{N,n}(z)$ as a polynomial of degree
$N-n$ in $z$.
Here the normalization is such that the polynomials
(\ref{basis-dual}) and (\ref{basis}) are related to each other
by the Mellin transformation (\ref{trans}).
The decomposition of the dual eigenfunction $\Phi(z_i)$ over the dual
basis has again the form  (\ref{expand1}) with the {\it same\/}
coefficients $u_n$
\begin{eqnarray}
\Phi(z_1,z_2,z_3) &=& (z_1-z_2)^N\sum_{n=0}^{N} i^n u_n f_n^{-1}
\varphi_{N,n}(z),
\labeltest{expand2}
\end{eqnarray}
with $z$ defined in (\ref{basis-dual}).

It is clear that the linear algebraic relations (\ref{Racah}),
(\ref{threediagonal}) and (\ref{Q-matrix}) satisfied by the polynomials
$\Psi^{(12)3}(x_i)$ remain valid for the dual polynomials
$\Phi^{(12)3}(z_i)$ provided that one changes the adjoint representation
of the $SL(2)$ generators (\ref{adjoint}) to the standard one in
(\ref{SL2-field}).

The function $\varphi_{N,n}(z)$ has two indices corresponding
to the total conformal spin of the system $h=N+3$ and the conformal
spin $j_{12}=n+2$ in the subchannel $(12)$.  In the sequel we will need
the asymptotic behavior of this function in the limit when any two
of the coordinates $z_i$ coincide, or
equivalently $z=0$, $1$ and $\infty$:
\begin{eqnarray}
\varphi_{N,n}(z) &\stackrel{z\to
\infty}{=}& f_n\,z^{N-n}\,\frac{(N+n+3)!(n+1)!}{(2n+2)!}\,,\quad
\nonumber\\
\varphi_{N,n}(z) &\stackrel{z\to 1}{=}&
f_n\, (2n+3)(N+1)!\,,
\labeltest{asymphi}
\\[3mm]
\varphi_{N,n}(z)
&\stackrel{z\to 0}{=}&
f_n\,(-1)^{N-n}\,(2n+3)(N+1)! ,
\nonumber
\end{eqnarray}
where the leading terms are kept only.

\section{Integrability}
\setcounter{equation}{0}

As was explained in Sect.~3.2, the Brodsky-Lepage evolution equations
(\ref{Sch}) have the form of the Schr\"odinger equations describing a
three-particle system with three degrees of freedom which we can
choose as quark momentum fractions $x_i$ or coordinates $z_i$ depending on
whether the `physical' or `dual' representation is used\footnote{Note that
the expressions for ${\cal H}^v$, ${\cal H}^e$ in Eqs.~(\ref{H32-part}),
(\ref{H12-part}) correspond to the dual representation.} for the `wave
functions'. Either way, the scale dependence of baryon distribution amplitudes
in QCD corresponds to a  {\it one-}dimensional quantum mechanical
3-body problem with very peculiar
Hamiltonians, (\ref{H32}), (\ref{H12}) and (\ref{H-SL2}),
determined by the underlying QCD dynamics. The conformal symmetry
allows to trade two degrees of freedom for
two quantum numbers corresponding to the total conformal spin $L^2$ and its
projection $L_0$ after which one is left
with one degree of freedom described by either
the set of coefficients $u_n$ in the conformal basis (\ref{expand1}) or,
equivalently, a function of a
single variable (\ref{tildepsi}) or (\ref{dual-cond}). The
original 3-body Schr\"odinger equation is reduced, accordingly,  to a
(complicated) one-body problem which is in general not possible
to solve analytically for arbitrary $N$.

The crucial observation is that the Hamiltonian ${\cal H}_{3/2}$
(but not ${\cal H}_{1/2}$) proves to be completely
integrable: The operator $Q$ defined in
(\ref{Q3}) commutes both with the Hamiltonian and with
generators of the $SL(2)$ group. The eigenvalues of $Q$ thus
provide us with the third quantum number allowing to specify
completely the three quark states with maximal helicity.
Existence of the nontrivial  `conserved charge'
implies that ${\cal H}_{3/2}$ is a (complicated)
function of two and only two mutually commuting operators $Q$ and
$L^2$. Therefore, instead of solving the Schr\"odinger equation (\ref{Sch})
directly, one can solve much simpler equations (\ref{O3})
supplemented by the additional condition
\begin{eqnarray}
Q\,\Phi(z_i) &\equiv&
-i \partial_{z_1}\partial_{z_2}\partial_{z_2} z_{12}z_{23}
z_{31}\Phi(z_i) = q\,\Phi(z_i),
\labeltest{Q3phi}
\\
\widehat Q\,\Psi(x_i) &\equiv & \phantom{-}i
\left(\partial_{x_1}-\partial_{x_2}\right)
                        \left(\partial_{x_2}-\partial_{x_3}\right)
                        \left(\partial_{x_3}-\partial_{x_1}\right)
   x_1 x_2 x_3 \Psi(x_i) = q\,\Psi(x_i)
\labeltest{Q3psi}
\end{eqnarray}
in the `dual' and the `physical' representations, respectively,
and find the spectrum of the Hamiltonian
${\cal H}_{3/2}={\cal H}_{3/2}(L^2,Q)$ by replacing
the operators by their corresponding eigenvalues.

Remarkably enough the Hamiltonian ${\cal H}_{3/2}$ is well known from
integrable generalizations of the Heisenberg spin magnet models \cite{TTF}.
Indeed, an inspection shows that the $SL(2)$ generators (\ref{SL2-field})
for quarks with conformal spin $j_k=1$ can be
interpreted as Lorentz spin $s=-1$ operators. In this way, we may
consider the Hamiltonian ${\cal H}_{3/2}$ as describing the
system of three interacting spins each acting on its internal
space labeled by the coordinates $z_k$. These spins carry the
index of the corresponding particles and form a one-dimensional
spin chain with three sites. This system coincides identically with
the celebrated one-dimensional XXX Heisenberg spin magnet of
noncompact spin $s=-1$ for which powerful Quantum Inverse
Scattering methods have been developed and a lot of results are
available \cite{K95}--\cite{K97}. Aim of this section is to elaborate on this connection
and adapt the existing results to the present context.
Some new results will be presented as well.

\subsection{The master recurrence relation}

By construction of the conformal basis, the eigenfunctions (\ref{expand1})
and (\ref{expand2}) obey the conditions (\ref{O3}) for arbitrary coefficients
$u_n$.  Using Eq.~(\ref{Q-matrix}) it is easy to derive that the equation
$Q\Psi = q\Psi$ is equivalent to the following three terms
recurrence relation for the
coefficients $u_n$, $(n=0,\ldots,N)$:
\begin{equation}
q \,u_n = f_n\left( u_{n+1}+u_{n-1} \right),
\labeltest{masterrec}
\end{equation}
with the `boundary' conditions
\be
u_{-1}=u_{N+1}=0,
\labeltest{boundary}
\ee
which follow from the properties of the coefficients $f_n$
(\ref{pn}). The overall normalization of  $u_n$ is
arbitrary and we choose for simplicity
\be
u_0=1\,.
\ee

The recurrence relations (\ref{masterrec}) represent the system of
$N+1$ linear homogeneous equations on the coefficients $u_k$. Solution
of this system is equivalent to diagonalization of a
$(N+1)\times(N+1)$ matrix with only two subleading
diagonals nonzero. The consistency condition for this system translates to
the characteristic polynomial of degree $N+1$ in $q$ whose zeros define
the $N+1$ quantized values of $q$.

It follows from the recurrence relations (\ref{masterrec}) that $u_n(q)$
($n=0,...,N$) form a system of (semiclassical) orthogonal polynomials
in a discrete variable $q$. Then, the boundary condition $u_{N+1}(q)=0$
implies that $N+1$ quantized values of $q$ have the properties of roots
of orthogonal polynomials, that is, they are real and simple, for
different $N$ the set of quantized $q$ are interlaced. The completeness
and orthogonality conditions for this system are given by the
Cristoffel-Darboux relations \cite{ER53}
\begin{eqnarray}
\sum_{n=0}^N \frac1{f_n\omega(q)} u_n(q)
u_n(q')&=&\delta_{qq'}
\,,
\\
\sum_q \frac1{f_n\omega(q)} u_n(q) u_m(q)&=&\delta_{nm},
\end{eqnarray}
where $\omega(q)= u_N(q)\partial_q u_{N+1}(q)$ and in the second line
the summation goes over $N+1$ quantized $q$.

The orthogonal polynomials $u_n(q)$ have an obvious parity property
\be
u_n(-q)=(-1)^n u_n(q),
\labeltest{parity}
\ee
where from it follows  that all nonzero eigenvalues of $q$ come in pairs:
If $q$ is an eigenvalue, then $-q$ is also an eigenvalue,
$u_{N+1}(-q)=0$. In addition, for any {\it even\/} $N$ there is
a single eigenvalue $q=0$ and the corresponding coefficients are given
by
\be
u_{2k}(q=0)=(-1)^k u_0\,,\qquad u_{2k-1}(q=0)=0
\labeltest{q=0-coeff}
\ee
for $k=1,...,N/2$.

\subsection{Permutation symmetry}

The Hamiltonian ${\cal H}_{3/2}$ is explicitly invariant under
cyclic permutations of the three particles.
We define the generator of the corresponding {\em discrete}
transformations ${\cal P}$ as
\begin{equation}
{\cal P} \, \Phi(z_1,z_2,z_3) = \Phi(z_2,z_3,z_1)\,,\quad
[{\cal P},{\cal H}_{3/2}]=[{\cal P},L^2]=[{\cal P},Q]=0\,,
\labeltest{P}
\end{equation}
where, for definiteness, we have chosen to use the dual
(coordinate space) representation.
Because of the symmetry, the eigenfunctions of ${\cal H}_{3/2}$
can simultaneously be chosen as eigenstates of ${\cal P}$:
\begin{equation}
{\cal P} \, \Phi(z_1,z_2,z_3)= \Phi(z_2,z_3,z_1)=
\theta\,\Phi(z_1,z_2,z_3)\,
\labeltest{lambda}
\end{equation}
with $\theta=\theta(N,q)$ being a function of quantum numbers.
Since ${\cal P}^3=1$, the possible eigenvalues $\theta$ are
given by three different
cubic roots of unity:
\begin{equation}
\theta=\e^{-i\phi(N,q)}\,,\qquad
\phi = 0, ~\frac{2\pi}{3}, ~\frac{4\pi}{3}.
\labeltest{roots}
\end{equation}

In addition, ${\cal H}_{3/2}$
is symmetric under permutations of quarks in the $(12)$ pair
\begin{equation}
{\cal P}_{12} \, \Phi(z_1,z_2,z_3) = \Phi(z_2,z_1,z_3)\,,\quad
[{\cal P}_{12},{\cal H}_{3/2}]=[{\cal P}_{12},L^2]=0\,.
\labeltest{P12}
\end{equation}
This implies that the eigenstates can be chosen to possess a definite
`parity' ${\cal P}_{12} = \pm 1$. In fact, the  spin and
isospin symmetry of the physical baryon distribution amplitudes introduced
in Sect.~2 lead to their definite parity properties, see  Eqs.~(\ref{12nuc}),
(\ref{12del}), so that expansion in parity eigenstates is natural.

One should stress that the operators ${\cal P}$ and ${\cal P}_{12}$ do not
commute and therefore the eigenstates of ${\cal P}$ do not
have, in general, definite parity, and vice versa. Nevertheless, the
symmetry of the Hamiltonian under both ${\cal P}$ and ${\cal P}_{12}$
immediately implies that the eigenvalues of $H_{3/2}$ with
$\theta\neq 1$ have to be (at least) double degenerate.%
\footnote{
To show this, consider an eigenstate of ${\cal H}_{3/2}$ which is
simultaneously an eigenstate of ${\cal P}$:
$
 {\cal H}_{3/2} \Phi = {\cal E}\Phi\,,$ ${\cal P}\Phi = \theta\Phi\,.
$
Acting on the first equation  by ${\cal P}_{12}$, one gets
$
  {\cal H}_{3/2} {\cal P}_{12} \Phi = {\cal E}{\cal P}_{12}\Phi\,,
$
so that either ${\cal P}_{12} \Phi$ is an independent eigenstate
with the same energy, or it is proportional to $\Phi$:~
${\cal P}_{12} \Phi = p\,\Phi$ and is therefore a parity eigenstate
with $p=\pm 1$. In the latter case, applying the identity
${\cal P}_{12}{\cal P}{\cal P}_{12}= {\cal P}^2$ to $\Phi$
one gets
$\theta =\theta^2$ where from necessarily $\theta=1$.
}

Integrability of ${\cal H}_{3/2}$ alias existence of the
conserved charge $Q$ increases the symmetry,
so that the $\theta=1$ eigenstates turn out to be
double degenerate as well, apart from the singular state corresponding
to $q=0$. To show this, note that
${\cal P}_{12}$  {\em anticommutes}
with  $Q$:
\begin{equation}
   {\cal P}_{12} Q=-Q {\cal P}_{12}.
\labeltest{anti}
\end{equation}
Since the Hamiltonian ${\cal H}_{3/2}$ commutes simultaneously
with ${\cal P}_{12}$ and $Q$, it should be an even function of
$Q$ and therefore the levels corresponding to nonzero $q$ and $-q$
have the same energy and are double degenerate.

It follows from (\ref{anti}) that permutation of quarks transforms
an eigenfunction of $Q$ into another eigenfunction with the opposite value
of $q$ and the same value of the energy:
\begin{eqnarray}
\Phi_q(z_2,z_1,z_3)&=&\Phi_{-q}(z_1,z_2,z_3)=\Phi_{q}^\ast(z_1,z_2,z_3)\,,
\nonumber\\
\Psi_q(x_2,x_1,x_3)&=&\Psi_{-q}(x_1,x_2,x_3)=\Psi_{q}^\ast(x_1,x_2,x_3)\,.
\labeltest{12symmetry}
\end{eqnarray}
These relations are an obvious consequence of (\ref{Q3phi}) and (\ref{Q3psi}).
For the corresponding functions of one variable one gets:
\begin{eqnarray}
{\rm Re}\, \widetilde\Phi(1-z) = {\rm Re}\, \widetilde\Phi(z)\,,&\qquad&
{\rm Im}\, \widetilde\Phi(1-z) = -{\rm Im}\, \widetilde\Phi(z)\,,
\\
{\rm Re}\, \widetilde\Psi(-x) = {\rm Re}\,  \widetilde\Psi(x)\,,&\qquad&
{\rm Im}\,  \widetilde\Psi(-x) = -{\rm Im}\,  \widetilde\Psi(x)\,.
\labeltest{psisymmetry}
\end{eqnarray}

Eqs.~(\ref{12symmetry}) suggest that real and imaginary parts
of the complex eigenfunctions $\Phi(z_i)$, $\Psi(z_i)$ have
definite parity with respect to the ${\cal P}_{12}$ permutations.
Define
\be
\Psi_q(x_i)=\Psi_q^{(+)}(x_i)+i\Psi_q^{(-)}(x_i)
\ee
with the real functions
\begin{eqnarray}
&&\Psi_q^{(+)}(x_i) = \frac12\left[\Psi_q(x_i) + \Psi_{-q}(x_i)\right],
\\
&&\Psi_q^{(-)}(x_i) = -\frac{i}2\left[\Psi_q(x_i) - \Psi_{-q}(x_i)\right].
\labeltest{eigen_parity}
\end{eqnarray}
Then $\Psi^{+}$ ($\Psi^{-}$) is
even (odd) with respect to permutations of the two first arguments:
\begin{equation}
\Psi_q^{(\pm)}(x_1,x_2,x_3) = \pm \Psi_q^{(\pm)}(x_2,x_1,x_3)\,.
\end{equation}
We recall that the eigenstates $\Psi_q^{(\pm)}$ correspond to the same
value of the energy but,  in contrast to $\Psi_{\pm q}$, they
do not correspond, in general, to any  definite eigenvalue $\theta(N,q)$.

The eigenvalues $\theta(N,q)$ of the cyclic permutation operator
${\cal P}$ can be expressed in terms of the solutions of the
recurrence relation. To this end, substitute $\Phi(z_i)$ in
(\ref{lambda}) by its expansion in Eq.~(\ref{expand2})
and take into account that the cyclic permutations
correspond to the following transformation rules for the
coordinate ratio $z=(z_3-z_2)/(z_1-z_2)$:
\begin{equation}
z \stackrel{{\cal P}}{\to} 1-\frac1{z}
\stackrel{{\cal P}}{\to} \frac1{1-z} \stackrel{{\cal P}}{\to} z\,.
\end{equation}
This gives
\be
(-1)^N z^{N} \sum_{n=0}^N i^n u_n f_n^{-1}
\varphi_n\left(1-\frac{1}{z}\right)
=\theta
\sum_{n=0}^N i^n u_n f_n^{-1}
\varphi_n(z),
\labeltest{lambda1}
\ee
which has to be valid for an arbitrary real $z$.
Consider the limit $z\to \infty$
or, equivalently, $z_1-z_2\to 0$. Taking into account Eqs.~(\ref{asymphi})
we compare the leading asymptotics of the both sides of
(\ref{lambda1}) to get
\begin{equation}
\theta(N,q) = \frac{2(-1)^Nu_0^{-1}}{(N+2)(N+3)} \sum_{n=0}^N i^n
u_n(q)(2n+3)=
\frac{\sum_{n=0}^N (-i)^n u_n(q) (2n+3)}
     {\sum_{n=0}^N i^n u_n(q)(2n+3)},
\labeltest{lambda-cn}
\end{equation}
where the second equality follows from the identity
$\theta=\theta^*/\theta$ and reality
of the coefficients $u_n$ as defined by the recurrence relation
with real coefficients. Comparing
(\ref{lambda-cn}) with (\ref{roots}) we end up with
\be
\phi(N,q) = 2\,{\rm arg}\,\left[
\sum_{n=0}^N i^n u_n(q) (2n+3) \right].
\labeltest{Quasi}
\ee
In terminology of integrable models this expression
defines the quasimomentum corresponding to the wave
function $\Phi(z_i)$.

Since $\phi(N,q)$ takes a discrete set of values (\ref{roots}),
Eq.~(\ref{Quasi}) suggests that eigenvalues of $Q$ can be
parameterized by an integer number $\ell$ and belong
to a one-parametric family of curves, $q=q(N,\ell)$.
We will elaborate on the physical interpretation
of such trajectories in Sect.~4.4.3 and construct them explicitly in
Sect.~4.5 using the WKB expansion.

\subsection{Duality}

The Hamiltonian ${\cal H}_{3/2}$ possesses a
duality symmetry~\cite{Lip98} which allows to establish the equivalence between
coefficient functions $\Psi(x_i)$ at $x_1+x_2+x_3=0$ and the dual
coefficient functions $\Phi(z_i)$. We recall that both functions
are related to each other through the integral transformation (\ref{trans})
that maps the momentum fractions $x_i$ into the light-cone coordinates $z_i$.

As a hint, observe that changing the variables
in (\ref{Q3psi}) as $x_1\to z_{12}$, $x_2\to z_{23}$ and $x_3\to z_{31}$
one can formally cast it into the form of (\ref{Q3phi}). In order
to establish a formal equivalence, define the duality
transformation $S$ as
\begin{eqnarray}
x_k
&\to&
S x_k S^{-1} = x_k-x_{k+1},
\nonumber
\\
\partial_{x_k}-\partial_{x_{k+1}}
&\to&
S \left(\partial_{x_k}-\partial_{x_{k+1}}\right) S^{-1}
=-\partial_{x_{k+1}},
\labeltest{S}
\\
\Psi(x_1,x_2,x_3)
&\to&
S\,\Psi(x_1,x_2,x_3) = \Psi(x_{12},x_{23},x_{31}),
\nonumber
\end{eqnarray}
with $x_{k+3}=x_k$ and $k=1,2,3$.
Here, the second relation follows from the remaining two. It is
easy to see that the constraint $x_1+x_2+x_3=0$ is required as
the consistency condition for these transformations.
Using the definition and taking into account that
$\Psi(x_1,x_2,x_3)$ is a homogeneous polynomial of degree $N$ it is easy to
check that
\begin{eqnarray}
&& S^2 \, x_k \, S^{-2} = -3\, x_{k+1}\,,
\\
&& S^2 \,\Psi(x_1,x_2,x_3) = \Psi(-3x_2,-3x_3,-3x_1) =
(-3)^N \Psi(x_2,x_3,x_1)\,,
\end{eqnarray}
which  allows to express $S^2$ in terms of the cyclic
permutation operator, ${\cal P}$, and the $SL(2)$ generator $L_0$ as
\be
S^2=(-3)^{L_0-3} {\cal P}\,.
\ee
Thus, the operator of the duality transformation $S$ is formally
proportional to the square root of the operator of cyclic permutations.

Applying the transformation (\ref{S}) to the conserved charge $Q$
in the adjoint representation (\ref{Q3psi}) we find the expression
\be
S \,\widehat Q \, S^{-1} =
-i\partial_{x_1}\partial_{x_2}\partial_{x_2} x_{12}x_{23}x_{31} = Q\,,
\ee
which, after the replacement of the momentum fractions by the coordinates,
$x_k\to z_k$, coincides with the operator $ Q$ acting on
the dual coefficient functions, $(\ref{Q3phi})$. For clarity, we have restored
the `hat' to indicate the adjoint representation, cf.~(\ref{adjoint}).
In the similar way, one can check that on the subspace $x_1+x_2+x_3=0$
\begin{eqnarray}
S\,\widehat L_0 S^{-1} =
S\,\sum_k (x_k \partial_{x_k}+1) S^{-1}&=& \sum_k (x_k \partial_{x_k}+1)
= L_0\,,
\\
S\,\widehat L^2 S^{-1} =
-S\,\sum_{j>k} (\partial_{x_k}-\partial_{x_j})^2 x_kx_j
S^{-1}&=& -\sum_{j>k} \partial_{x_k}\partial_{x_j} (x_k-x_j)^2
= L^2\,.
\end{eqnarray}
In other words, the duality transformation $S$ maps
the conserved charge $Q$
and the $SL(2)$ generators, $L_0$ and $L^2$, in the standard and the adjoint
representations, (\ref{SL2-field}) and (\ref{adjoint}), one into another.
Since in both descriptions they form a complete set of
mutually commuting operators, it follows that the eigenfunctions
must transform one into another as well, up to a numerical factor:
\be
\Psi_{N,q}(x_1,x_2,x_3)\bigg|_{x_k=z_{k}-z_{k+1}}=  {\cal C}\cdot\,
\Phi_{N,q}(z_1,z_2,z_3)\,
\labeltest{S-dual}
\ee
with ${\cal C}={\cal C}(N,q)$ being a normalization constant. Its
value can be found by examining the asymptotics of the both sides
as $z_1-z_3\to 0$, or equivalently $z\to 1$. Using (\ref{expand1}),
(\ref{basis-dir})
and (\ref{expand2}) it is straightforward to get
\be
{\cal C}^{-1} = i^N \frac{\theta u_0}{2u_N}
\frac{(N+1)!(N+2)!(N+3)!}{(2N+3)!}\,.
\ee

The duality relation (\ref{S-dual}) is highly nontrivial and it is easy
to see that this relation does not hold
for the basis functions
$\Psi^{(12)3}$ and $\Phi^{(12)3}$. The reason for this is that
the defining relations (\ref{basis2}) and (\ref{basis-dual}) are
not mapped one into another by the duality transformation since
\be
S \, \widehat L_{12}^2 \, S^{-1} =
- S \, (\partial_{x_1}-\partial_{x_2})^2 x_1 x_2
\, S^{-1} = -\partial_{x_2}^2 x_{12} x_{23} \neq -\partial_{x_1}\partial_{x_2}
x_{12}^2\,.
\ee
This also explains why the expansions (\ref{reduc}) and (\ref{expand2})
involve the same coefficients $u_n$ but different special functions.

Going over from $\Psi(x_i)$, $\Phi(z_i)$
to the corresponding functions of one variable, $\widetilde\Psi(x)$ and
$\widetilde\Phi(z)$, the duality relation (\ref{S-dual}) takes the form
\begin{equation}
    (1-z)^{N+3}\widetilde\Psi\left(\frac{1+z}{1-z}\right) =-{\cal C}\cdot
   z(1-z)\,\widetilde\Phi(z)
\labeltest{sduality1}
\end{equation}
or, equivalently
\begin{equation}
  \widetilde\Psi(x) = {\cal C}\cdot\,(-1)^{N}\theta\, \frac{1-x^2}{4}\,
     \widetilde\Phi\left(\frac{1+x}{2}\right),
\labeltest{sduality2}
\end{equation}
where we have used that $\widetilde\Phi(z)$ transforms to
$(-z)^N\widetilde\Phi(1-1/z)$ under cyclic permutations.

\subsection{Energy spectrum: Exact solution}

\subsubsection{Calculation of the energy}
\labeltest{tra}

The set of coefficients $u_k \equiv u_k(N,q)$ uniquely defines the
eigenfunction (\ref{expand2}) corresponding to the pair of quantum
numbers $h=N+3$ and $q$. Once the eigenfunction is known,
the corresponding value of the energy ${\cal E}(N,q)$
can in principle be found by `brute force' as the expectation value
of the Hamiltonian.
As we show in this section, there exists a  simpler and much more elegant
way to calculate the energy ${\cal E}(N,q)$ by
using the cyclic permutation symmetry. To this end,
is proves convenient to work in the dual representation.

The calculation is based on a simple identity
$
{\cal H}_{23}^v={\cal P}{\cal H}_{12}^v {\cal P}^{-1}
={\cal P}{\cal H}_{12}^v {\cal P}^{2}
$
which allows to rewrite the Hamiltonian as
\begin{equation}
{\cal H}_{3/2} = {\cal H}_{12}^v + {\cal P} {\cal H}_{12}^v {\cal P}^2
+ {\cal P}^2{\cal H}_{12}^v {\cal P}\,.
\end{equation}
Applying the wave function $\Phi(z_i)$ to the both sides of
this relation and using Eqs.~(\ref{lambda}) and (\ref{expand2}) we get
\begin{equation}
{\cal E} \Phi(z_i)
=
\left(1+\theta {\cal P}^2 +\theta^2 {\cal P}\right){\cal H}_{12}^v
\Phi(z_i)
=
\sum_{n=0}^{N}i^n u_n\, f_n^{-1} \varepsilon(n)
\left(1+\theta {\cal P}^2 +\theta^2 {\cal P}\right)
\Phi^{(12)3}_{N,n}(z_i).
\labeltest{pre-E}
\end{equation}
Here, $\varepsilon(n)$ denotes the energy of two-particle Hamiltonian
(\ref{H-SL2}) defined as
\begin{equation}
{\cal H}_{12}^v\,\Phi^{(12)3}_{N,n}=\varepsilon(n)
\Phi^{(12)3}_{N,n}\,,\qquad
\varepsilon(n)= 2\left[\psi(n+2)-\psi(2)\right]\,.
\labeltest{twospectrum}
\end{equation}
Using the explicit expressions (\ref{basis-dual}) for $\Phi^{(12)3}_{N,n}(z_i)$,
one can rewrite  (\ref{pre-E}) as
\begin{equation}
0=\sum_{n=0}^N i^n u_n f_n^{-1}
\left[
[{\cal E}-\varepsilon(n)] \varphi_n(z)
- \theta^2(-z)^{N} \varphi_n\left(1-\frac1{z}\right)
- \theta (z-1)^N \varphi_n\left(\frac1{1-z}\right)
\right],
\end{equation}
which has to be valid for an arbitrary real $z$. Taking the limit $z\to 0$
and using the relations (\ref{asymphi}), we get
\begin{equation}
{\cal E}(N,q)=
\frac{4(-1)^Nu_0^{-1}}{(N+2)(N+3)}
\,{\rm Re}\!\left[ \frac1{\theta}
\sum_{n=0}^N {i^n}\, u_n\, \varepsilon(n)\,(2n+3)
\right]\,.
\end{equation}
Finally, taking into account Eq.~(\ref{lambda-cn})
we obtain the following expression for the energy
\begin{equation}
{\cal E}(N,q)=4\,{\rm Re}\frac
{\sum_{n=0}^{N}i^n\, u_n(q)\, \left[\psi(n+2)-\psi(2)\right]
\,(2n+3)}
{\sum_{n=0}^{N}i^n\, u_n(q)\, (2n+3)}\,.
\labeltest{energy1}
\end{equation}

To summarize,
the recurrence relations (\ref{masterrec}) combined with the expression for
the energy (\ref{energy1}) and the eigenfunctions (\ref{expand2})
provide one with the {\em exact\/} solution to the Schr\"odinger equation for
the Hamiltonian ${\cal H}_{3/2}$.

An immediate consequence of (\ref{energy1}) and the parity property
(\ref{parity}) is that
\begin{equation}
{\cal E}(N,q) = {\cal E}(N,-q),\qquad
\theta(N,q) = 1/\theta(N,-q).
\labeltest{degeneracy}
\end{equation}
Thus, the energy levels corresponding to nonzero values of quantized
$q$ are double degenerate.

The resulting spectra of the conserved charge $q$ and the energy
${\cal E}_{3/2}$ are shown in Fig.~\ref{figure2} for $N\le 30$. As we
are going to argue in Sect.~4.4.3, the eigenvalues form the set of
trajectories a few of which are shown in Fig.~\ref{figure2} by solid
curves.

\begin{figure}
\centerline{\epsfxsize12.0cm\epsffile{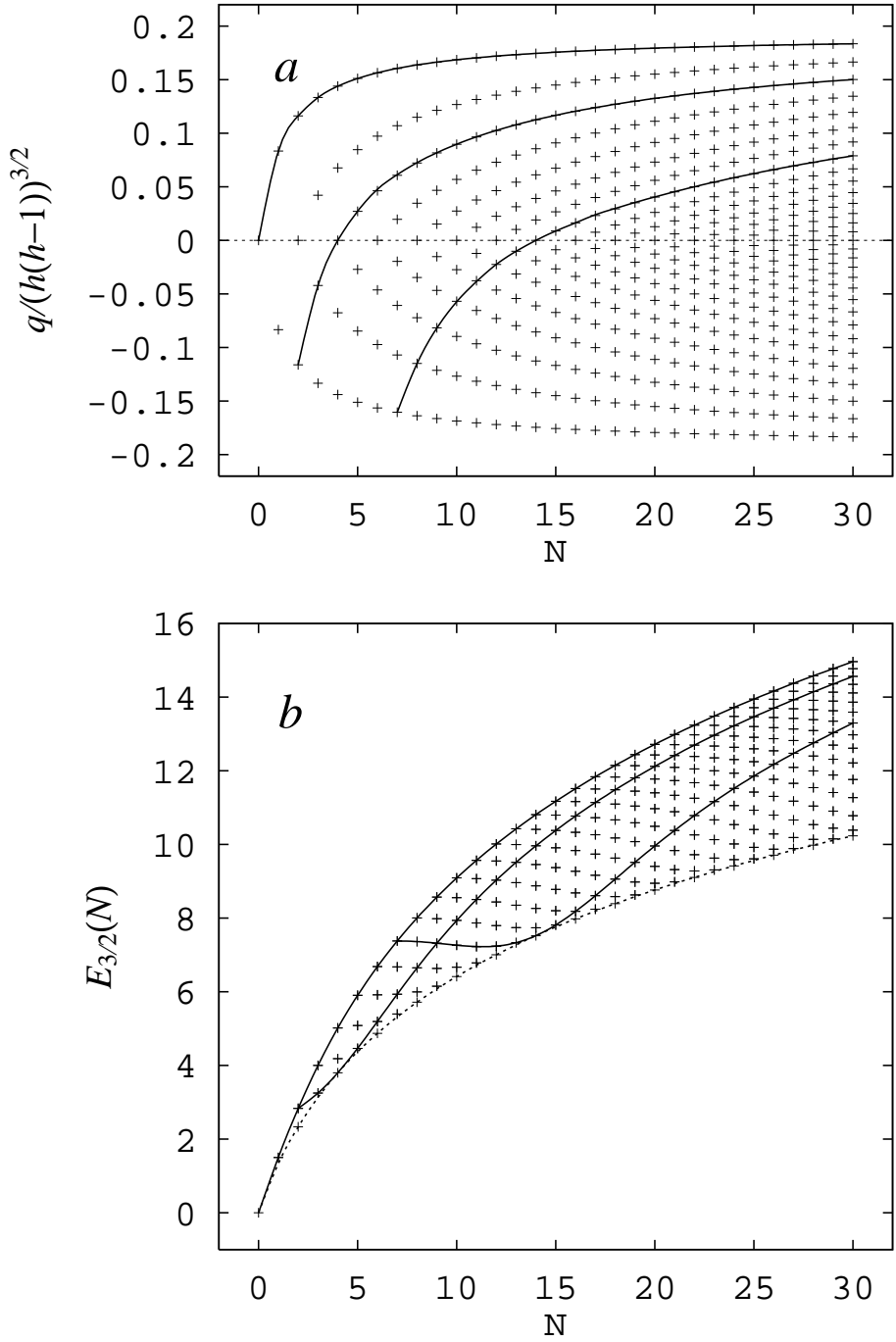}}
\caption[]{\small The spectrum of eigenvalues for the conserved charge $Q$ (a)
and for the helicity-3/2 Hamiltonian $H_{3/2}$ (b), see text.
}
\label{figure2}
\end{figure}

\subsubsection{The exact solution for $q=0$}

The energy and the eigenfunction of the state with $q=0$ can be calculated
explicitly. We recall that this state exists  for even $N$ only and its
expansion coefficients in terms of conformal polynomials are given by (\ref{q=0-coeff}).
Their substitution into (\ref{energy1}) and (\ref{lambda-cn}) yields
\begin{equation}
   {\cal E}(N,q=0) = 4\Psi(N+3)+4\gamma_E-6\,,
   \qquad
   \theta(N,q=0)=1\,.
\labeltest{e0}
\end{equation}
The curve corresponding to this expression for the energy is shown in
Fig.~\ref{figure2}b by dots. We observe that for even $N$ the state with $q=0$ is
the ground state of the Hamiltonian ${\cal H}_{3/2}$. According
to (\ref{12symmetry}), the corresponding wave function,
$\Psi_{q=0}(x_i)$, is a completely symmetric real function of $x_i$. Its
explicit expression can easily be obtained directly from (\ref{Q3psi}),
without an expansion over the conformal basis. It is straightforward
to verify that for $q=0$ the only solution to (\ref{Q3psi}) and (\ref{O3})
with the required symmetry is (up to an overall normalization factor)
\begin{eqnarray}
\lefteqn{ x_1 x_2 x_3 \Psi_{N,q=0}(x_1,x_2,x_3) =}
\\
&=& x_1(1-x_1)C_{N+1}^{3/2}(1-2x_1) +x_2(1-x_2)C_{N+1}^{3/2}(1-2x_2)
    +x_3(1-x_3)C_{N+1}^{3/2}(1-2x_3),
\nonumber
\label{ground32}
\end{eqnarray}
where $x_1+x_2+x_3=1$. This translates to
\begin{equation}
\widetilde\Psi_{N,q=0}(x) = 1-[(1+x)/2]^{N+3}-[(1-x)/2]^{N+3}.
\label{ground32tilde}
\end{equation}
Note that $\widetilde\Psi_{N,q=0}(x)$
does not have zeros on the interval $-1<x<1$ and vanishes at the end points.

\subsubsection{The Baxter equation, Bethe ansatz and analytic structure of
the spectrum}

Numerical solutions shown in Fig.~\ref{figure2}
exhibit remarkable
regularity. To understand their properties we develop the WKB expansion
of the energy ${\cal E}(N,q)$ and the conserved charge $q$ at large $N$.

The strategy is in many respects similar to the
Bohr's description  of the hydrogen atom.
The Hamiltonian ${\cal H}_{3/2}$ describes the system
of three particles with the coordinates $z_i$.
The scale of the energy is fixed by the conformal spin $N+3$ which
plays the r\^ole of the inverse Planck constant, $\hbar \sim
1/N$, in the corresponding Schr\"odinger equation.
The size of quantum fluctuations decreases with
$N$ and at large $N$ the quantum mechanical motion of
three particles is confined to their classical trajectories
that can be shown to have a finite period.
We then quantize the system
semiclassically by imposing Bohr-Sommerfeld quantization conditions
on the periodic classical trajectories.
 This procedure corresponds to the WKB solution
of the Schr\"odinger equation $\Phi(z_i) =\exp[iNS_{\rm eik}(z_i)]$ which,
as we will show in the next section, gives a good quantitative description
of the system.
Our aim in this section is develop a physical interpretation of the WKB
solutions, and to this end we have to introduce some methods
of integrable models.

The classical analog of the Hamiltonian is obtained by replacing
the derivatives by the momenta, $-i\partial_{z_k} \to p_k$, and the
commutators by the Poisson brackets. One gets ${\cal H}_{3/2}$ as a
function of the conserved charges, $L_-$, $L^2$ and $Q$,
each of which describes certain modes of the classical motion which
will later be quantized giving rise to a complete set of quantum numbers.
Note that $L_-=i(p_1+p_2+p_3)$ is the total momentum of the system.
The condition (\ref{HW}) then implies that the center-of-mass
stays at rest, $z_1+z_2+z_3=0$.  Similarly,
$L_0=-i(z_1p_1+z_2p_2+z_3p_3)+3=N+3$
generates dilatations of the coordinates and its eigenvalue fixes
the overall scale of coordinates and momenta.

The classical motion driven
by the conserved charge $Q$ is, however, very nontrivial.
It generates a collective motion of all the three particles which
represents a wave packet (or solitonic wave) propagating on the periodic
chain with three sites~\cite{KK}
\be
z_n(t)= \Theta(\kappa n + \omega t),
\labeltest{soliton}
\ee
where $\Theta(\varphi)$ is a $2\pi$ periodic
function of the argument,
and $t$ is the evolution `time' conjugate to the `Hamiltonian' $Q$:
$\partial z_n(t)/\partial t=\{Q,z_n\}$.
$\omega$ and $\kappa=\frac{2\pi}{3}\ell$ are the
proper frequency and the quasimomentum of this wave, respectively, both
depending on $q$ and $N$. The periodicity
condition $z_{n+3}(t)=z_n(t)$ leads to quantization of the quasimomentum
$\ell={\rm integer}$.
The explicit expressions~\cite{KK}
 can be derived by applying the methods of the theory of the
finite-gap soliton solutions but are of no relevance for
what follows. The eigenvalue of the Hamiltonian ${\cal
H}_{3/2}(L^2,Q)$ defines the energy of the soliton wave ${\cal E}={\cal
E}(N,q)$.

Quantization of the charge $Q$ appears as the result of imposing the
Bohr-Sommerfeld
quantization conditions on the periodic classical trajectories (\ref{soliton}).
To this end, one has to identify the corresponding action-angle variables which
in turn are constructed through the separation of variables.

The definition of the separated variables for the Hamiltonian ${\cal H}_{3/2}$
is known thanks to the similar construction for the XXX Heisenberg magnet of
spin $s=-1$ \cite{Skl}. It amounts to the unitary transformation of the operators and
the wave function under which the original coordinates $z_i$ are replaced by
new collective separated coordinates $\xi_i$ and the wave function $\Phi(z_i)$
is transformed into the wave function having a factorized dependence
on each of new coordinates
\be
\Phi(z_i) \to  
{\bf Q}(\xi_1) {\bf Q}(\xi_2)\, \xi_3^{-N-3}\,.
\labeltest{WF-sep}
\ee
Explicit expressions for the transformation $z_i\to \xi_i$ can be found in
\cite{Skl,K95}.
The last factor in (\ref{WF-sep}) carries conformal
spin of the state and has a trivial dependence on the coordinate $\xi_3$.
The original 3-body Schr\"odinger equation for ${\cal H}_{3/2}$ is translated into the
Schr\"odinger equation on the wave function
${\bf Q}(\xi)$ and is given by \cite{FK}:
\be
-\left(\frac{(N+3)(N+2)}{\xi^2}+\frac{q}{\xi^3}
\right) {\bf Q}(\xi) = {\bf Q}(\xi+i)+{\bf Q}(\xi-i)-2 {\bf Q}(\xi).
\labeltest{Baxter}
\ee
This equation is known as the Baxter equation for the XXX Heisenberg magnet of
spin $s=-1$. In the WKB approach the wave function ${\bf Q}(\xi)$ describes
the wave function of the semiclassically quantized soliton wave in the separated
coordinates.

We will later show that the Baxter equation is equivalent to the
recurrence relations (\ref{masterrec}).
Main advantage of considering the ${\bf Q}$-function instead of the
set of coefficients $u_n$ is that it has all
intuitive properties of the wave function that
one is used to, whereas for $u_n$ it is difficult to
invoke any physical intuition.

As such, ${\bf Q}(\xi)$ should have a finite number of zeros in the
classically allowed region on the real $\xi-$axis whose position and the total
number is determined by the quantum numbers $N$ and $q$. The only
`physical' solution to the Baxter equation (\ref{Baxter}) satisfying
this condition defines ${\bf Q}(\xi)$ as a polynomial
in $\xi$ of degree $N+3$
\be
{\bf Q}(\xi)={\rm const}\times\prod_{k=1}^{N+3} (\xi-\lambda_k).
\labeltest{Roots}
\ee
Replacing ${\bf Q}(\xi)$ in (\ref{Baxter}) by this expression
we immediately find that the charge $q$ is quantized. Moreover,
putting $\xi\to\lambda_k$ in the Baxter equation it is easy to see that
for $q\neq 0$
one of the roots, $\lambda_{N+1}=\lambda_{N+2}=\lambda_{N+3}=0$,
is three times degenerate and the remaining $N$ roots satisfy the Bethe
equations corresponding to the spin $s=-1$ XXX Heisenberg magnet
\be
\left(\frac{\lambda_n+i}{\lambda_n-i}\right)^3=
\prod_{k=1, k\neq n}^{N}
\frac{\lambda_n-\lambda_k-i}{\lambda_n-\lambda_k+i}\,,\quad
n=1\,,...\,,N\,.
\ee
It can be shown \cite{K95} that the
solutions to the Bethe equations define the set of {\it real\/}
roots $\{\lambda_k\}$ which have the properties of the roots of
orthogonal polynomials and uniquely determine the ${\bf Q}-$function (\ref{Roots})
as well as the quantized values of the charge
\be
q = -\frac{{\bf Q}(i)+{\bf Q}(-i)}{\lim_{\xi\to 0} \xi^{-3} {\bf Q}(\xi)}
= - 2\,{\rm Im}\, \prod_{k=1}^N\left( 1-\frac{i}{\lambda_k} \right).
\ee
The explicit expression for the polynomial solution can be found for $q=0$
\be
{\bf Q}(\xi)\bigg|_{q=0}\equiv {\bf Q}_{N+1}(\xi)=
i^{N+1} (N+3)(N+2) \xi^2 \, {}_3F_2\left(
{N+4,-N-1,1-i\xi \atop 2,2}\bigg | 1
\right)\,.
\labeltest{Q-exp}
\ee
This expression defines the so-called Hanh orthogonal polynomials~\cite{K95}
and in the sequel we will use some of their properties:
\begin{eqnarray}
& &
\xi {\bf Q}_n(\xi) = -\frac{(n+2)(n+1)}{2(2n+3)} \left[
{\bf Q}_{n+1}(\xi)+{\bf Q}_{n-1}(\xi)\right],
\nonumber
\\
& &
{\bf Q}_n(\pm i) = -(\mp i)^n (n+2)(n+1)\,,
\labeltest{Q-prop}
\\
& &
\bigg(\ln {\bf Q}_n(\pm i)\bigg)' = \mp 2i \left[\psi(n+2)-\psi(1)\right],
\nonumber
\end{eqnarray}
where the prime denotes a derivative with respect to $\xi$.
Since the functions ${\bf Q}_n(\xi)$ form the complete set of orthogonal polynomials
we may seek for the general polynomial solution to (\ref{Baxter}) for $q\neq 0$
as an expansion over the $q=0$ solutions
\begin{equation}
{\bf Q}(\xi) = \sum_{n=0}^{N+1} u_n(q) f_n^{-1} \, {\bf Q}_n(\xi)
= -\frac{\xi}{q} \sum_{n=0}^N \frac{2(2n+3)}{(n+1)(n+2)}
u_n(q)\, {\bf Q}_n(\xi)\,.
\nonumber
\end{equation}
It is easy to check using (\ref{Q-exp}) and (\ref{Q-prop}) that thus defined
function ${\bf Q}(\xi)$ satisfies the Baxter equation (\ref{Baxter}) provided that
the coefficients $u_n$ satisfy the recurrence relations (\ref{masterrec}).
Thus, the analysis of the recurrence relations (\ref{masterrec}) is equivalent
to finding the polynomial solutions to the Baxter equation (\ref{Baxter}).
Moreover, comparing the relations (\ref{Q-exp}) and (\ref{expand2}) we observe
that the transition to the separated coordinates amounts to the replacement
 $i^n \varphi_n(z)\to {\bf Q}_n(\xi)$ in the expansion of the wave functions $\Phi(z_i)$
and ${\bf Q}(\xi)$, respectively.

It is worthwhile to note that
lengthy expressions for the spectrum of the Hamiltonian ${\cal H}_{3/2}$,
Eqs.~(\ref{energy1}) and (\ref{lambda-cn}),
take a remarkably simple form in terms of the ${\bf Q}-$function. In particular,
\begin{eqnarray}
\theta &=&-\frac{{\bf Q}(i)}{{\bf Q}(-i)}=\prod_{k=1}^{N}
\frac{\lambda_k-i}{\lambda_k+i},
\\
{\cal E} &=&
i\frac{{\bf Q}'(i)}{{\bf Q}(i)}
-
i\frac{{\bf Q}'(-i)}{{\bf Q}(-i)}-6
=\sum_{k=1}^{N}
\frac2{\lambda_k^2+1}.
\end{eqnarray}

To summarize, the Baxter equation (\ref{Baxter}) takes
the form of a finite-difference Schr\"o\-dinger equation with the conformal
spin $N+3$ playing the r\^ole of the (inverse) Planck constant.
Applying the standard WKB analysis
one can find the asymptotic expressions for the solutions
corresponding to classical
soliton waves propagating on the chain of 3 particles.
The quasimomentum of the soliton is characterized by an integer
number $\kappa = \frac{2\pi}{3}\ell$ and the proper frequency $\omega$
is a (complicated) function of conformal spin $N$.
Changing $N$ continuously with $\ell$ fixed  amounts to the adiabatic
deformation of the soliton solution.
This suggests that quantized values of energy and conserved charge $q$
in Fig.~2 belong to trajectories parameterized by the integer $\ell$
defining the quasimomentum in (\ref{Quasi}) and (\ref{soliton}).
One important property
of this deformation which is responsible for the analyticity of the
trajectories is that it does not destroy the wave packet but rather
induces the flow of its parameters with $N$ known as the Whitham flow \cite{K97}.
The precise definition of $\ell$ will be given below.

\subsection{Energy spectrum: WKB expansion}
\label{WKB-sect}

The WKB solution to the eigenvalue problem for ${\cal H}_{3/2}$ can be
based on the asymptotic behavior
of the recurrence relations (\ref{masterrec}) at large
$N$. In this limit it is convenient to introduce the scaling variables
\be
\eta=\sqrt{h(h-1)} = N+\frac52+\CO(1/N)
\,,\qquad
x=\eta^{-1}(n+\frac32)
\,,\qquad
\bar q=\eta^{-3} q
\ee
such that $x$ takes continuous values on the interval $0\le x \le 1$. At
large $N$, the $n-$dependent coefficients entering the recurrence relations
(\ref{masterrec}) become functions of $x$, which we define as
\be
f_n \equiv \eta^3 f(x)\,,\qquad
u_n \equiv u(x)\,.
\ee
{}From  the definition (\ref{pn}) we find the scaling function $f(x)$:
\begin{equation}
f(x)=\frac{1}{4x}\left[1-x^2+\frac{1}{4\eta^2}\right]
   \left[x^2-\frac1{4\eta^2}\right].
\end{equation}
Notice that there is no  $\CO(1/\eta)$ term
with our  definition of the scaling variables. At large $N$ the recurrence
relations (\ref{masterrec}) take the form of the second-order finite
difference equation
\be
u(x+\eta^{-1})+u(x-\eta^{-1})-2 u(x)=
\left(\frac{\bar q}{f(x)}-2\right)u(x)\,.
\labeltest{disc-Schr}
\ee
It has to  be supplemented by the boundary conditions (\ref{boundary})
which can be written as
\be
u\left(\frac1{2\eta}\right)=u\left(\sqrt{1+1/4\eta^2}\right)=0\,.
\labeltest{x-boundary}
\ee
The parity property (\ref{parity})
allows to restrict our consideration to positive values of
$\bar q$ only.

 It is convenient to interpret Eq.~(\ref{disc-Schr})  as a
discretized Schr\"odinger equation with $\eta^{-1}$ playing
the r\^ole of the Planck constant and $2-{\bar q}/f(x)$
the effective potential. It is then clear that the
`wave function' $u(x)$ has different behavior depending on the
sign of $2-{\bar q}/f(x)$. The interval of $x$, on which
$f(x) \le \bar q/2$, corresponds to the classically forbidden region
where $|u(x)|$ is a monotonous (decreasing or increasing) function of $x$.
The crucial observation is that for $\bar q\ge 0$ the
equation $f(x)={\bar q}/2$ has two real roots, $x_-$ and $x_+$,
on the interval $[0,1]$:
\be
f(x_\pm)=\frac12 \bar q
\labeltest{end-points}
\ee
for $\bar q \le 1/\sqrt{27}$ and
none for $\bar q > 1/\sqrt{27}$. In the latter case, $|u(x)|$
is a monotonous function of $x$ throughout the whole interval
$0\le x \le 1$ and the only way to satisfy the boundary conditions
(\ref{x-boundary}) is to put $u(x)=0$. Therefore, the recurrence relations
have nontrivial solutions satisfying (\ref{x-boundary}) in the former
case only, leading to the constraint on possible values of the charge $q$
\be
-\frac1{\sqrt{27}} \le \bar q \le \frac1{\sqrt{27}}\,,
\labeltest{range}
\ee
which one readily verifies using Fig.~\ref{figure2}a.
For the values of $\bar q$ in this range,  $u(x)$ grows (decreases)
on the interval $[0,x_-]$ ($[x_+,1]$) and has a local maximum(s)
on the interval $[x_-,x_+]$. The interval $[x_-,x_+]$ corresponds
to the classically allowed region for the Schr\"odinger equation
(\ref{disc-Schr}).

\subsubsection{Upper part of the spectrum}

We first consider the `upper' part of the spectrum
$\bar q \to 1/\sqrt{27}$. In this case, $x_\pm \to 1/\sqrt{3}$
with $x_+-x_-=\CO((\bar q-1/\sqrt{27})^2)$ and the interval
$[x_-,x_+]$ shrinks to a point. Assuming that $u(x)$
is a smooth function of $x$ on this interval, $|u'(x)/u(x)| \ll \eta$,
we replace Eq.~(\ref{disc-Schr})
in the leading $N\to \infty$ limit by the second-order differential
equation
\begin{equation}
-\frac{1}{\eta^2}
\frac{d^2 }{dz^2}u^{(0)}(z)+9z^2 u^{(0)}(z) = -\frac{2}{\eta}q^{(1)} u^{(0)}(z)\,,
\qquad
\labeltest{harmonic}
\end{equation}
with $z=x-\frac1{\sqrt{3}}$ and
\ba
& &u(x)=u^{(0)}(z)+ \CO(\eta^{-1/2})\,,
\labeltest{u0}
\\
& &\bar q=\frac1{\sqrt{27}}
\left[1+\eta^{-1} q^{(1)}+\CO(\eta^{-2})\right]\,,
\labeltest{q-exp}
\ea
which one recognizes as the Schr\"odinger equation for the
harmonic oscillator. Thus, we get readily
the quantized values of the charge
\be
q^{(1)}=-3\left(\ell+\frac12\right) \,,\qquad \ell=0\,,1\,, \ldots
\labeltest{oscillator}
\ee
and the coefficient function
\be
u^{(0)}(z)=H_\ell(\sqrt{3\eta}z)\,\exp{\left(-\frac{3\eta}{2} z^2\right)},
\labeltest{Hermite}
\ee
where $H_\ell(z)$ are the Hermite polynomials.

The following comments are in order. The solution (\ref{Hermite})
was found under the assumption that $u(x)$ is a smooth
function, $|u'(x)/u(x)|\ll \eta$. We verify that it is
satisfied indeed provided that $z\ll 1$ and $\ell\ll N$.
For higher excited states, $\ell\sim N$, we are approaching
the region $\bar q\to 0$, in which the solution is expected
to oscillate rapidly, cf. (\ref{q=0-coeff}), and the
above approximation does not work.

The quantized values of the charge, (\ref{oscillator}) and (\ref{q-exp}),
are enumerated by a nonnegative integer $\ell$ which counts
levels of the harmonic oscillator (\ref{harmonic}). Using
(\ref{oscillator}) and (\ref{q-exp}) as the definition of the
family of curves $q=q(N,\ell)$ for
continuous $N$ and discrete $\ell$ one obtains the trajectories
shown in Fig.~\ref{figure2}a. Namely, the largest values of the quantized $q$ for
any $N$ belong to the same trajectory with $\ell=0$, the
next-to-largest values ---  to the trajectory with $\ell=1$ and so on.

The integer $\ell$ has a simple interpretation in terms of the
solutions (\ref{Hermite}). As  expected,  $u(x)$
oscillates on the interval $[x_-,x_+]$. The integer $\ell$ counts
the number of its zeros and the solutions belonging to the same
trajectory for different $N$ all share this number.
Recall that considering properties of the exact solutions
we have found that they are parameterized by a discrete quantum number
which is the eigenvalue of the cyclic permutation operator (\ref{Quasi}).
An explicit calculation gives
\begin{equation}
  \theta(N,q) = e^{-i\phi(N,q)},~~~~\phi(N,q) = \frac{2\pi}{3}(N+\ell)\,.
\end{equation}

The approximation (\ref{q-exp}) can be systematically improved by taking
into account $\CO(1/\eta)$ corrections to Eq.~(\ref{harmonic}). This
allows to evaluate nonleading corrections to the spectrum (\ref{q-exp})
and (\ref{Hermite}).
We obtain
\begin{eqnarray}
 \bar q(N,\ell)&=&\frac1{\sqrt{27}}
 \left[1+ \frac{q^{(1)}}{\eta}+\frac{q^{(2)}}{\eta^2}+
    \ldots\right],
\nonumber\\
q^{(1)} &=& -3 (\ell+\frac{1}{2}),
\nonumber\\
q^{(2)} &=& \phantom{-}2\ell^2 + {2}\ell -\frac{13}{24},
\nonumber\\
q^{(3)} &=& -\frac{1}{9}(2\ell^3+3\ell^2+23 \ell +11),
\nonumber\\
q^{(4)} &=& \phantom{-}\frac{1}{31104}(3840 \ell^4+
     7680 \ell^3-112800 \ell^2-116640 \ell-90899).
\label{WKB-q}
\end{eqnarray}
Explicit expressions  up to
$\CO(\eta^{-8})$ can be found in \cite{K96,K97}.

\subsubsection{Lower part of the spectrum}

In the limit $\bar q\to 0$ the classical `turning points' $x_\pm$
are approaching the end-points
$x_-\to 0$ and $x_+\to 1$ where $u(x)$ must vanish.
The WKB analysis is not applicable in the
vicinity of these points and one has to solve the recurrence
relation (\ref{masterrec}) for small $n$ and $N-n$ directly,
by expanding $f_n$ in powers of $n/N$ and $(N-n)/N$, respectively:
\begin{eqnarray}
\bar q u_n&=&\eta^{-1}\frac{(n+1)(n+2)}{2(2n+3)}[u_{n+1}+u_{n-1}]\,,
\quad u_{-1}=0\,,
\qquad
\mbox{$n \ll N$},
\labeltest{small-n}
\\
\bar q u_n&=&\frac12\eta^{-1}(N-n)[u_{n+1}+u_{n-1}]\,,
\quad u_{N+1}=0\,,
\qquad
\mbox{$N-n \ll N$}.
\labeltest{large-n}
\end{eqnarray}
Quantization of $q$ appears as the condition for these two solutions
to match the WKB asymptotics in the regions $1 \ll n \ll N$
and $1 \ll N-n \ll N$, respectively, in which the WKB analysis is
still applicable.

For $x$ away from the end-point region, $x_- \ll x \ll x_+$ we look
for the WKB solutions to Eq.~(\ref{disc-Schr}) in the form
\be
u(x)=\rho \cos(\eta S(x))\,,
\ee
where  $\rho$ is a real normalization factor. Substitution of this ansatz
into (\ref{disc-Schr}) yields in the leading large $N$ limit the following
equation on the eikonal phase
\be
\cos(S'(x))=\frac{2\bar q}{x(1-x^2)}\,.
\ee
Solving it at small $\bar q\ll 1$ we obtain the leading WKB solution
for $n \gg 1$ and $N-n \gg 1$ as
\be
u_n=\rho\cos
\left(
\frac{\pi}2 n -\varphi -\bar q N \ln \frac{n^2}{N^2-n^2}
\right)
= \rho\,{\rm Re}\left[ i^n \e^{-i\varphi}\, \left(\frac{x^2}{1-x^2}
\right)^{-i\bar q N}
\right],
\labeltest{Sol-inter}
\ee
with $\varphi$ being an integration constant.
Using Eq.~(\ref{lambda-cn}) and replacing the sum over $n$
by the integral over $x$ one gets
\be
\e^{-i\varphi}=\theta\,,\qquad
\rho =(-1)^N |\Gamma(1-i\bar q N)|^{-2}.
\ee

The solution to the recurrence
relations for $n\ll N$ can be obtained by noticing the striking
similarity of (\ref{small-n}) with the
first relation in (\ref{Q-prop}) that describes properties of  the solution of the
Baxter equation for $q=0$, so that
\be
u_n= {\bf Q}_n(-\bar q\eta),
\labeltest{Sol-small}
\ee
with  ${\bf Q}_n$ as defined in (\ref{Q-exp}).
Finally, for $(N-n)\ll N$ it is easy to verify
that the solution to (\ref{large-n}) is given by
\be
u_{N-n}=i^{N-n} 
\int_{-1}^1 dx (1-x)^{i\bar q\eta -1} (1+x)^{-i\bar q\eta -1} x^{N-n}.
\labeltest{Sol-large}
\ee

The three expressions in  (\ref{Sol-inter}), (\ref{Sol-small}) and
(\ref{Sol-large}) correspond to the solution of the the recurrence relation
in the three different regions which overlap
however, for $1\ll n \ll N$ and $1 \ll N-n \ll N$.
Requiring that (\ref{Sol-inter}) can be sewed
with (\ref{Sol-small}) for $1\ll n \ll N$ and
with (\ref{Sol-large}) for $1 \ll N-n \ll N$,
we find the quantization condition on $q$
\be
q N^{-2}\ln N -{\rm arg}\, \Gamma(1+iqN^{-2}) +\CO(1/N) = \frac{\pi}6(N-2\ell),
\labeltest{LO-quant}
\ee
where  $\ell$ is  an integer.
 This result is valid to $\CO(1/N)$ accuracy
for small $q/N^3\ll 1 $ and
can significantly be improved by taking into account nonleading corrections
to (\ref{LO-quant}).
In this way one gets
\begin{equation}
\pi f(q,N)/3\equiv
q^{*}\eta \ln{\eta}-{\rm arg}\,\Gamma(1+i\eta q^{*})
+\frac{q^{*}}{6\eta}
-\frac{11}{6}\eta  {q^{*}}^{3}
 +{\cal O}(\eta{q^{*}}^{4})
 =\frac{\pi}{6}(N-2\ell),
\labeltest{q-low}
\end{equation}
where
$$
q^{*}=\bar q\, (1+2\bar q^{2})
\,.
$$
For given $\ell$, the quantized values of $q$ belong to the
$\ell-$th trajectory
$q=q(N,\ell)$ which depends analytically on $N$. It follows from (\ref{q-low})
that the function $q(N,\ell)$ has the reflection symmetry
\be
q(N,\ell)=-q(N,N-\ell)\,,
\labeltest{q->-q}
\ee
which maps positive values of $q$ on the $\ell-$th trajectory into the
negative $q$ on the $(N-\ell)-$trajectory.
The $\ell-$th trajectory crosses the
zero $q=0$ at even $N=2\ell$ and rises towards larger values of
$q$ corresponding to the `upper' part of the spectrum.

To illustrate this property,
we evaluate the function $f(q,N)$ for the exact numerical
values of $q$ belonging to the same trajectory shown in Fig.~\ref{figure2}a,
and plot it against $N$ as shown in Fig.~\ref{figure3}. It is seen that
the linear behavior in $N$ continues to the upper part of the spectrum
where  Eq.~(\ref{q-low}) can be matched with the WKB expansion in
(\ref{WKB-q}). In this way we can check that the definitions of $\ell$
in Eqs.~(\ref{WKB-q}) and (\ref{q-low}) do match each other and
describe the same trajectory.

Solving Eq.~(\ref{q-low}) for small $N-2\ell$ and large $\eta$ one gets
\begin{equation}
q/\eta^2
=\frac{\pi}{6\ln(\eta{\rm e}^{\gamma_E})}(N-2\ell )+\CO((\ell/\ln\eta)^3)
\labeltest{small-q}
\end{equation}
so that a few lowest eigenvalues of $Q$ are of order $\CO(\eta^2/\ln\eta)$.

\begin{figure}
\centerline{\epsfig{file=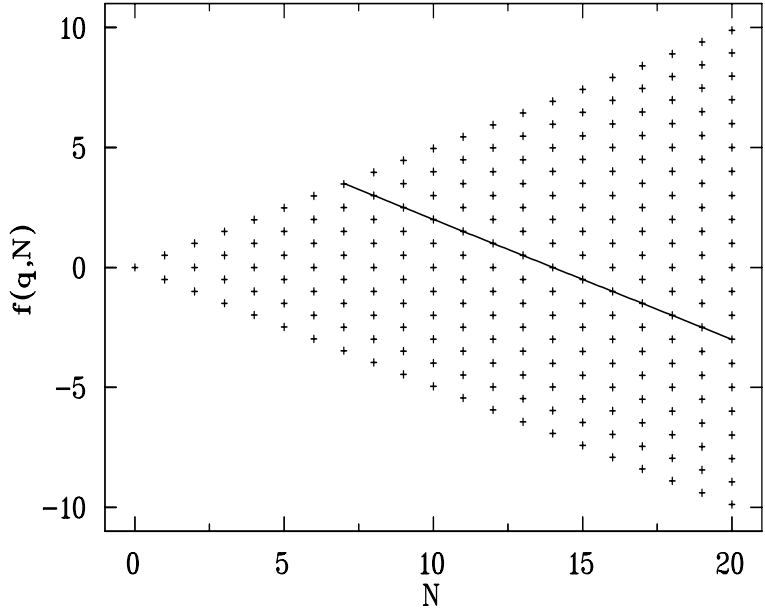, height=10.0cm, width=8.cm,
angle=90}}
\caption[]{\small
The trajectories for $q$ as given by Eq.~(\protect{\ref{q-low}}).
The crosses give  the values of the function $f(q,N)$ calculated using
 the exact eigenvalues $q$.}
\label{figure3}
\end{figure}

\subsubsection{Asymptotic expansion of the energy}

Let us use the WKB solutions to the recurrence relations to obtain the
asymptotic expressions for the energy.

For the `lower' part of the spectrum we substitute (\ref{Sol-inter}) into the
exact expression for the energy, Eq.~(\ref{energy1}), to get after the integration
\be
{\cal E}_q = 4\ln (N+3) - 6 + 6\gamma_E + 2{\rm Re}\, \psi(1+i N^{-2} q)\,.
\labeltest{Energy-low}
\ee
This expression is valid for $q=\CO(N^2)$ up to
corrections suppressed by powers of $1/N$ and, in particular, for
$q=0$ it reproduces the exact result (\ref{e0}).

A more accurate and general expression can be obtained
\cite{K96,K97} by asymptotic expansion of the Baxter equation:
\be
{\cal E}_q = 2\ln 2 - 6 + 6\gamma_E +
2 {\rm Re} \sum_{k=1}^3 \psi(1+i\eta^3\delta_k) + \CO(\eta^{-6}),
\labeltest{Energy}
\ee
where
$\delta_k$ are defined as roots of the following cubic
equation:
\be
2\delta_k^3 - \delta_k - \bar q=0
\ee
and $\bar q$ satisfies the condition (\ref{range}).
It is easy to see that for $q$ belonging to the interval $(\ref{range})$
all roots $\delta_k$ are real. The expression in (\ref{Energy})
is valid with high accuracy for the whole spectrum.
 For $q=\CO(N^2)$ both expressions,
(\ref{Energy-low}) and (\ref{Energy}), coincide.

\begin{figure}[th]
\centerline{\epsfxsize12cm\epsfysize10cm\epsffile{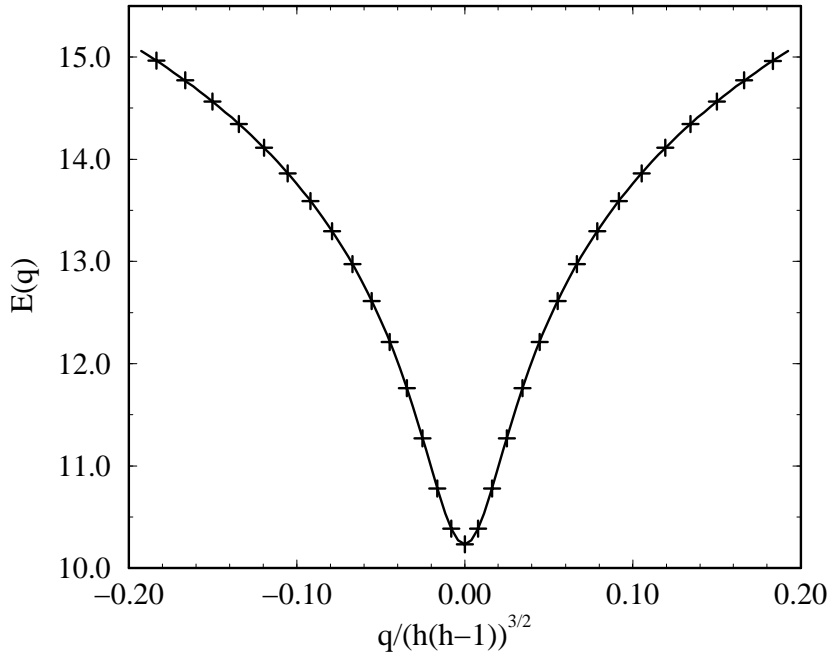}}
\caption[]{\small The dependence of the energy ${\cal E}$ on the
charge $q$ for $N=30$. The solid curve is calculated
using Eq.~(\protect{\ref{Energy}}) and the exact values of energy
for quantized $q$ are shown by crosses.}
\label{figure4}
\end{figure}

The resulting dependence of the energy ${\cal E}_q$ on the charge $q$ for
$N=30$ is shown in Fig.~\ref{figure4}.
We find from (\ref{Energy-low}) and (\ref{Energy}) that
the energy is quadratic in $q$ close to the origin $q=0$
\be
{\cal E}_q=4\ln(N+3) -6 + 4 \gamma_E + \frac{2\zeta(3)}{N^4} q^2,
\ee
with $\zeta(3)=1.20205690$, whereas at large $q=\CO(N^3)$ the asymptotic
behavior of the energy is given by
\be
{\cal E}_q = 2\ln q -6 + 6\gamma_E + \CO(N^{-2}).
\ee
We would like to stress that the expression (\ref{Energy}) defines the
dependence of the Hamiltonian on the conserved charges
${\cal H}_{3/2}={\cal H}_{3/2}(Q,L^2)$ for large eigenvalues of
$L^2$. To find the spectrum of the energy ${\cal E}$ one should replace
$q$ in (\ref{Energy}) by their quantized values.

Calculating the quantized values of the energy ${\cal E}_{q}$
we find that each trajectory
$q(\ell,N)$ is mapped into the corresponding trajectory for the energy
${\cal E}(\ell,N)$ as shown in Fig.~\ref{figure2}b. In particular,
the $\ell-$th trajectory starts at $N=\ell$, approaches the `Fermi
surface' ${\cal E}_{q=0}$ at $N=2\ell$, gets repelled from it and
monotonously grows to infinity at large $N$. The corresponding
asymptotic expression for the energy reads~\cite{K96,K97}:
\begin{eqnarray}
 {\cal E}(N,\ell) &=& 6\ln\eta-3\ln 3 -6 +6\gamma_E -\frac{3}{\eta}(2\ell+1)
- \frac{1}{\eta^2}\left(5\ell^2+5\ell -7/6\right)
\nonumber\\
&&{} - \frac{1}{72\eta^3}\left(464\ell^3+696\ell^2-802\ell -517\right)
+\ldots
\labeltest{upperE}
\end{eqnarray}
at large $N\gg \ell$, and
\begin{eqnarray}
{\cal E}(N,\ell)= 4\ln(N+3)+4\gamma_E-6+\frac{\pi^2\zeta(3)}
{18\ln^2(\eta\e^{\gamma_E})} (N-2l)^2
\labeltest{middleE}
\end{eqnarray}
in the vicinity of $N=2\ell$. To find the behavior around $N=\ell$ one
has to use Eqs.~(\ref{q->-q}) and (\ref{degeneracy}) to get
\begin{equation}
{\cal E}(N,\ell) = {\cal E}(N,N-\ell)
\labeltest{leftE}
\end{equation}
and substitue the expression in the r.h.s.\ by  (\ref{upperE}) with $\ell$
replaced by $N-\ell$.

The relations (\ref{upperE}), (\ref{middleE}) and (\ref{leftE}) define the
asymptotic expansion
for the energy levels of the Hamiltonian ${\cal H}_{3/2}$ parameterized
by the integer $\ell$. We observe that for given $N$ the distribution
of levels is different in the lower, $\ell\sim N/2$ or equivalently
$q\to 0$, and the upper part of the spectrum, $\ell \ll N$ or
$q\to 1/\sqrt{27}$. Using (\ref{upperE}) and (\ref{middleE}) we find the
corresponding level spacings as
\begin{equation}
\delta {\cal E}(N,\ell) \stackrel{q\to 0}{=}
   \CO\left(\frac1{\ln^2\eta}\right)\,,
\qquad
\delta {\cal E}(N,\ell) \stackrel{q\to 1/\sqrt{27}}{=}
 \CO\left(\frac1{\eta}\right)\,.
\labeltest{levelspacing}
\end{equation}

\subsection{Analytical continuation and the parton model}

Each polynomial eigenstate of the Hamiltonian ${\cal H}_{3/2}$ corresponds
to a multiplicatively renormalizable local operator and thus an
independent nonperturbative parameter in the distribution amplitude 
(\ref{expansion}). If, as usually assumed, the sum in (\ref{expansion}) is 
uniformly convergent pointwise in $x_i$, then the baryon distribution 
amplitude is restored uniquely from this expansion. 
The assumption of uniform convergence ensures that the distribution 
function vanishes as $x_1x_2x_3$ at the end points $x_i\to 0$ 
and implies that the nonperturbative reduced matrix elements 
decrease sufficiently fast for large conformal spins. If the  initial 
condition to the Brodsky-Lepage evolution equation (distribution amplitude 
at a low scale) decreases for $x_i\to0$ at a slower rate, then the series in 
(\ref{expansion}) diverges close to the end points and the scaling
behavior has to be defined  by a (infinite) resummation of the dominant
contributions of large conformal spins. This resummation can be performed
in the standard way by replacing an infinite sum over $N$ by an integral 
over complex $N$. To this end the analytic continuation in $N$ becomes 
necessary and, in particular, the anomalous dimensions 
$\gamma_N$ ought to be analytical functions of $N$.
It is this situation that occurs in the study of the $x\to1$ and $x\to0$ 
limits  of  an inclusive process with exchange of baryon quantum numbers 
in which case forward matrix elements of 
baryon operators contribute and the expansion in moments leads  to the
expansion of the corresponding generalized parton distribution
in derivatives of the $\delta$-function at $x=0$. 

As familiar from studies of deep inelastic scattering, restoration of parton
distributions from known values of the moments calculated within the
framework of the operator product expansion involves, first, an analytic 
continuation of the anomalous dimensions from integer positive $N$ corresponding
to spins of composite operators into the complex $N$ plane and, second,
decomposition of the distribution amplitude $\phi(x_i;\mu)$ for
arbitrary $x_i$ into irreducible components having an appropriate analytical 
(spectral) properties and admiting the parton model interpretation.
This procedure is well understood for leading twist parton distributions, 
see e.g. \cite{CFP81}, but, to our knowledge, has never been discussed for 
three-particle distributions. 

Mathematically, the first task consists of defining anomalous
dimensions as analytic functions of $N$ such that their values at 
positive integer $N$ are given by the eigenvalues of the evolution
kernel ${\cal H}_{3/2}$ and the asymptotics at infinity is such that
$\gamma_N \sim \exp(-\delta |N|)$ with $\delta < \pi$~\cite{Carlson}.
The anomalous dimensions of the different components of the distribution
amplitude do not necessary coincide and it is known from the 
studies of the deep inelastic scattering that one may need to consider 
analytic continuation from odd and even $N$ separately,
which in general correspond to contributions of operators
with different parity. 

The set of trajectories shown in Fig.~2
presents a legitimate analytic continuation and
reflects the highly nontrivial analytic structure of the integrable
model. This set is complimentary to a simpler and more general
analytic continuation
corresponding to ordering of the anomalous dimensions
from below, see Fig.~\ref{fig:anal}. 
\begin{figure}[th]
\centerline{\epsfxsize16cm\epsfysize8cm\epsffile{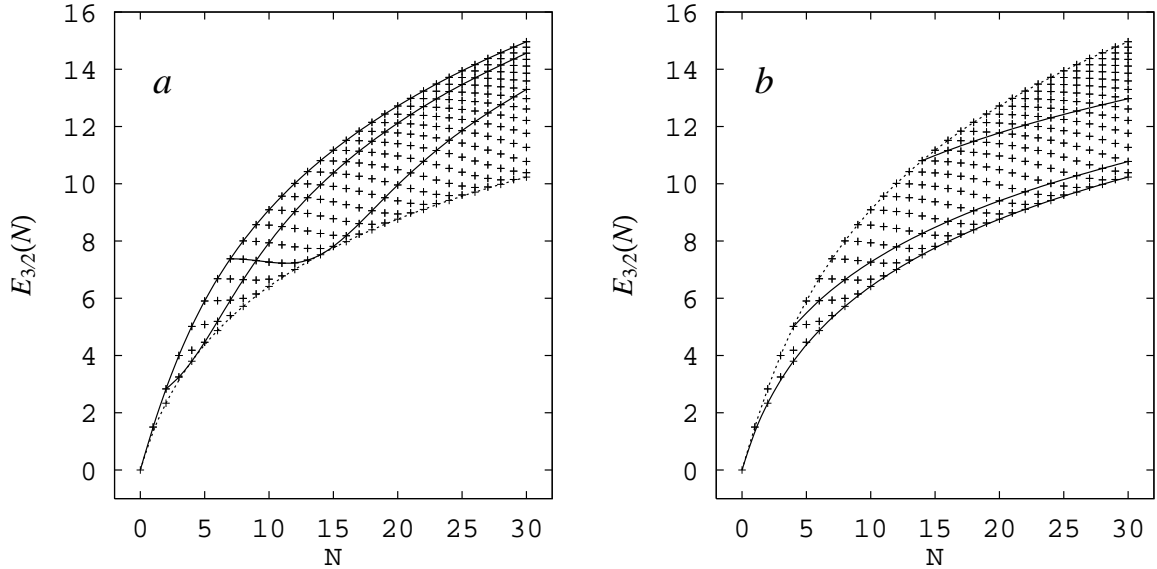}}
\caption[]{\small Two different analytic continuations
of eigenvalues  for the integrable Hamiltonian ${\cal H}_{3/2}$, see text.}
\label{fig:anal}
\end{figure}
 The trajectories shown in Fig.~\ref{fig:anal}a are copied from
Fig.~\ref{figure2}b. They are enumerated by an integer number $\ell$
defined in (\ref{oscillator}) the  physical interpretation of which
is discussed above at length. Going over to the trajectories
shown in Fig.~\ref{fig:anal}b corresponds to the rearrangement
of eigenvalues according to a different integer number $\bar\ell$
which is related to $\ell$
by a formal substitution
\begin{equation}
     \bar\ell = [N/2]-\ell,
\end{equation}
where $[N/2]$ denotes the integer part of $N/2$. The expression on
the right hand side of Eqs.~(\ref{LO-quant}),~(\ref{q-low}) then
becomes $\pi/3(\bar\ell+\delta_N/2)$ where $\delta_N=0$ and $\delta_N=1$
for even and odd values of $N$, respectively.
The assignment of the eigenvalues
to trajectories becomes, therefore,
different for odd and even $N$. The three trajectories shown in
Fig.~\ref{fig:anal}b correspond to $\bar\ell =0,2,7$ and correspond to
the analytic continuation from even $N$.
In particular,
the $\bar\ell=0$ trajectory going through the lowest eigenvalues
at even $N$ is given by Eq.~(\ref{e0}). Note that the
corresponding eigenstates all have positive parity.
The two (degenerate)
lowest trajectories for odd $N$ formally
correspond to $\bar\ell =0$ and $\bar\ell =-1$ and can further be rearranged
in contributions of definite parity. Note that, in contrast to
Fig.~\ref{fig:anal}a, each trajectory in Fig.~\ref{fig:anal}b
corresponds to a fixed eigenvalue $\theta$ of the cyclic
permutation operator. Another important difference is that each
trajectory in Fig.~\ref{fig:anal}a behaves as $\sim 6 \ln N$ at $N\to\infty$
while each trajectory in Fig.~\ref{fig:anal}b grows as $\sim 4 \ln N$ only.
The upper boundary $\sim 6\ln N$ (shown by dots)
arises in this case because new trajectories are being built on the top
of the spectrum starting at each integer $N$. In both cases the 
asymptotics of the anomalous dimension at large $N$ does not exhibit
an exponential growth and an analytical continuation to complex $N$
is unique.

We emphasize that both sets of trajectories define
legitimate analytic continuations and the one shown in Fig.~\ref{fig:anal}a
is made possible by an additional `hidden' symmetry on the integrable
Hamiltonian. The choice between them is defined by the 
process in which the baryon distributions are measured or, equivalently,
by the way in which the end-point region in $(x_1,x_2,x_3)-$space is 
approached. It remains to be studied which analytic continuation ensures
the true asymptotic behaviour of the distribution amplitudes
at the end points. This question goes beyond the
tasks of the present paper. 

\subsection{Eigenfunctions}

In this section we find an explicit expression for the eigenfunctions
$\tilde\Psi(x)$ in the large $N$ limit.
Requiring $Q\Psi = q\Psi$ yields a third-order differential equation
\begin{equation}
  (h-1)(h-2)\widetilde\Psi'(x)-2(h-2)x
  \widetilde\Psi''(x)-(1-x^2)\widetilde\Psi'''(x)
= -2i q\,\frac{1}{1-x^2}\widetilde\Psi(x),
\labeltest{fuchs}
\end{equation}
where $h=N+3$, which is  symmetric under the transformations
corresponding to cyclic permutations
$x_1\to x_2\to x_3\to x_1$
\begin{equation}
  x \stackrel{\cal P}{\longrightarrow}\frac{x-3}{x+1},~~~~~~~
  \widetilde\Psi(x)\stackrel{\cal P}{\longrightarrow}
    (-1)^h \left(\frac{x+1}{2}\right)^h
 \widetilde\Psi\left(\frac{x-3}{x+1}\right).
\labeltest{monodromy}
\end{equation}
Note that the cyclic permutation symmetry maps the interval
$[-1,1]\stackrel{\cal P}{\longrightarrow}[-\infty,-1]
\stackrel{\cal P}{\longrightarrow} [1,\infty]$.
It is sufficient, therefore,  to consider the region $-1<x<1$ only
since  the function $\tilde\Psi$ outside this interval can be
recovered by the transformation (\ref{monodromy}).

\subsubsection{WKB solution}

Eq.~\ref{fuchs} can be solved at large values of $\eta=\sqrt{h(h-1)}$
by the WKB expansion.
To this end we write
\begin{equation}
 \widetilde \Psi(x) = \exp\Big[ \eta\,S_0(x)+S_1(x)+{\cal O}(1/\eta)\Big],
\labeltest{WKB1}
\end{equation}
with $S_0,S_1,\ldots$ being $\eta$-independent.
We are going to use that values of the quantized charge
$\bar q = q/\eta^3$  are smaller than
$\bar q^2\leq 1/27$, see (\ref{range}). This allows to
expand the functions $S_0,S_1,\ldots$ in powers of $\bar q \ll 1$.

In particular, substituting the WKB ansatz (\ref{WKB1}) into the
differential equation  (\ref{fuchs}),
one gets in the leading $\eta\to\infty$ limit the following equation
for $S'_0\equiv\partial_x S_0(x)$:
\begin{equation}
(1-x^2)S_0'(x)[1-(1+x)S_0']\,[1+(1-x)S_0'] = -2 i\, \bar q.
\end{equation}
This cubic algebraic equation has three independent solutions
related to each other by the symmetry transformation (\ref{monodromy}).
Therefore, it is sufficient to consider only one of them, the one which
vanishes as $\bar q\to0$. The first few terms of its expansion in powers
of $\bar q$ are given by
\begin{equation}
  S'_0(x) = -\frac{2i\bar q}{1-x^2}-\frac{8\bar q^2\,x}{(1-x^2)^2}+
             8i\bar q^3 \frac{1+7x^2}{(1-x^2)^3}
    +160\bar q^4\,x\frac{1+3x^2}{(1-x^2)^4}+{\cal O}(\bar q^5).
\end{equation}
Integrating this result
and substituting $S_0(x)$ into (\ref{WKB1}) one gets the leading WKB asymptotics
\begin{equation}
\widetilde\Psi(x) \sim \left(\frac{1+x}{1-x}\right)^{-i\eta\bar q}
    \exp\left[ -\frac{4\eta\,\bar q^2}{1-x^2}\right],
\end{equation}
where the ${\cal O}(\bar q^3)$ and ${\cal O}(\bar q^4)$ terms are omitted
for simplicity.
To find the first nonleading correction to this expression, one further
expands the differential equation and keeps the ${\cal O}(1/\eta)$ terms.
In this way, one finds the expression for $S_1'(x)$ in terms of
$S_0'(x)$ and its higher derivatives that one integrates to get
\begin{equation}
 S_1(x) = {\mbox{\rm const}}
-\frac{4i\bar q x}{1-x^2}
+\frac{i\bar q}{2} \ln\frac{1+x}{1-x}.
\end{equation}
Finally, one obtains the following WKB approximation for the
eigenfunction
\begin{eqnarray}
 \widetilde\Psi^{\rm WKB}(x)=\mbox{\rm const}\times\!
     \left(\frac{1+x}{1-x}\right)^{-i\eta \bar q}\!\!\!\!\!
    \exp\!\left[ -\frac{4\eta\,\bar q^2}{1-x^2}-\frac{4i\bar q\,x}{1-x^2}
+{\cal O}(\eta\bar q^3)+{\cal O}(1/\eta)\right]\!.
\labeltest{PsiWKB}
\end{eqnarray}
The constant fixes the normalization of $\tilde\Psi$
and is otherwise arbitrary.
Applying the symmetry transformation (\ref{monodromy}) to this expression
one can construct the two remaining fundamental solutions to the
differential equation. The general solution is given by their
linear combination with two arbitrary constants. The latter can be fixed
from the requirement for $\tilde\Psi$ to be an eigenfunction of the
permutation operator. In this way, one finds the final expression
for the eigenfunction as
\begin{eqnarray}
\tilde\Psi(x) &=& \widetilde\Psi^{\rm WKB}(x)+\theta_q
(-1)^h \left(\frac{x+1}{2}\right)^h
 \widetilde\Psi^{\rm WKB}\left(\frac{x-3}{1+x}\right)
\nonumber\\&&{}
+\theta_q^2
  (-1)^h \left(\frac{1-x}{2}\right)^h
 \widetilde\Psi^{\rm WKB}\left(\frac{x+3}{1-x}\right).
\labeltest{sym}
\end{eqnarray}
This approximation is valid for large $\eta$ and for all real $x$ except
in the vicinity of the singular points $x=-1,1,\infty$.
Note that the two added terms in (\ref{sym})
represent `quantum' corrections to the WKB solution in (\ref{PsiWKB})
in the region $-1<x<1$,
which are not seen to all orders in the $1/\eta$ expansion.

\begin{figure}[t]
\centerline{\epsfxsize8.0cm\epsfysize8.0cm\epsffile{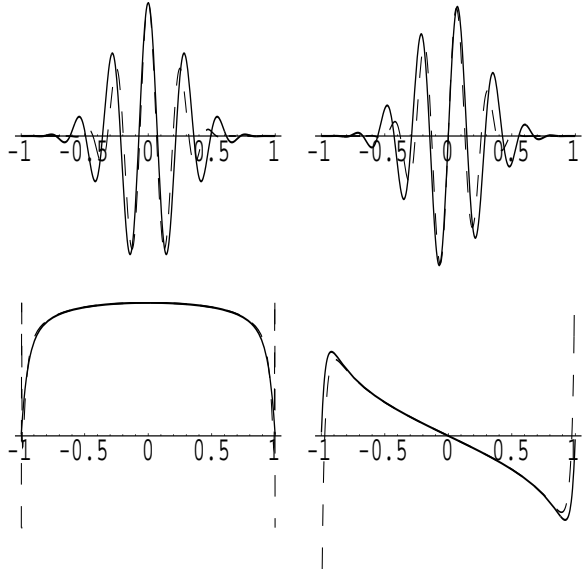}}
\caption[]{\small Exact (solid lines) and WKB (dashed lines) eigenfunctions
$\widetilde\Psi(x)$ for $N=60$. The figures in the first and the second
row correspond
to the maximum value of $q$ (alias energy) and minimum nonzero value
of $q$, respectively.
The left and the right figures show real and imaginary parts of
$\widetilde\Psi(x)$, respectively..}
\label{figure5}
\end{figure}

\subsubsection{Resummation of leading 
corrections}

As seen from the above, the expansion in powers of $\bar q$
actually proves to be the expansion in $4\bar q/(1-x^2)$ and
is compromised close to the end points. One can improve the WKB expansion
in the previous subsection by making a  resummation of the leading
singular $[\bar q/(1-x)^2]^k$ terms to all orders.
To this end, we consider the limit $x\to1$ of the
differential equation Eq.~(\ref{fuchs}):
\begin{equation}
 S_0'[1-2S_0']=-i\bar q/(1-x),
\end{equation}
where from
\begin{equation}
 S_0' = \frac{1}{4}\left(1-\sqrt{1+\frac{8i\bar q}{1-x}}\right).
\end{equation}
Integrating this relation and adding a similar contribution from
$x\to-1$ we get
\begin{eqnarray}
\lefteqn{S_0^{\rm res}(x)= -i\bar q\ln\left(\frac{1+x}{1-x}\right)
  -2i\bar q\ln\left[\frac{1+\sqrt{1-8i\bar q/(1+x)}}
{1+\sqrt{1+8i\bar q/(1-x)}}\right]}&&
\nonumber\\
&&{}-\frac{1+x}{4}\left(1-\sqrt{1-8i\bar q/(1+x)}\right)
  -\frac{1-x}{4}\left(1-\sqrt{1+8i\bar q/(1-x)}\right).
\end{eqnarray}
Adding the less singular terms and collecting everything, we get the
leading-order resummed WKB eigenfunction:
\begin{eqnarray}
 \widetilde\Psi_{\rm res}^{\rm WKB}(x)&=&\mbox{\rm const}\times
     \left(\frac{1+x}{1-x}\right)^{-i\eta\bar q(1+2\bar q^2)}
     \left[\frac{1+\sqrt{1-8i\bar q/(1+x)}}
          {1+\sqrt{1+8i\bar q/(1-x)}}\right]^{-2i\eta\bar q}
\nonumber\\&&{}\times
\exp\left\{-\eta\frac{1+x}{4}\left(1-\sqrt{1-8i\bar q/(1+x)}\right)\right\}
\nonumber\\&&{}\times
\exp\left\{-\eta\frac{1-x}{4}\left(1-\sqrt{1+8i\bar q/(1-x)}\right)\right\}
\nonumber\\&&{}\times
\exp\left\{-\frac{4i\bar q(1+\eta \bar q^2)x}{1-x^2}
   -\frac{40\eta\bar q^4}{(1-x^2)^2}
+{\cal O}[\eta\bar q^5(1-x^2)^{-3}]\right\}
\labeltest{PsiWKBres}
\end{eqnarray}
which has to be inserted in (\ref{sym}) and
presents our final result.

The numerical comparison of the exact and the WKB eigenfunctions
in presented in Fig.~\ref{figure5} for $N=60$. This large value
of $N$ is chosen to illustrate that the eigenfunctions corresponding
to large energy eigenvalues (the two upper figures)
have a typical wave packet structure:
The size of the packet is of order $1/\sqrt{N}$ and the oscillation
frequency of order $\sim 1/N$. The eigenfunctions corresponding
to lowest eigenvalues are, on the contrary, smooth functions
(the two lower figures) for which the WKB approximation works very well.

\section{Helicity $\lambda=1/2$ distribution amplitudes}
\setcounter{equation}{0}
\labeltest{Pert}

The scale dependence of the $\lambda=1/2$ distribution amplitudes is driven
by the Hamiltonian ${\cal H}_{1/2}$ defined in (\ref{H12}), which
differs from ${\cal H}_{3/2}$ by the two terms corresponding to gluon
exchange between quarks of opposite chirality, see Fig.~\ref{figure1}:
\begin{equation}
{\cal H}_{1/2} = {\cal H}_{3/2} + {V}\,,\qquad
{V}= -\left(\frac1{L_{12}^2}+\frac1{L_{23}^2}\right).
\labeltest{V}
\end{equation}
The spectrum of eigenvalues of ${\cal H}_{1/2}$ corresponds to the
spectrum of anomalous dimensions in the evolution equation.
The Hamiltonian ${\cal H}_{1/2}$ is not integrable and the
corresponding eigenproblem cannot be solved exactly. For a given $N$,
the spectrum and the eigenfunctions can most efficiently
be calculated by a numerical diagonalization of the mixing
matrix for the additional --- exchange interaction --- terms evaluated in the
basis of the exact eigenstates of  ${\cal H}_{3/2}$. The reason is that this
matrix is strongly peaked at the diagonal: Matrix elements
$\langle \Psi_{q'}|1/L_{ik}^2|\Psi_q\rangle$ between the
${\cal H}_{3/2}$ eigenstates labeled by the
values of the conserved charge $q$ and $q'$ decrease rapidly with $|q-q'|$,
see Fig.~\ref{fig:pipka}.
\begin{figure}
\centerline{
   \epsfig{file=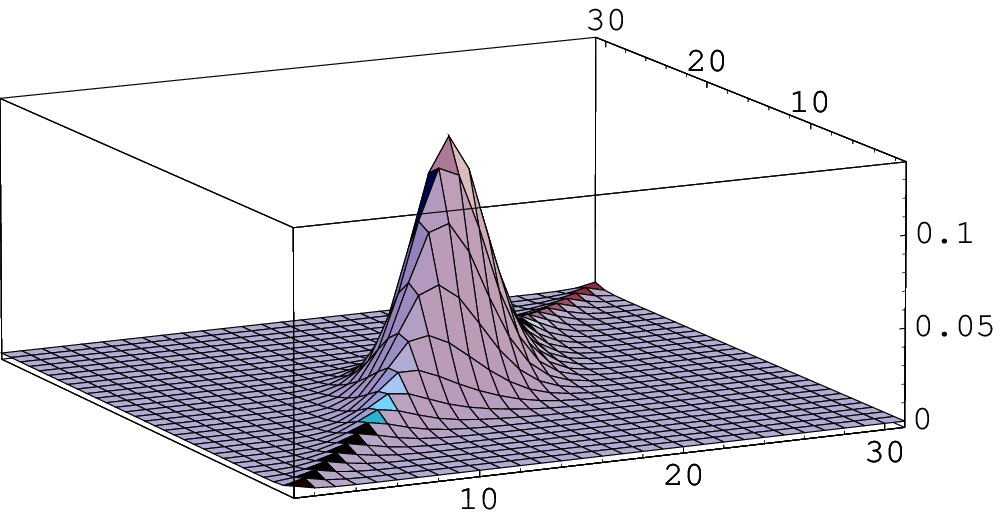, height=7.5cm, width=6.5cm}
   \epsfig{file=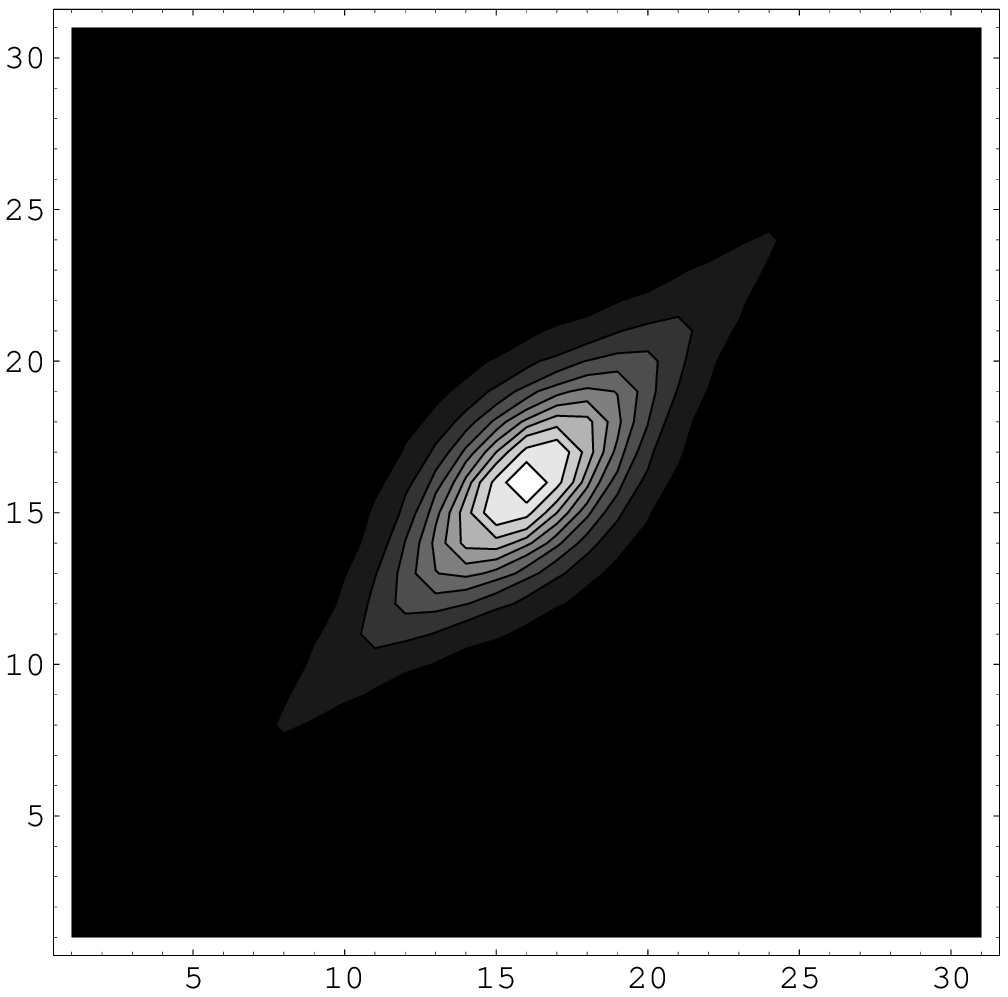, height=7.5cm, width=6.5cm}
           }
\caption[]{\small
Matrix elements of the exchange interaction
$|\langle \Psi_{q(\ell')}|1/L_{12}^2|\Psi_{q(\ell)}\rangle|$  evaluated between the
${\cal H}_{3/2}$ eigenstates labeled by the
integer $\ell,\ell'=0,\ldots N+1$ which
enumerates the quantized values of $q$ and $q'$
from the above, see Sect.~4.5, for $N=30$. The picture to the right shows
the contour plot of the 3-dimensional plot.
}
\label{fig:pipka}
\end{figure}
In contrast to ${\cal H}_{3/2}$,
the Hamiltonian ${\cal H}_{1/2}$ is not invariant under cyclic permutations
but, still, is symmetric  under the interchange of the first and the third
quarks, $[{\cal H}_{1/2},{\cal P}_{13}]=0$. This allows to choose its
eigenstates to have definite parity with respect to the ${\cal P}_{13}$
permutations
\begin{equation}
{\cal H}_{1/2} \Psi^{(\pm)}_{\ell,N}(x_i) = {\cal E}^{(\pm)}_{1/2}(N,\ell)
     \Psi^{(\pm)}_{\ell,N}(x_i)\,,\qquad
{\cal P}_{13} \Psi^{(\pm)}_{\ell,N}
(x_i) = \pm \Psi^{(\pm)}_{\ell,N}(x_i)\,.
\end{equation}
Here, $N$ refers to the total number of derivatives and $\ell$ numerates
the energy eigenstates. It is therefore natural
to decompose the eigenfunctions $\Psi^{(\pm)}(x_i)$
over the basis of eigenstates of
${\cal H}_{3/2}$ with definite parity\footnote{Throughout this
section we define parity with respect to permutation of the
first and the third quark, instead of the first and the second quark
in Sect.~4. To account for this change, we also use the basis of
the `permuted' ${\cal H}_{3/2}$ eigenstates
${\cal P}_{23} \Psi_q^\pm$, with $\Psi_q^\pm$ defined in
(\ref{eigen_parity}).
We usually  omit ${\cal P}_{23}$ and use the same notation $\Psi_q^\pm$ to
simplify the presentation.}
$\Psi_{q,N}^{(\pm)}(x_i)$ and $q\ge 0$,
defined as in (\ref{eigen_parity}).
Using  the identity
$
L_{12}^{-2}+L_{23}^{-2} = {\cal P}^2 L_{13}^{-2} {\cal P} +
{\cal P} L_{13}^{-2} {\cal P}^2
$
and taking into account that the states $\Psi_q$ diagonalize the
cyclic permutation operator ${\cal P}$ we obtain
\begin{equation}
\langle \Psi_{q',N}^{(\pm)} | {V} | \Psi_{q,N}^{(\pm)} \rangle
=
-2 \sqrt{{\cal N}_q{\cal N}_{q'}}
\sum_{m=0}^N
\frac{f_m^{-1}\,u_m(q')u_m(q)}{(m+1)(m+2)}
\left[
\cos(\phi_q-\phi_{q'})\pm (-1)^m\! \cos(\phi_q+\phi_{q'})
\right],
\labeltest{matr_elem}
\end{equation}
where $u_n(q)$ correspond to the expansion coefficients
of the ${\cal H}_{3/2}$ eigenfunctions in the $\Psi_{n,N}^{(31)2}$
basis and
\begin{equation}
\langle \Psi_{q',N}^{(\pm)} | {V} | \Psi_{q,N}^{(\mp)}
\rangle = 0\,.
\end{equation}
The factor
\begin{equation}
  {\cal N}^{-1}_q = \sum_{m=0}^N {f_m^{-1}u_m(q)u_m(q)}
\labeltest{normalization}
\end{equation}
comes from the normalization condition for the states, which we
assume in this section to be $
\langle \Psi_{q,N}^{(\pm)} | \Psi_{q,N}^{(\pm)} \rangle =1$.

The explicit calculation gives the spectrum shown
in Fig.~\ref{fig:e12}.
\begin{figure}
\centerline{\epsfxsize10cm\epsffile{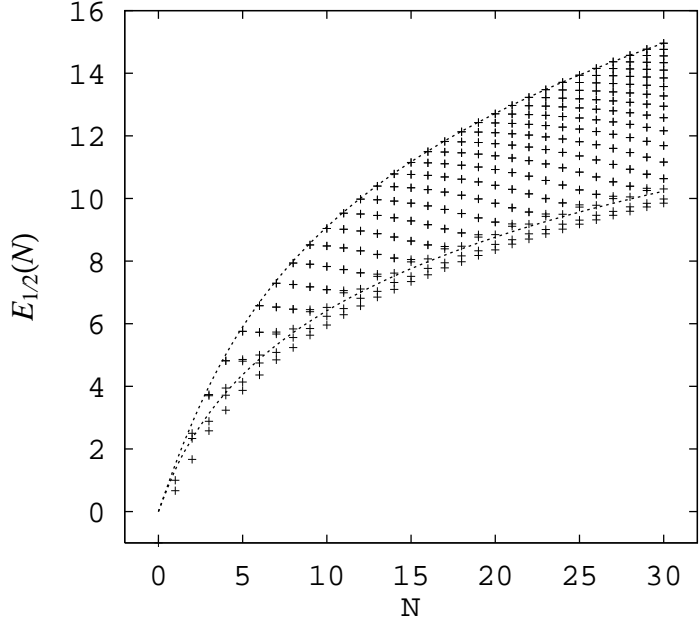}}
\caption[]{\small The spectrum of eigenvalues for the
Hamiltonian ${\cal H}_{12}$. The lines of the largest and the smallest
eigenvalues of ${\cal H}_{32}$ are indicated by  dots
for comparison.}
\label{fig:e12}
\end{figure}
The lines of the largest and the smallest
eigenvalues of ${\cal H}_{3/2}$ are indicated by  dots for comparison.

As seen from the figure, the spectra of ${\cal H}_{1/2}$ and ${\cal H}_{3/2}$
are very similar in the upper part, for larger eigenvalues, and at the same
time the two lowest levels of the ${\cal H}_{1/2}$ Hamiltonian appear to
be special and `dive' considerably below the line of lowest
eigenvalues of ${\cal H}_{3/2}$, given by Eq.~(\ref{e0}).
Our goal in this section is to explain this structure and to get
the quantitative description of the ${\cal H}_{1/2}$ spectrum in the large
$N$ limit. Note that at $N\to\infty$ the spectrum of
${\cal H}_{3/2}$ becomes
very dense, see Eqs.~(\ref{levelspacing}),
and approaches a continuos spectrum inside the band of the
width $\sim 2\ln N$:
$4 \ln N - 6 < {\cal E}_{3/2}< 6 \ln N -6 -3\ln 3$.
We will demonstrate that
exactly the same band of the continuos spectrum is
formed for the ${\cal H}_{1/2}$. Inside the band, the
distribution of levels is
perturbed by corrections
at most $\CO(1/ N^2)$  and $\CO(1/\ln^2 N)$ at the upper  and the lower
boundary, respectively. In addition, the two lowest eigenstates
of ${\cal H}_{1/2}$ (one for each parity) fall below the `Fermi surface'
and are separated from the bottom of the band by a finite constant.
Existence of such a `mass gap' presents our main result
in this section and its formation will be interpreted as due to binding
of quarks with opposite chirality by the exchange interaction and
formation of scalar diquarks. The eigenfunctions
of the `bound states' and the value of the `mass gap' will be
estimated.

To visualize both the similarities and the differences between
the spectra of ${\cal H}_{3/2}$ and ${\cal H}_{1/2}$ and to
trace formation of the `mass gap' for ${\cal H}_{1/2}$,
it proves convenient to introduce a somewhat more general Hamiltonian
\begin{equation}
{\cal H}(\epsilon) = {\cal H}_{3/2} + \epsilon {V}\,,\qquad
\labeltest{Heps}
\end{equation}
with $\epsilon$ being a new coupling constant. ${\cal H}(\epsilon=0)$
reproduces ${\cal H}_{3/2}$ whereas ${\cal H}(\epsilon=1)$ coincides
with the Hamiltonian
${\cal H}_{1/2}$.
Thus, the spectra of ${\cal H}_{3/2}$ and ${\cal H}_{1/2}$ are related to
each other through the flow of the energy levels of ${\cal H}(\epsilon)$
from $\epsilon=0$ to $\epsilon=1$, see Fig.~\ref{figure7}.
\begin{figure}
\centerline{
   \epsfig{file=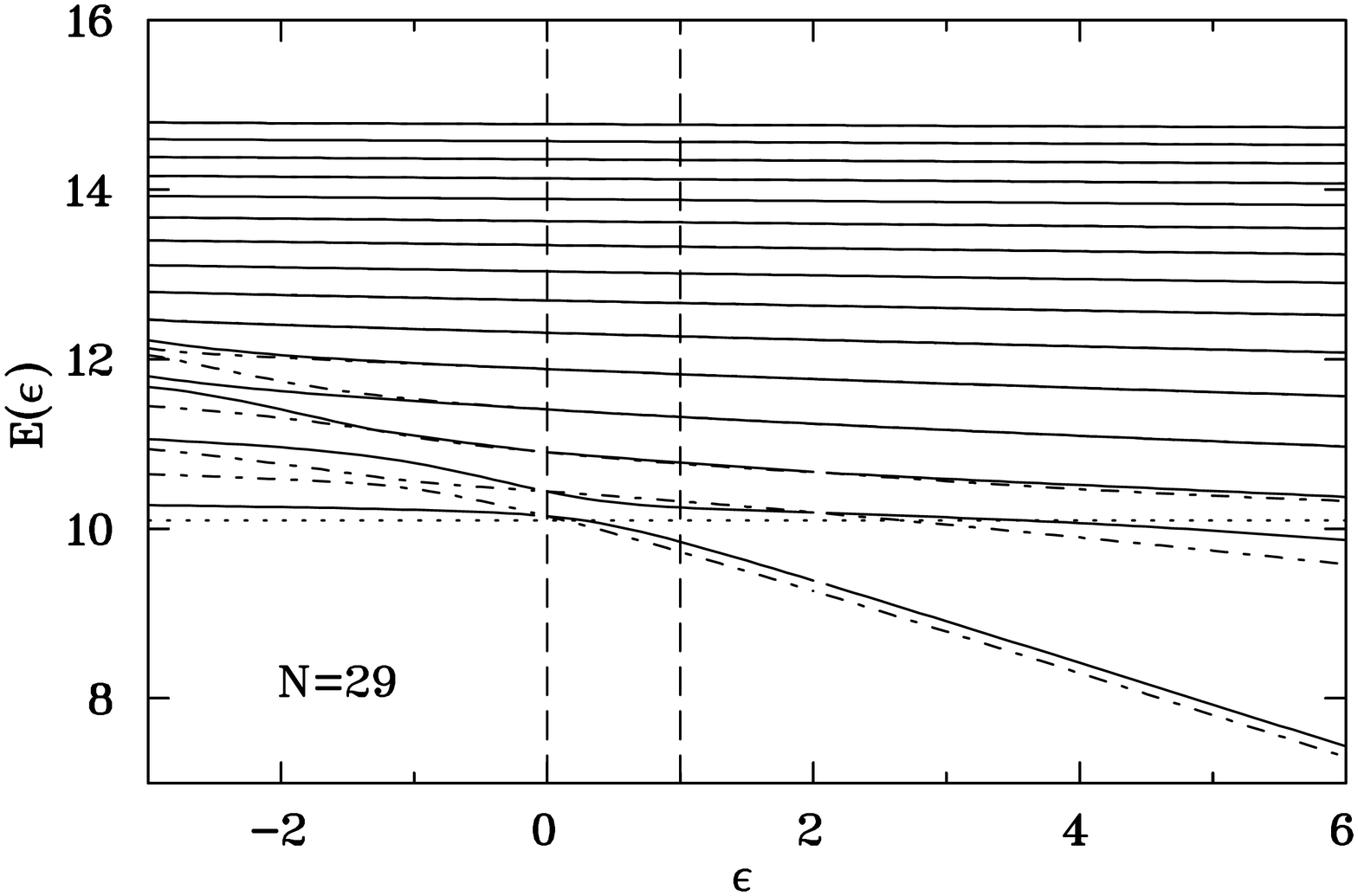, height=7.5cm, width=6.5cm, angle=90}
   \epsfig{file=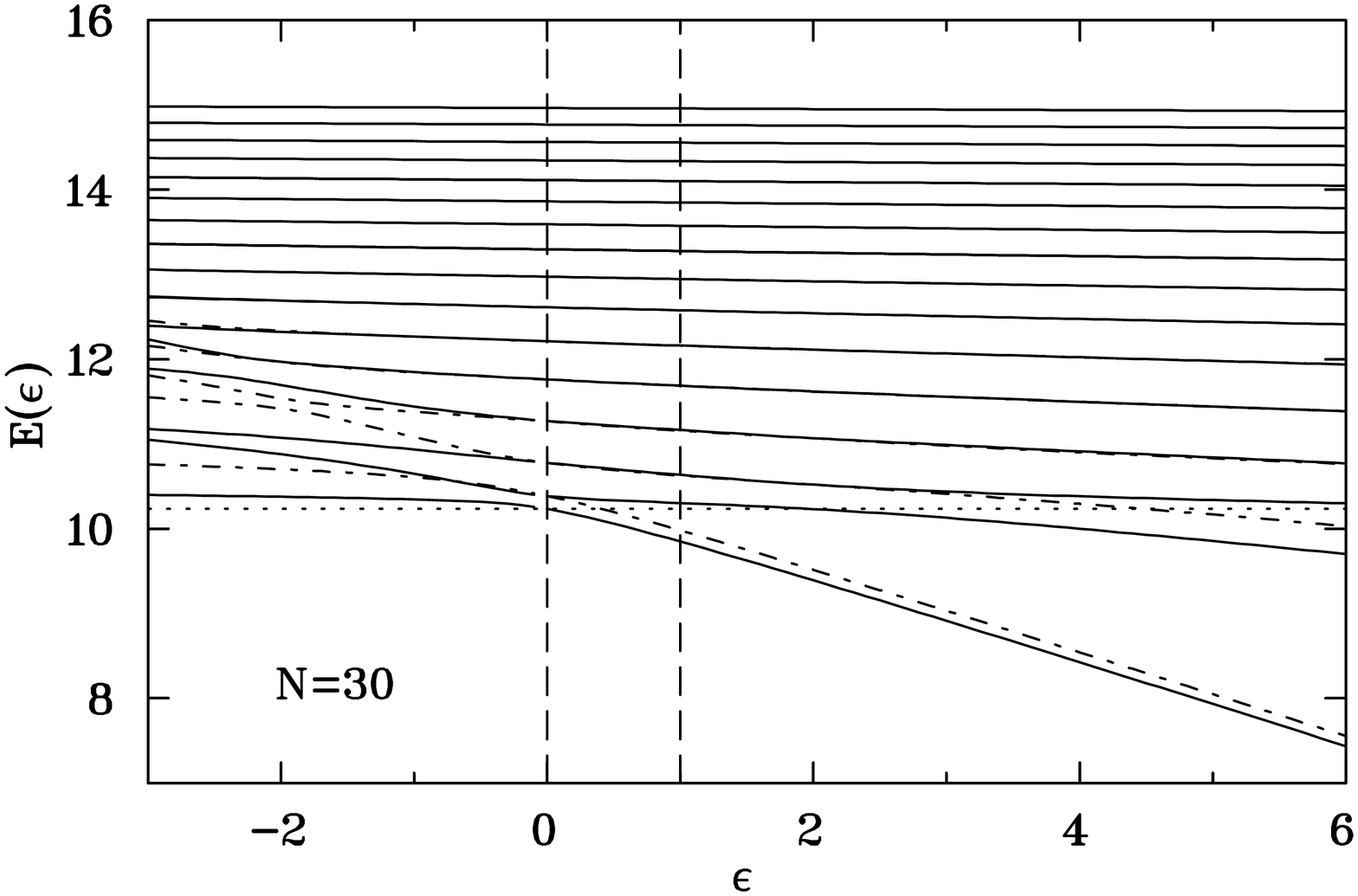, height=7.5cm, width=6.5cm, angle=90}
           }
\caption[]{\small The flow of energy eigenvalues for the Hamiltonian
   ${\cal H}(\epsilon)$ for $N=29$ and $N=30$, see text.
The solid and the dash-dotted curves show
the parity-even and parity-odd levels, respectively.
The two vertical dashed lines
indicate ${\cal H}_{3/2}\equiv{\cal H}(\epsilon=0)$ and
${\cal H}_{1/2}\equiv{\cal H}(\epsilon=1)$, respectively. The horizontal
dotted line shows position of the `ground state' given by
Eq.~(\protect{\ref{e0}}).
}
\label{figure7}
\end{figure}
Note that for  $\epsilon\neq 0$ the Hamiltonian ${\cal H}(\epsilon)$
is neither integrable, nor cyclic symmetric.
It is still invariant  under conformal transformations
and under permutations of the first and the third quarks,
$[{\cal H}(\epsilon),{\cal P}_{13}]=0$, but the degeneracy
between parity-odd and parity-even eigenstates is lifted and,
in fact, the flow of levels with different parity is completely
independent from one another.

The spectra in Fig.~\ref{figure7} exhibit the following characteristic
features:
\begin{enumerate}
\item[--] In the upper part of the spectrum
the effect of $\epsilon-$proportional terms on the
spectrum of the `unperturbed'
Hamiltonian ${\cal H}(\epsilon=0)={\cal H}_{3/2}$ is very mild.
While at $\epsilon=0$ the energy levels are double degenerate,
their splitting
at $\epsilon\neq 0$ remains  (exponentially, as we will argue)
small for large $N$.
\item[--] For $\epsilon>0$, the two lowest
levels are decoupled from the rest of the spectrum and fall off
with $\epsilon$ almost linearly.
For $\epsilon<0$, the levels with different parity start to
cross each other, whereas the flow of the levels with the same parity
follows the pattern well known in
quantum mechanics as the `repulsion of levels' \cite{QM}.
\end{enumerate}
This structure suggests that the difference
$\epsilon {V} = {\cal H}(\epsilon) - {\cal H}(\epsilon=0)$
can be considered as a perturbation for most
of the levels, but not for
the few lowest ones (for large $N$).
To formalize the argument, one has to evaluate the matrix elements
in (\ref{matr_elem})
and compare them with
the energy splittings for the `unperturbed' Hamiltonian.
The explicit calculation (see below) gives
\begin{equation}
\langle \Psi_{q',N}^{(\pm)} | {V} | \Psi_{q,N}^{(\pm)} \rangle
 \stackrel{q,q'\to 0}{=}
   \CO\left(\frac1{\ln N }\right)\,,
\qquad
\langle \Psi_{q',N}^{(\pm)} | {V} | \Psi_{q,N}^{(\pm)} \rangle
\stackrel{q,q'\to 1/\sqrt{27}}{=}
 \CO\left(\frac1{N^2}\right)\,.
\labeltest{levelperturbation}
\end{equation}
Comparing this result with the level spacings in Eq.~(\ref{levelspacing}),
we conclude that the perturbation theory in ${V}$ is justified
for large $N$ (or for small $\epsilon \ll 1$ and arbitrary $N$) for the
upper part of the spectrum, while several (of order
$\sim \ln N$ as we will find) lowest energy eigenstates are affected
strongly and have
to be rediagonalized (unless $\epsilon \ll 1/\ln N$).
In the sequel, we are going to consider the two different regions
separately in more detail.

\subsection{Upper part of the spectrum}

Eq.~(\ref{matr_elem}) becomes, for diagonal transitions $q'=q$
\begin{equation}
\rvev{V}_{q,N}^{(\pm)}\equiv
{\langle \Psi_q^{(\pm)}|{V}
|\Psi_q^{(\pm)}\rangle}
= -2{\cal N}_q \sum_{m=0}^N
\frac{f_m^{-1}u_m^2(q)}{(m+1)(m+2)}
\left[1\pm (-1)^m \cos(2\phi_q)\right]\,.
\labeltest{upper_shift}
\end{equation}
Each term in the sum is explicitly positive so that
the matrix element $\rvev{V}_{q,N}^{(\pm)}$ is always negative
meaning that for all levels
${\cal E}_{3/2}-{\cal E}_{1/2} > 0$.
According to (\ref{Hermite}), the coefficients $u_m(q)$ are smooth functions
of the scaling variable
$x=m/N$, peaked around $x=1/\sqrt{3}$ and rapidly decreasing outside
the region $(x-1/\sqrt{3})^2\le \eta^{-1}/3$. Splitting the sum
in (\ref{upper_shift}) into contributions of even and odd $m$ one finds
that the terms proportional
to $\cos(2\phi_q)$ tend to cancel each other and their total
contribution is approximately given
by the sum of two boundary terms.
This contribution is negligible compared to the sum of the phase-independent
terms, but at the same time it defines the splitting between energy
levels with the different parity
\begin{equation}
{\cal E}_{1/2}^{(+)}-{\cal E}_{1/2}^{(-)}\sim
\frac{(-1)^mf_m^{-1}u_m^2(q)}{(m+1)(m+2)}
\bigg|_{m_{\rm min}}^{m_{\rm max}}\,.
\end{equation}
Here $m_{\rm min}/N\ll 1$ and $1-m_{\rm max}/N\ll 1$ define
the interval, $m_{\rm min} \le m\le m_{\rm max}$, on which the WKB
expansion (\ref{Hermite}) is applicable. Applying similar arguments to
the sum entering the matrix element
\begin{equation}
 0 = \langle \Psi_q | \Psi_{-q}\rangle \sim \sum_{m=0}^N (-1)^m f_m^{-1}
u_m^2(q)\,,
\end{equation}
we conclude that the values of $(-1)^m f_m u_m^2$ have to coincide
at the end points so that
\begin{equation}
{\cal E}_{1/2}^{(+)}-{\cal E}_{1/2}^{(-)}\sim
\frac{(-1)^{m_{\rm min}}f_{m_{\rm min}}^{-1}u_{m_{\rm min}}^2(q)}
{(m_{\rm min}+1)(m_{\rm min}+2)}
\sim \eta^{-3+\ell} \exp(-\eta)\,.
\label{split+-}
\end{equation}
Thus, the splitting between the energy levels of different parity is
governed by the tail of the wave function (\ref{Hermite}) and, as a
consequence, is exponentially small at large $N$.

Neglecting exponentially small terms and, in particular, the
level splitting ${\cal E}_{1/2}^{(+)}-{\cal E}_{1/2}^{(-)}$,
we can replace the sum in (\ref{upper_shift}) by the integral over the
scaling variable $x=m/N$ and substitute the coefficients $u_m$ by
their WKB expansion (\ref{u0})
with the leading term given by (\ref{Hermite}).
Substituting this expansion into (\ref{normalization}) and (\ref{upper_shift})
one can calculate
the leading and the next-to-leading corrections to the energy (\ref{upper_shift})
as%
\footnote{To calculate the next-to-leading correction to this expression one
needs to know the $\CO(\eta^{-1/2})-$term in the WKB expansion (\ref{u0}) of $u(x)$.
It can be obtained by substituting (\ref{u0}) into (\ref{disc-Schr}) and comparing the
coefficients in front of the nonleading power of $1/\eta$.}
\begin{equation}
{\cal E}_{1/2}^{(\pm)}(\ell)-{\cal E}_{3/2}(\ell) \simeq
\rvev{V}_{q,N}
= - \frac{6}{\eta^2}\left[1+\frac{2(2\ell+1)}{\eta}\right]
+\CO(1/\eta^4),
\labeltest{upper_PT}
\end{equation}
verifying the estimate in (\ref{levelperturbation}).
We remind that $\eta = \sqrt{(N+3)(N+2)}$.

\subsection{Lower part of the spectrum}

The analysis of the low part of the spectrum is considerably
more involved.

We start with calculation of the normalization factor ${\cal N}_q$
defined in (\ref{normalization}). Assuming $\bar q \equiv q/\eta^3 \ll 1$
one can use the WKB approximation (\ref{Sol-inter}) for $u_m(q)$ to get
\begin{equation}
{\cal N}_q=|\Gamma(1-i\bar qN)|^4 \left(\sum_{n=0}^{[N/2]} 1/f_{2n}\right)^{-1}
= \frac23 \eta^{2}\ln^{-1}(\eta \e^{\gamma_E})\,
\left(\frac{\pi \bar q N}{\sinh(\pi \bar q N)} \right)^2\,.
\labeltest{N-factor}
\end{equation}
This expression defines ${\cal N}_q$ to be an exponentially decreasing
function of $|\bar q|N$. Because of this factor, most of the elements
of the $(N+1)\times(N+1)$ matrix ${\langle \Psi_q^{(\pm)}|{V}
|\Psi_q^{(\pm)}\rangle}$ (\ref{matr_elem}) are very small, see
Fig.~\ref{fig:pipka}. It is clear that the off-diagonal matrix elements
of this matrix are those responsible for the mixing of different
energy levels of the `unperturbed' Hamiltonian and, therefore,
strong mixing can only occur if $\bar q N,\bar q'N \sim \CO(1)$.
Using the large-$N$ approximate expressions for quantized $q$
(\ref{small-q}) we may
expect that the number of such levels $k_{\rm max}$ is of order $\ln \eta$.

To calculate the matrix elements ${\langle \Psi_q^{(\pm)}|{V}
|\Psi_q^{(\pm)}\rangle}$ given by (\ref{matr_elem})
we observe that, unlike in the calculation of ${\cal N}_q$,
the sum over $m$ is saturated by the contribution of the first few
terms, $m\ll N$. Hence  $u_m(q)$ can be replaced at small $\bar qN$ by
 $u_{2m}=(-1)^m+\CO(\bar qN)^2$ and
$u_{2m+1}=\CO(\bar q N)$, corresponding to the solutions to the
recurrence relations (\ref{small-n}). This gives
\begin{eqnarray}
\langle \Psi_{q'}^{(\pm)}|V|\Psi_{q}^{(\pm)} \rangle
&=&-2{\cal N}_q\sum_{n={\rm even}}\frac{f_n^{-1}}{(n+1)(n+2)}
\left[\cos(\phi_q-\phi_{q'})\pm \cos(\phi_q+\phi_{q'})\right]
\nonumber
\\
&=&
-\frac{\pi^2}9 \ln^{-1}(\eta\e^{\gamma_E})
\left[\cos(\phi_q-\phi_{q'})\pm \cos(\phi_q+\phi_{q'})\right]\,.
\labeltest{Mr12}
\end{eqnarray}
The $q-$dependence enters  this expression through the
phases $\phi_q$ and $\phi_q'$ which take quantized values defined
in (\ref{roots}). It is easy to see that the possible values of
$\cos(\phi_q\pm\phi_{q'})$ are $1$ and $-1/2$
depending on whether  the phases $\phi_q$ and $\phi_{q'}$ coincide.
For the present purpose it turns out to be more convenient to use
the basis of eigenstates $\Psi_q$  with fixed $q$ rather than fixed parity
$\Psi^{(\pm)}_q \sim \Psi_q \pm \Psi_{-q}$ and write (\ref{Mr12})
in matrix form, introducing an integer $k=[N/2]-\ell$,
$k = 0,\pm 1, \pm2,\ldots$ to numerate
quantized values of $q$ starting from the ones with the lowest absolute
value.
We get:
\begin{equation}
\langle \Psi_{q'}|\epsilon V| \Psi_q \rangle =
-g
\Lambda_{k'k}\,,\qquad
\Lambda_{k'k}=\left(
\begin{array}{rrrrr}
1&-\frac12&-\frac12&1&\ldots\\
-\frac12&1&-\frac12&-\frac12&\ldots\\
-\frac12&-\frac12&1&-\frac12&\ldots\\
1&-\frac12&-\frac12&1&\ldots\\
\vdots&\vdots&\vdots&\vdots&\ddots
\end{array}
\right),
\labeltest{potential}
\end{equation}
where, to our accuracy,
\begin{equation}
    q = k \frac{\pi\eta^2}{6\ln(\eta\e^{\gamma_E})}\,,
\label{q(k)}
\end{equation}
and the dependence on $\epsilon$ and $\eta$ is absorbed
in the `effective coupling'
\begin{equation}
    g= \frac{\epsilon\pi^2}{9\ln(\eta\e^{\gamma_E})}.
\end{equation}
The approximation in (\ref{potential}) is justified
 for $|k| \ll k_{\rm max}=\CO(\ln\eta)$.

Imagine, for a moment, that  all the
participating  levels, $|k| < k_{\rm max}$ of the `unperturbed'
Hamiltonian ${\cal H}_{3/2}$ were degenerate, {\it i.e.}\ their
energy splitting negligible compared to the
interaction in (\ref{potential}). The true energy
eigenstates would coincide then with the eigenstates of
the mixing matrix $\Lambda_{kk'}$ (of  the size $k_{\rm max}$), and
the corresponding eigenvalues would define the energies.
Remarkably, the spectrum of $\Lambda_{kk'}$ is extremely simple.
One can easily convince oneself that $\Lambda_{kk'}$ has two and only
two nonzero eigenvalues which both are equal to $k_{\rm max}$.
Remembering that $g \sim 1/\ln \eta$ and $k_{\rm max} \sim \ln \eta$,
this implies that two energy levels will get shifted by the finite amount
$g k_{\rm max} = \CO(\ln^0\eta)$ while all the other ones remain
exactly degenerate to this accuracy. Since
(\ref{potential}) was derived up to corrections of order
$\sim 1/\ln^2 \eta$, this implies that the true energy
shift for all levels apart from the lowest two ones
is at most $\sim 1/\ln^2 \eta$.
This simple heuristic observation explains the pattern observed
in Figs.~\ref{fig:e12}, \ref{figure7}.

The real situation is certainly much more complicated. The splitting
between lowest energy levels of ${\cal H}_{3/2}$ cannot be neglected,
and, in fact, it
is precisely this energy splitting which determines the
number of lowest states $k_{\rm max}$
that can effectively be
considered as degenerate\footnote{So that for
the state with the number $k_{max}$ this energy splitting is of order
of the largest eigenvalue of the matrix $\Lambda$.},
and the precise value of the
`mass gap'.
For small values of $\bar qN$ we can calculate
the energies ${\cal E}_{3/2}$ using the asymptotic expression
in (\ref{middleE}) that we rewrite as
\begin{equation}
{\cal E}_{3/2}(q)={E}_0+\frac{2\zeta(3)\pi^2}{9\ln^2(\eta\e^{\gamma_E})}
\left(k+\frac{\delta_N}{2}\right)^2+\CO((\bar q N)^4)\,,\qquad
\bar qN = k \frac{\pi}{6\ln(\eta\e^{\gamma_E})},
\labeltest{kinetic}
\end{equation}
where $k$ is defined as above,
${E}_0$ is the ground state energy given by (\ref{e0})
and $\delta_N=0\,,1$ for even and odd $N$, respectively.
Combining together (\ref{kinetic}) and (\ref{potential}) we obtain that
in the lowest part of the spectrum,
corresponding to $\bar qN$, $\bar q'N<1$, the
Hamiltonian ${\cal H}(\epsilon)$ can be represented by the following matrix:
\begin{eqnarray}
\langle\Psi_{q'}|{\cal H}(\epsilon)-{E}_0|\Psi_q\rangle
& =& g\left\{\frac{1}{2m} \left[\frac{2\pi}{3}\left(k+\frac{\delta_N}2\right)
\right]^2\delta_{k'k}
-\Lambda_{k'k}\right\}\,,
\labeltest{kin+pot}
\\
m&=&\frac{\pi^2}{9\zeta(3)}\epsilon \ln(\eta\e^{\gamma_E})\,.
\nonumber
\end{eqnarray}

The corresponding eigenvalue problem is solved in Appendix B.
The idea of the solution is to interpret the integer $k$ as a
discrete momentum variable.
Then, the expression in (\ref{kin+pot}) can be considered as an effective
Hamiltonian for the low `frequency' $|k| < k_{\rm max}=\CO(\ln\eta)$ modes
of ${\cal H}(\epsilon)$ and
the two terms in the r.h.s.\ of (\ref{kin+pot})
can be identified as the kinetic
energy and the periodic potential for a particle on a line.
The corresponding wave functions in configuration space correspond to
Bloch-Floquet waves and the resulting Schr\"odinger equation
turns out to be a generalization of the famous Kroning-Penney model
of a single particle in a periodic $\delta$-function
potential\footnote{Inclusion of a few first corrections in $\bar qN$
to the matrix $\Lambda$ effectively amounts to
smearing of the $\delta$-function potentials.}.
The solution then follows the classical procedure \cite{QM}.

One has to keep in mind, however, that the effective
Hamiltonian in (\ref{kin+pot})
presents an approximation to ${\cal H}(\epsilon)$  up to corrections of order
$\CO(\bar q N)$ and one has to check whether values of $\bar q N$
are small on the solutions. It is possible to show that this
condition is indeed satisfied for small $\epsilon \ll 1 $ (and this is
the place where introducing $\epsilon$ as a new parameter starts to
play a r\^ole), see Appendix B. From this analysis we obtain, therefore, a
quantitative description of the spectrum of ${\cal H}(\epsilon)$
in the region
\begin{equation}
   1/\ln \eta \ll \epsilon \ll 1\,,
\label{region2}
\end{equation}
where the lower bound comes from the condition that the interaction
$\epsilon V$ is sufficiently strong to excite many levels.

In agreement with the heuristic argument given above,
we find two bound states and the continuum spectrum.
The levels in the continuum are $\epsilon$-independent, in the
small-$\epsilon$ limit, and are given by
\begin{equation}
{\cal E}_k= {E}_0+\frac{\zeta(3)\pi^2 k^2}{8\ln^2(\eta\e^{\gamma_E})},
\label{others}
\end{equation}
with $1\le k < \ln\eta$. The two bound states are degenerate
up to $1/\eta^2-$corrections that we neglected from the beginning
and the binding energy (which we identify as a `mass gap') is given by
\begin{equation}
 \Delta (\epsilon) = {\cal E}_{\rm bound}-{E}_0
     =-\epsilon^2\frac{\pi^4}{72\zeta(3)}.
\label{gap1}
\end{equation}
Comparing this expression with the nonzero eigenvalue of the
perturbaton in (\ref{potential}), $gk_{\rm max}$,
we calculate the number of excited
levels of the unperturbed Hamiltonian ${\cal H}_{3/2}$ as $k_{\rm max}=
\epsilon \pi^2\ln\eta/(8\zeta(3))$,
which is in agreement with our expectations.

The {\it low-frequency part\/} of the wave functions of the
two bound states coincide
for $\ln \eta\to\infty $ with the two lowest energy eigenfunctions
of $V$, that are given by the parity even and parity odd
combinations of the basis functions (\ref{basis}) diagonalizing
the Casimir operators $L_{12}^2$ and $L_{23}^2$:
\begin{eqnarray}
 \Psi^{(\pm)}_{\rm bound}(x_i) &\sim&
\left[\Psi_{N,n=0}^{(12)3}(x_i) \pm \Psi_{N,n=0}^{(23)1}(x_i) \right]
 +\CO(1/\ln\eta)
\nonumber\\
 & \stackrel{\sum x_i=1}{=} &
  \left[P^{(3,1)}_N(1-2x_3)\pm P^{(3,1)}_N(1-2x_1)\right]
 +\CO(1/\ln\eta)\,.
\label{ground12}
\end{eqnarray}
 This can be seen from the fact that in the limit $\ln\eta \to \infty$ the
`mass' $m$ entering (\ref{kin+pot}) becomes large and the
kinetic term irrelevant.
As seen from (\ref{ground12}), the wave functions have
a two-particle structure corresponding to the relative
motion of the conformal spin $j=1$
quark with momentum fraction $x_1$ (or $x_3$) with positive helicity
(the first or the third quark,
in notations of (\ref{B1/2}))
and an effective particle with momentum fraction $x_2+x_3$
(or $x_1+x_2$) and
conformal spin $j=2$, which is
easily recognized as a scalar diquark.
This is in striking contrast to the structure of the lowest energy
state for the $\lambda=3/2$ baryons  (\ref{ground32}),
and in fact the corresponding wave functions are mutually orthogonal
in the large$-N$ limit:
\begin{equation}
   \langle \Psi^{1/2}_{\rm bound} |\Psi^{3/2}_0\rangle \sim 1/\ln \eta.
\end{equation}
We will further elaborate on the physical interpretation and consequences
of this structure in Sect.~6.

Extension of these results to case $\eps=1$ is nontrivial
since the higher order corrections in $\bar q N$ become
dominant, and presents a typical strong coupling problem.
{}From Fig.~\ref{figure7} one observes, however, that the
quadratic in $\epsilon$ behavior of the bound state energy, (\ref{gap1}),
is replaced by
the linear asymptotics starting already from $\epsilon\sim 0.3-0.4$.
This suggests to study the energy spectrum
of ${\cal H}(\epsilon)$ in the large-$\epsilon$ limit
\begin{equation}
   \epsilon \gg 1
\label{region3}
\end{equation}
and try to find an approximate value of the mass gap at $\epsilon=1$ by
matching the small-$\epsilon$ and the large-$\epsilon$ expansions.
This program is carried out in Sect.~5.3 below.

For completeness, we quote  here the results
for  very small $\epsilon$
\begin{equation}
  \epsilon \ll 1/\ln \eta\,,
\label{region1}
\end{equation}
for which case a simple perturbation theory is again valid
and the energy shifts are given (up to $\CO(\epsilon^2)$) by matrix elements
of $\epsilon V$ over the eigenstates of the nonperturbed Hamiltonian.
For the states with $\phi_q=0$ (but $q\neq 0$)
we find, for low lying levels
\begin{equation}
{\cal E}^{(+)}(\epsilon)-{\cal E}_{3/2} =  -2 g
\,,\qquad
{\cal E}^{(-)}(\epsilon)-{\cal E}_{3/2} =   0,
\labeltest{shift1}
\end{equation}
while for the states
with $\phi_q=\pm 2\pi/3$ one gets
\begin{equation}
{\cal E}^{(+)}(\epsilon)-{\cal E}_{3/2} =  -\frac12 g
\,,\qquad
{\cal E}^{(-)}(\epsilon)-{\cal E}_{3/2} =  -\frac32 g\,,
\labeltest{shift2}
\end{equation}
respectively.
Eq.~(\ref{Mr12}) is not applicable for calculating the
correction to the ground state energy with $q=0$ since this
state is not degenerate. A direct calculation of the matrix
element $\langle \Psi_{q=0}|V|\Psi_{q=0}\rangle$ based on the
exact solutions (\ref{q=0-coeff}) gives
\begin{equation}
 {\cal E}(\epsilon)-{E}_0 = - g\,.
\labeltest{gap2}
\end{equation}
This relation is exact (to $\CO(\epsilon)$)  whereas Eqs.~(\ref{shift1})
and (\ref{shift2}) are valid up to  $O(1/\ln^3{h})$ corrections.

\subsection{Large$-\epsilon$ expansion}

Assuming $\epsilon \gg 1$ is a large parameter, it is natural
to invert the logic which we have accepted up to now, and
consider ${\cal H}_{3/2}$
as a perturbation of the spectrum of the Hamiltonian $V$ defined
in Eq.~(\ref{V}).

At the first step, therefore, we have to study the spectrum and the
eigenfunctions of $V$
itself. Although the Hamiltonian $V$ is not integrable
and cannot be diagonalized exactly, it
can be studied in the large $N$ limit using the techniques
developed in Refs.~\cite{DM,DKM} for Hamiltonians of similar form.
We find that in the leading large$-N$ approximation
 the eigenstates in the lower
part of the spectrum are given by a linear combination of the states
diagonalizing the Casimir operators $L_{12}^2$ and $L_{23}^2$. Their
relative coefficient is fixed by the requirement for the eigenstates of $V$
to have definite parity with respect to ${\cal P}_{13}$ permutations
\begin{eqnarray}
\Psi_{V,k}^{(\pm)}&=&\frac1{\sqrt{2}}\left[
\Psi_k^{(12)3}(x_1,x_2,x_3)\pm\Psi_k^{(23)1}(x_1,x_2,x_3)
\right]+\CO(\eta^{-2})
\\
&=&\frac1{\sqrt{2}}\left[
\Psi_k^{(12)3}(x_1,x_2,x_3)\pm\Psi_k^{(12)3}(x_3,x_2,x_1)
\right]+\CO(\eta^{-2})
\,,
\labeltest{L2-states}
\end{eqnarray}
where $\Psi_k^{(12)3}(x_i)$ are functions of the conformal basis
defined  in (\ref{basis3}), which we assume here to be
normalized as $\langle \Psi_k^{(12)3}|\Psi_n^{(12)3}\rangle=\delta_{kn}$.
 The corresponding eigenvalues,
${\cal E}^{(\pm)}_V=\langle \Psi_{V,k}^{(\pm)} |\epsilon V
|\Psi_{V,k}^{(\pm)}\rangle$ can be evaluated using (\ref{Racah}) and
(\ref{propt}) as
\begin{equation}
\epsilon^{-1}\,{\cal E}_V^{(\pm)}(k) =
-\frac{1}{(k+1)(k+2)}
-\sum_{m=0}^N \frac{\Omega_{k m}^2\Omega_{mk}}{(m+1)(m+2)}
\mp \frac{\Omega_{kk}}{(k+1)(k+2)}
+\CO(\eta^{-4})\,.
\end{equation}
Taking into account Eqs.~(\ref{Omega-as}) and (\ref{Omega-small}) we find
\begin{equation}
{\cal E}_V^{(\pm)}(k) =
-\frac{\epsilon}{(k+1)(k+2)}
-\epsilon\eta^{-2}\left[1\pm(-1)^{N+k}\right](2k+3)
+\CO(\eta^{-4})\,.
\labeltest{L2-spec}
\end{equation}
Here, the integer $k=0\,,1\,,2,\ldots$ enumerates the levels of
$V$ with definite parity starting from the lowest one. We observe
that for given conformal spin $N$ the nonleading $\CO(\eta^{-2})$
corrections vanish for each second level leading to
${\cal E}_V^{(\pm)}(k)=-\epsilon/[(k+1)(k+2)]$
provided that $k+N={\rm even~(odd)}$ for levels with
positive (negative) parity. It can be shown
that this result is {\it exact\/} to all orders in $1/\eta$.

The nonleading $\CO(\eta^{-2})$ corrections remove  degeneracy of
the levels
\begin{equation}
{\cal E}_V^{(+)}(k)-{\cal E}_V^{(-)}(k) =- 2\epsilon\,
(2k+3)(-1)^{N+k}\eta^{-2}
+\CO(\eta^{-4})\,.
\labeltest{Shift}
\end{equation}
Thus, in the lowest part of the spectrum, $k\ll N$, the eigenvalues of the
Hamiltonian $V$ belong to the two trajectories,
\begin{eqnarray}
{\cal E}_{V,{\rm up}}(k)&=&-\frac{\epsilon}{(k+1)(k+2)}\,,\qquad
\nonumber
\\
{\cal E}_{V,{\rm down}}(k)&=&{\cal E}_{V,{\rm up}}(k)-2\epsilon(2k+3)\eta^{-2}
+\CO(\eta^{-4})
\end{eqnarray}
parameterized by a nonnegative integer $k$. Parity of the eigenstates
alternates along each trajectory.

At the second step, we evaluate the matrix elements of ${\cal H}_{3/2}$
over the eigenstates of $V$ defined in Eq.~(\ref{L2-states}).
Using (\ref{matrix-3/2}) and (\ref{L2-states}) one can write
\begin{eqnarray}
\frac12\left[ \langle \Psi_{V,k}^{(+)}| {\cal H}_{3/2}|
     \Psi_{V,k}^{(+)} \rangle
+\langle \Psi_{V,k}^{(-)}| {\cal H}_{3/2}| \Psi_{V,k}^{(-)} \rangle
\right]
&=&\varepsilon(k)+2\sum_{m=0}^N (-1)^{m+k} \Omega_{k m}\varepsilon(m)
\Omega_{mk}\,,
\\
\frac12\left[ \langle \Psi_{V,k}^{(+)}| {\cal H}_{3/2}|
\Psi_{V,k}^{(+)} \rangle
-\langle \Psi_{V,k}^{(-)}| {\cal H}_{3/2}| \Psi_{V,k}^{(-)} \rangle
\right]
&=& 2 \varepsilon(k) \Omega_{kk}
+\sum_{m=0}^N \Omega_{k m}\varepsilon(m)
\Omega_{mk},
\end{eqnarray}
where $\varepsilon(m)=2[\psi(m+2)-\psi(2)]]$ is the two-particle energy. The
sums over $m$ are dominated by contributions of $x\equiv m/N=\CO(1)$.
Therefore, replacing $\varepsilon(m)=2\ln(Nx)-2\psi(2)$ and using
properties of the Racah $6j-$symbols, (\ref{propt}), (\ref{Omega-as}) and
(\ref{Omega-small}), we get
for the parity-averaged spectrum of ${\cal H}(\epsilon)$
\begin{eqnarray}
{\cal E}^{(\pm)}(\epsilon) &=&
   {\cal E}_V^{(\pm)}(k) +
\langle \Psi_{V,k}^{(\pm)}| {\cal H}_{3/2}| \Psi_{V,k}^{(\pm)}
\rangle +\CO(1/\epsilon),
\labeltest{L2-pert}
\\
\langle \Psi_{V,k}^{(\pm)}| {\cal H}_{3/2}| \Psi_{V,k}^{(\pm)}\rangle
 &=&
 4\ln\eta-6\psi(2) +6\psi(k+2)-4\psi(2k+4) +\!\frac{2}{k+2}
+\!\frac{2}{2k+3} +\!\CO(\eta^{-1}),
\nonumber
\end{eqnarray}
while the energy splitting between the eigenstates with opposite parity equals
\begin{eqnarray}
\lefteqn{
\hspace*{-1cm}
  {\cal E}^{(+)}(\epsilon) - {\cal E}^{(-)}(\epsilon) =
    {\cal E}^{(+)}_V(\epsilon) - {\cal E}^{(-)}_V(\epsilon)~+}
\nonumber\\
& + & 4\left[
\ln\eta+2\psi(k+2)-3\psi(2)\right]
\eta^{-2} (-1)^{N+k}
(2k+3)(k+1)(k+2)\,.
\labeltest{L2-split}
\end{eqnarray}
For large $\eta$ the correction terms in both cases are dominated
by the first term $\sim \ln\eta$ which can be related
to the `ground state'  energy ${E}_0$
of the Hamiltonian ${\cal H}_{3/2}$ defined in (\ref{e0}). It provides
an overall shift of all levels and can be  absorbed into the
definition of the nonperturbed Hamiltonian $V$. Then,
comparing the matrix elements (\ref{L2-pert}) with the
level spacing in the $V$-spectrum in (\ref{L2-spec}),
we conclude that the large$-\epsilon$ expansion is well-defined
for $\epsilon\gg 1$.

In particular, for the lowest level $k=0$ we obtain:
\begin{equation}
\Delta(\epsilon) =
{\cal E}(\epsilon)-{E}_0 =
  -\frac12\epsilon +\frac13 +\CO(\epsilon^{-1}).
\label{gap3}
\end{equation}
One should stress that all the above expressions are only
valid for the lowest levels $k \ll N$ in the spectrum.

It is instructive to
examine the flow of the energy levels defined by the perturbative expressions
(\ref{L2-spec}) and (\ref{L2-pert}). Varying $\epsilon$ from $\epsilon=\infty$
towards $\epsilon=1$ we find that the energy levels are changing linearly
in $\epsilon$, with the slope depending on integer $k$
and on the parity of the level. The perturbative correction generates
the $\epsilon-$independent shift of the trajectories whose amount again
depends on the parity and $k$. The `critical' values of
$\epsilon$ at which the linear trajectories cross the ground state
energy ${E}_0 = {\cal E}_{3/2}(q=0)$
set up the low boundary for $\epsilon$ such that
the $1/\epsilon-$expansion is applicable. Using (\ref{L2-spec}) and
(\ref{L2-pert}) we find the corresponding intersection points as
\begin{equation}
\epsilon_{\rm crit}(k=0)=0.66 \,,\qquad
\epsilon_{\rm crit}(k=1)=5.60 \,,\qquad
\epsilon_{\rm crit}(k=2)=16.97 \,.\qquad
\end{equation}
These values are in a good agreement with the numerical solutions
shown in Fig.~\ref{figure7}.

Finally,  notice that the $1/\epsilon-$expansion can also be applied to
describe the flow of energy levels  for $\epsilon < 0$ but its range of the
applicability is in this case only $-\infty\ll \epsilon \ll -\ln\eta$.
The main difference between positive and negative $\epsilon$ is that
in the latter case the lowest energy
levels of the Hamiltonian ${\cal H}(\epsilon)$ are rapidly approaching
each other for $-\ln\eta\ll\epsilon < 0$.
As a consequence, the naive $1/\epsilon-$expansion becomes
divergent due to small denominators and  should be replaced by
the so-called `degenerate perturbative expansion'.

\subsection{Estimate of the mass gap}

Having derived the asymptotic expressions for the mass gap $\Delta(\epsilon)$
in the two limits $\epsilon \ll 1$ and $\epsilon \gg 1$,
Eqs.~(\ref{gap1}) and (\ref{gap3}), respectively, we can make an estimate
for $\epsilon\sim 1$ by matching
the two expansions. To this end, we design an interpolating
Pad\'e formula:
\begin{equation}
 \Delta(\epsilon) = - \frac{\epsilon^2(\kappa+\alpha\epsilon)}
   {1+2(\kappa+2\alpha/3)\epsilon+2\alpha\epsilon^2},
\end{equation}
with $\kappa = \pi^4/(72\zeta(3)) = 1.12549$,
which reproduces (\ref{gap1}) and (\ref{gap3}) in the appropriate
limits and contains one free parameter $\alpha$ that
has to be positive in order to avoid spurious singularities.
 Allowing $\alpha$ to vary within the extreme limits $0 <\alpha<\infty$
we get
\begin{equation}
 \Delta_{1/2} \equiv \Delta(\epsilon =1) = - (0.30000 - 0.34620) \,,
\label{Delta}
\end{equation}
which compares very well with the result of the direct
numerical calculation of the parity average energy of the
lowest eigenstates at $N=300$:
\begin{equation}
  \frac{1}{2}\Big[{\cal E}_{1/2}^{(+)}+{\cal E}_{1/2}^{(-)}\Big]
  -{E}_0 = -0.32097\,.
\label{n260}
\end{equation}

The results of the numerical calculation of the lowest few
eigenvalues of ${\cal H}_{1/2}$ are shown in Fig.~\ref{fig:largeN}.
\begin{figure}[ht]
\centerline{
   \epsfig{file=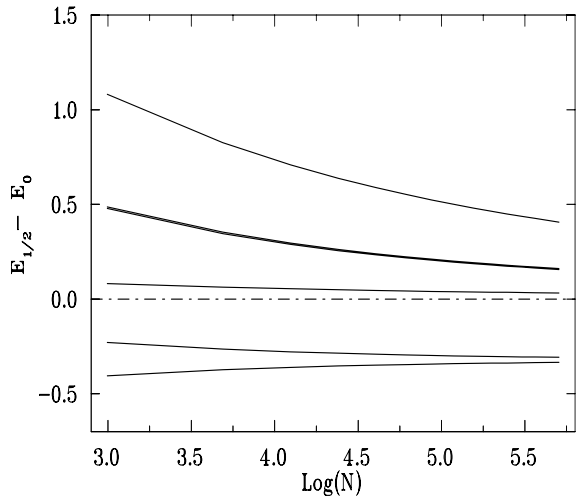, height=7.5cm, width=6.5cm, angle=90}
           }
\caption[]{\small The difference ${\cal E}_{1/2}-{E}_0$
for the first few energy levels.}
  \label{fig:largeN}
\end{figure}
It is seen clearly that the distance between the two lowest eigenvalues
and the `vacuum energy' ${E}_0$ approaches a constant $\sim 0.3$,
while for higher levels this distance decreases as $1/\ln^2 N$.
Notice, however, that the distance to  ${E}_0$
and the level splittings for higher levels are still
quite large for $\ln N \sim 5$ ($N \sim 10^{2}$). The reason for this is that
the expansion parameter for the upper part of the spectrum
proves to be $1/\ln N$ (rather than  $1/N$) and the asymptotic large$-N$
limit is, therefore, approached very slowly.

\section{Distribution amplitudes $\phi^{3/2}_\Delta$,
 $\phi^{1/2}_\Delta$, $\phi^{1/2}_N$: Summary of results}
\setcounter{equation}{0}

In this section we give a short summary of the results of
phenomenological relevance for the physical baryon distribution amplitudes
defined in Sect.~2.1 and discuss an overall physical picture that emerges.

The Brodsky-Lepage evolution equation for the helicity-$3/2$
distribution amplitude $\phi^{3/2}_\Delta$ is exactly integrable and
is considered in much detail in Sect.~4. The physical interpretation
of integrability is  that we are able to identify a
new `hidden' quantum number which distinguishes components in the
$\Delta$-resonance with different scale dependence.
The anomalous dimensions and the eigenfunctions can be calculated in this
case exactly using a simple three-term recurrence relation given
in Eq.~(\ref{masterrec}). The coefficients $u_n$ define the expansion
coefficients (\ref{expand1}) for the eigenfunctions of the evolution equation
over the complete set of mutually orthogonal conformal polynomials
(\ref{DAbasis}) and the corresponding anomalous dimensions are
given in terms of the same coefficients by Eqs.~(\ref{energy1})
and (\ref{anomal}). Alternatively, we have derived a systematic
WKB-type expansion for large values of $N$ which provides one
with a systematic expansion of the eigenvalues, [see Eqs.~
(\ref{WKB-q}), (\ref{q-low}), (\ref{small-q}), (\ref{Energy})]
and the eigenfunctions
[see Eqs.~(\ref{PsiWKB}), (\ref{PsiWKBres}), (\ref{sym})],
in powers of $1/N$.

The case of $\phi^{3/2}_\Delta$ is still specific as compared to the
general treatment in Sect.~4 in that neglecting tiny $SU(2)$-flavor
violation effects due to quark masses $\phi^{3/2}_\Delta$
is totally symmetric in all three arguments.
As a consequence, only one third of the existing multiplicatively
renormalisable operators have nonvanishing matrix elements, namely,
those corresponding to the unity eigenvalue $\theta=1$ of the
cyclic permutation operator. Note that  the value of $\theta$ alternates
along the trajectories shown in Fig.~\ref{figure2} so that each third
of the eigenvalues gives a relevant contribution.

Recall that each eigenvalue in Fig.~\ref{figure2} (except for the
lowest one for each $N$) is double degenerate.
The two degenerate eigenstates can be chosen either as eigenstates of the
$Q$-operator with opposite sign eigenvalues $q$ and $-q$, or as states
with definite parity, defined in Eq.~(\ref{eigen_parity}).
The latter choice is more convenient since the parity eigenfunctions are
real and contributions to $\phi^{3/2}_\Delta$ of the operators with
negative parity vanish identically. One is left with the sum over positive
parity eigenstates with real coefficients.

The most interesting result concerns the structure of the eigenstates
with the lowest eigenvalue (anomalous dimension) for each $N$ which present,
therefore, the leading contributions to the distribution amplitude
in the formal $\mu^2\to\infty$ limit:
\begin{eqnarray}
\lefteqn{ x_1 x_2 x_3 \Psi^{3/2}_{N}(x_1,x_2,x_3) =}
\label{32a}
\\
&=& x_1(1-x_1)\,C_{N+1}^{3/2}(1-2x_1) +x_2(1-x_2)\,C_{N+1}^{3/2}(1-2x_2)
    +x_3(1-x_3)\,C_{N+1}^{3/2}(1-2x_3)\,,
\nonumber
\end{eqnarray}
see Fig.~\ref{fig:WF0}.
\begin{figure}[ht]
\centerline{\epsfxsize7cm\epsfysize6.2cm\epsffile{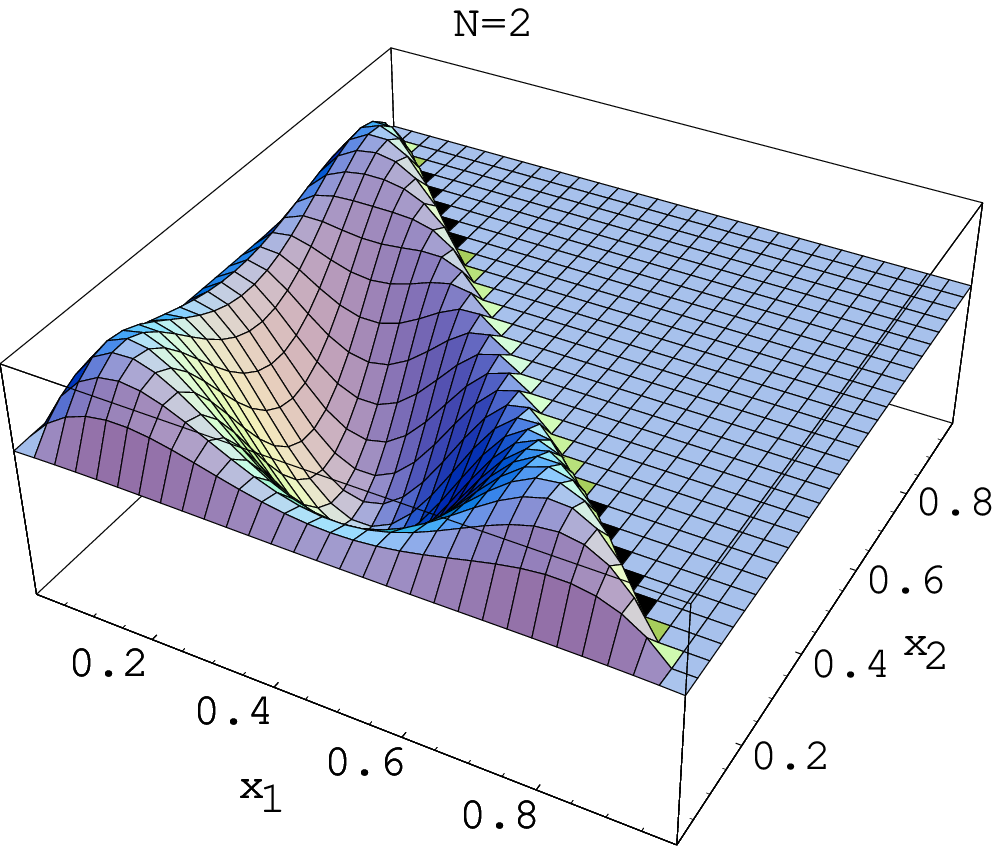}~~~
            \epsfxsize7cm\epsfysize6.2cm\epsffile{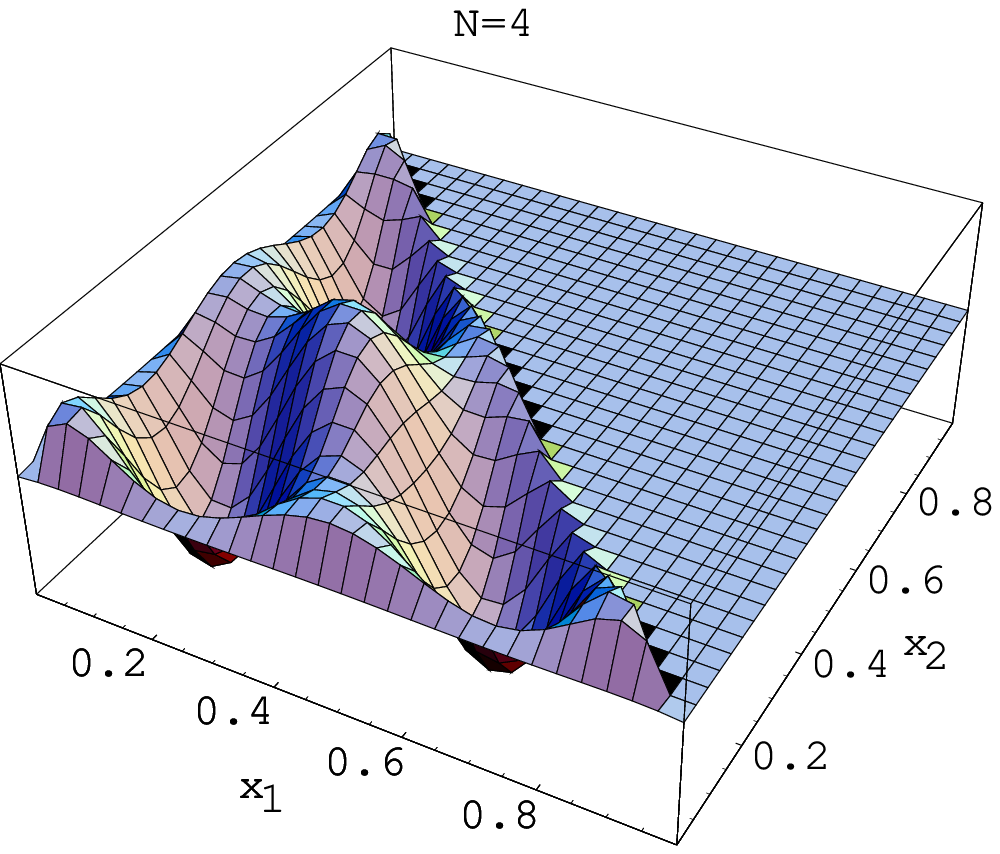}
            }
\caption[]{\small Contributions to the $\lambda=3/2$ distribution amplitude
$\phi_\Delta^{3/2}(x_i) $ with
lowest anomalous dimensions for $N=2$ and $N=4$. The normalization is
arbitrary.}
\label{fig:WF0}
\end{figure}
The corresponding eigenvalues are known exactly and are given in
 Eq.~(\ref{e0}).

The physical interpretation of such `ground states' is most transparent
in coordinate space.  Neglecting the  operators with total
derivatives, which amounts to going over from (\ref{32a}) to
the distribution function $\widetilde\Psi$ of one variable
in Eq.~(\ref{ground32tilde}), one can represent
the three-quark `ground state' in a concise form as the
nonlocal light-cone operator \cite{ABH}
\begin{equation}
B_{3/2}^{(q=0)}(z_1,z_2,z_3)=\frac12
\sum_{\alpha,\beta=1,2,3 \atop \alpha\neq \beta} 
\varepsilon^{ijk}\int_0^1\!dv\, \!\not\!n q_{i}^\uparrow (z_\alpha n)
   \!\not\!n q_{j}^\uparrow([vz_\alpha+(1-v)z_\beta]n)
   \!\not\!n q_{k}^\uparrow (z_\beta n)
\labeltest{Blow32}   
\end{equation}   
The Tailor expansion of the {\it forward\/} matrix elements of
(\ref{Blow32}) at short distances, $z_{12}, z_{32} \to 0$, generates
the series of local multiplicatively renormalizable three-quark
local operators with the lowest anomalous dimension for each even $N$
\begin{equation}
B_{3/2}^{(q=0)}(z_1,z_2,z_3)=\sum_{N={\rm even}} \frac{
z_{12}^N+z_{23}^N+z_{31}^N}{(N+1)!}\
 \Psi_{N,q=0}^{3/2}(\partial_1,\partial_2,\partial_3)\
B(z_1,z_2,z_3)\bigg|_{z_\alpha=0\,,
\atop \partial_1+\partial_2+\partial_3=0}
\end{equation}
Note integration in (\ref{Blow32}) with unit weight over the position of 
the quark in the middle that goes in between  the light-cone positions
 of the other two quarks, up to permutations.  If renormalization of the
operator is interpreted as interaction, integration
with the unit weight can in turn be interpreted as the statement that
 the quark in the middle is effectively
`free': In the `ground state' with the lowest `energy', the interaction
of the quark in the middle with its right and left neighbours exactly
compensate each other.

The evolution equations for helicity-1/2 distribution amplitudes
$\phi^{1/2}_\Delta$ and  $\phi^{1/2}_N$ differ from the
evolution equation for $\phi^{3/2}_\Delta$ by the additional
contribution of gluon exchange between the quarks with opposite
helicity, see Eq.~(\ref{H12}) and Fig.~\ref{figure1}. The added
terms destroy exact integrability, but, as we found, can be considered
as a small perturbation for the upper part of the spectrum. As a consequence,
there is a direct correspondence between eigenoperators and
anomalous dimensions for helicity-3/2 and helicity-1/2 distributions
and the corrections can be calculated to $1/N^3$ accuracy using the
standard quantum-mechanical perturbation theory, see Eq.~(\ref{upper_PT}).
The splitting between the eigenstates with opposite parity proves to be
exponentially small at large $N$, see Eq.~(\ref{split+-}).

For low-lying levels the situation turns out
to be dramatically different. We find that the two lowest eigenvalues
(anomalous dimensions) decouple from the rest of the spectrum and
in the limit $\ln N\to\infty$ are separated from the other eigenvalues
by a finite constant $\Delta \sim -0.3$ (\ref{Delta})
that we call the `mass gap'.
The corresponding contributions to the distribution amplitudes
are given by
\begin{equation}
 x_1 x_2 x_3 \Psi^{1/2}_{N\to\infty}(x_1,x_2,x_3) =
  x_1 x_2 x_3 \left[P^{(3,1)}_N(2x_3-1)\pm P^{(3,1)}_N(2x_1-1)\right]
\label{12a}
\end{equation}
where $P^{(3,1)}_N$ are Jacobi polynomials, and correspond, in the
same sense as above, to the contribution of the nonlocal
light-cone operator
\begin{eqnarray}
\lefteqn{  B(z_1,z_2,z_3)=}
\nonumber\\
&=& \varepsilon^{ijk}\left[ (\!\not\!n q_{i}^\uparrow
   \!\not\!n q_{j}^\downarrow)(z_1n)
   \!\not\!n q_{k}^\uparrow (z_3n)\delta(z_2-z_1)
\pm
   \!\not\!n q_{i}^\uparrow(z_1n)
   (\!\not\!n q_{j}^\downarrow)
   \!\not\!n q_{k}^\uparrow)(z_3n)\delta(z_2-z_3)\right]
\labeltest{Blow12}
\end{eqnarray}
Formation of the mass gap in the spectrum of anomalous dimensions is,
therefore, naturally interpreted as due to binding of the quarks
with opposite helicity and forming scalar diquarks.

Note that while the expression for the eigenfunction in (\ref{32a})
is exact, the result in (\ref{12a}) is only valid in the
asymptotic $\ln N\to\infty$ limit. In the coordinate space picture,
the restriction to large $N$ is translated to the condition that
the light-cone separation between the same helicity quarks
is very large to allow for the formation of a diquark. In momentum
space, the result means that at sufficiently large scales $Q^2$
the quark carrying a very large momentum fraction is more often with the same
helicity as of the parent baryon. This observation seems
to be in qualitative agreement with phenomenological models
of baryon distribution amplitudes derived from QCD sum rules~\cite{CZ84,FZOZ88}.

\begin{figure}[ht]
\centerline{\epsfxsize6.0cm\epsffile{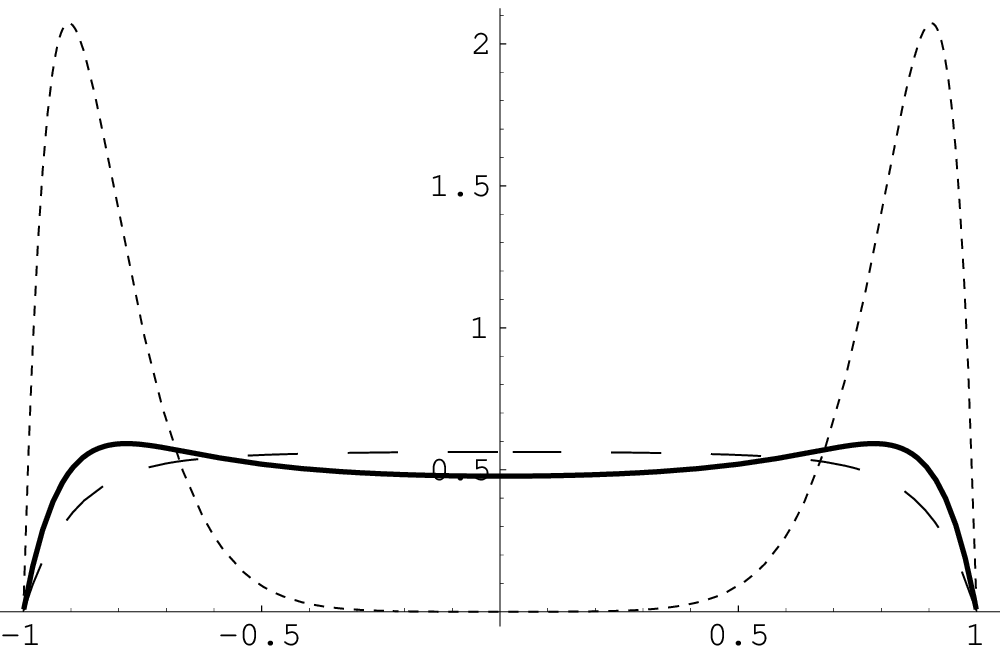}\hspace*{0.5cm}
            \epsfxsize6.0cm\epsffile{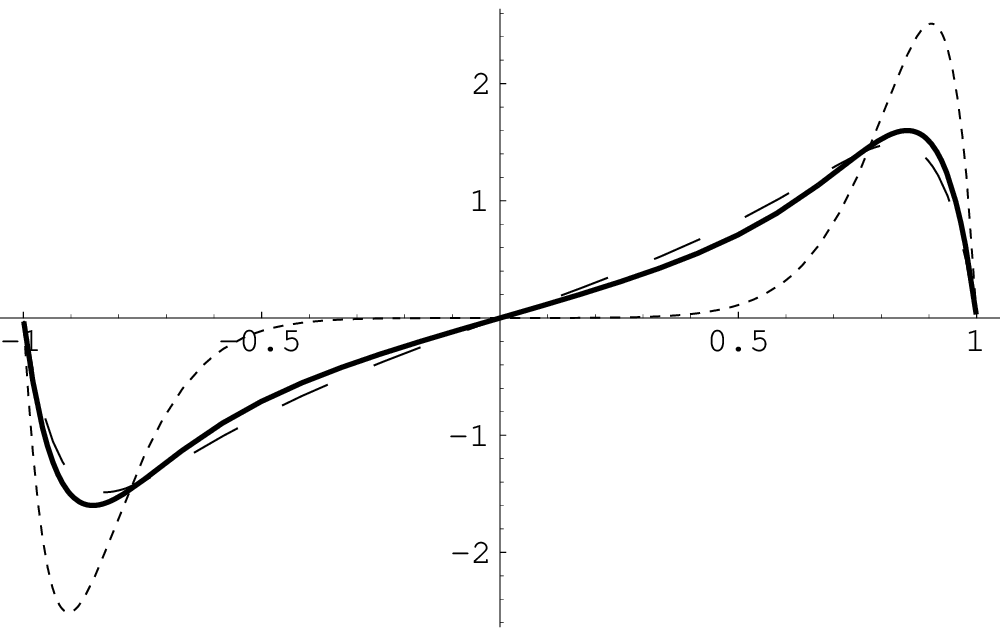}}
\caption[]{\small The `ground state'  eigenfunctions $\widetilde\Psi(x)$
  for the Hamiltonians
 ${\cal H}_{1/2}$ (solid), ${\cal H}_{3/2}$ (long dashes) and
$V$ (short dashes) for $N=19$. The normalization is to unit integral
$\int\! dx\,\widetilde\Psi(x) = 1$ and to the unit first moment
$\int\! dx\,x\,\widetilde\Psi(x) = 1$ for the symmetric and
the antisymmetric wave functions, respectively.
 }
\label{ma-gap}
\end{figure}

One has to keep in mind, however, that the diquark picture
of the states with the lowest anomalous dimensions only becomes
quantitative for very large $N$. To illustrate this point,
we plot in Fig.~\ref{ma-gap} the exact eigenfunctions $\widetilde\Psi$
for ${\cal H}_{1/2}$ at $N=19$ (with positive and negative parity)
corresponding to the lowest anomalous dimensions (solid curves),
and compare them both with the  lowest-level eigenfunctions
for ${\cal H}_{3/2}$ (dashes) and with the `diquark' eigenfunction
corresponding to Eq.~(\ref{12a}) (dots). It is seen that
the eigenfunctions of ${\cal H}_{1/2}$ for this value of $N$
are still very close to ${\cal H}_{3/2}$ and only start to
develop small `horns' close to the end points, characteristic for the
diquark picture. As mentioned above, the true large-$N$ limit is
approached very slowly since parameter of the expansion is in this
case $1/\ln N$ rather than $1/N$.

\section{Conclusions}
\setcounter{equation}{0}

To summarize, in this paper we have developed a new theoretical
framework for the description of baryon distribution amplitudes
in QCD, based on integrability of the helicity--3/2 Brodsky-Lepage
evolution equation. The mathematical structure of the evolution equations
reflects a clear physical structure of the distribution amplitudes that
we tried to emphasize. A lot of analytic results is obtained,
in different limits.

The formalism proposed in this paper is rather general and can be
applied, as indicated in \cite{BDM}, to the studies of
quark-antiquark-gluon and, possibly,
three-gluon distributions.

Three general questions related to the theory of three-particle
distributions are not covered in this work and deserve further
attention. First, as we have indicated, analytic continuation
of the spectrum of anomalous dimensions of three-particle operators
to the complex angular momentum plane is intrinsically ambiguous.
One has to study whether this mathematical ambiguity is resolved
by imposing certain physical conditions on the amplitudes.
Second, the solution of the evolution equations for three-particle 
distributions depends on the nonpertrubative initial conditions. 
Depending on the choice of the three-particle 
distribution amplitudes at low scales, there is a possibility that
at large evolution times the solutions to the evolution equation
become independent on the initial conditions and are governed 
entirely by perturbative evolution. Such perturbatively driven 
distribution amplitudes would generalize the GRV partonic 
distributions \cite{GRV} which prove to be successful in the phenomenology 
of hard inclusive processes. Finally, the integrability of the evolution
equations reveals an additional hidden symmetry of QCD and its
close relation to exactly solvable statistical models. Remarkably
enough the same symmetry has been observed in the studies of the
Regge asymptotics of three-particles distributions. 
These properties are not seen at the level of the QCD Lagrangian 
and their origin needs to be understood better.

\subsection*{Acknowledgments}

V.B., S.D. and A.M. would like to thank L.N.~Lipatov, 
and G.K. is grateful to A.B.~Kaida\-lov for
useful discussions. Our special thanks are due to 
A.V.~Belitsky for the critical reading of the manuscript.
This work was supported in part by NORDITA through the
Baltic Fellowship program funded by the Nordic Council of
Ministers, by Russian Fond of Basic Research, Grant 97--01--01152
(S.D. and A.M.) and by the EU networks
`Training and Mobility of Researchers',
contracts EBR FMRX--CT96--008 (V.B.) and  FMRX--CT98--0194 (G.K.).

\appendix
\renewcommand{\theequation}{\Alph{section}.\arabic{equation}}
\setcounter{table}{0}
\renewcommand{\thetable}{\Alph{table}}

\section{Appendix: Racah $6j$-symbols of the $SL(2)$ group}
\labeltest{app:a}
\setcounter{equation}{0}

The functions $\Psi^{(12)3}_{N,n}(x_i)$ introduced in Sect.~3.4
have a simple group theory
interpretation. They define the addition rules for the sum of three
conformal spins $j=1$ each corresponding to the $SL(2)$ generators
$L_{\alpha,k}$ ($k=1\,,2\,,3$) acting on the light-cone coordinates of
three quarks and the superscript $(12)3$ indicates the order in which
the tensor product of three $SL(2)$ representations has been decomposed
into the irreducible components for which $\Psi^{(12)3}_{N,n}(x_i)$ is
the highest weight. According to (\ref{basis3}) and (\ref{basis2}),
$\Psi^{(12)3}_{N,n}(x_i)$ describes the irreducible component for which
the total conformal spin is $h=N+3$ and the conformal spin in the channel
$(12)$ is equal to $j_{12}=n+2$. Changing the order in which the
spins are added one obtains an equivalent basis of functions
$\Psi^{1(23)}_n(x_i)$ that are linear related to
$\Psi^{(12)3}_n(x_i)$ through the Racah decomposition (\ref{Racah}).
The coefficients $\Omega_{kn}$ thus define the
 $6j-$symbols for the discrete positive series of the $SL(2)$ group.
It is easy to see that the corresponding basis functions
are transformed into each other by cyclic permutations
\begin{equation}
\Psi^{1(23)}_{N,n}(x_1,x_2,x_3) = \Psi^{(12)3}_{N,n}(x_2,x_3,x_1)
\end{equation}
and, therefore, the $(N+1)\times(N+1)$ matrix $\Omega$
represents the operator of cyclic permutations ${\cal P}$ (\ref{P}) in the
conformal basis
\be
\langle \Psi_m^{(12)3}|\Psi_n^{1(23)}\rangle
 = \langle \Psi_m^{(12)3}|{\cal P}|\Psi_n^{(12)3}\rangle
 = \Omega_{nm} \frac{60f_m}{(2N+5)},
\ee
where we used the expression for the norm of the basis vectors
(\ref{norm-basis}). Since $\Psi_n^{(12)3}$ is a real function of $x_i$,
the
calculation of the scalar product leads to real matrix elements
\be
\Omega_{nm}^*= \Omega_{nm} \,,
\qquad
\sum_{n=0}^N f_n \Omega_{ln} \Omega_{mn} = f_m \delta_{ml}\,,
\labeltest{reality}
\ee
where the second relation is the unitarity condition.
Since ${\cal P}^3=1$ and
${\cal P}^2={\cal P}_{12}{\cal P} {\cal P}_{12}={\cal P}^{-1}$,
the matrix $\Omega$ has to satisfy the following conditions:
\be
\Omega^3_{nm}=\delta_{nm}\,,\qquad
\Omega^2_{nm}=(-1)^{n+m} \Omega_{nm} = \Omega^{-1}_{nm}\,,
\labeltest{propt}
\ee
where in the last relation we used the identity
\be
{\cal P}_{12}\Psi^{1(23)}_{N,n}(x_1,x_2,x_3)
=\Psi^{1(23)}_{N,n}(x_2,x_1,x_3)=(-1)^n
\Psi^{1(23)}_{N,n}(x_1,x_2,x_3)\,.
\ee

Explicit expressions for the matrix elements $\Omega_{kn}$ can
be obtained in terms of the ${}_4F_3-$hypergeometric series of unit
argument \cite{GZ93}. To show this, consider the defining relation
(\ref{Racah}) and choose $x_1=-x_2=x$ and $x_3=1$ so that
$x_1+x_2+x_3=1$ and $x_1+x_2=0$ and $\Psi_n^{(12)3}(x_i)$ reduces
to $x^n$ up to a numerical factor. Rewriting the Gegenbauer
and Jacobi polynomials in terms of hypergeometric functions, one brings
Eq.~(\ref{Racah}) to the form
\begin{equation}
{}_2F_1\lr{{-m,-m-1 \atop 2}\bigg|x}\,
{}_2F_1\lr{{-N+m,N+m+5 \atop 2m+4}
\bigg|x}
\rho_m
=\sum_{n=0}^N \Omega_{mn} x^n \sigma_n
\labeltest{genfun}
\end{equation}
where
\begin{eqnarray}
\sigma_n &=&
(-1)^{N-n}\frac{(N+n+4)!}{(n!)^2(N-n)!}\frac{1}{(2n+2)(2n+3)},
\nonumber\\
\rho_m&=&\frac12(-1)^m(m+1)(m+2)\frac{(N+m+4)!}{(N-m)!(2m+3)!}.
\end{eqnarray}
The two hypergeometric functions entering (\ref{genfun}) are polynomials
in $x$ of degree $m$ and $N-m$, respectively. Their product defines
the generating function for the matrix elements $\Omega_{mn}$.
Expanding the l.h.s.\ of \ref{genfun}
in powers of $x$, one can write the coefficient
 of $x^n$ as the  ${}_4F_3-$function
of unit argument and identify it with $\Omega_{mn}$.
Explicit expressions for arbitrary $N$ and $n$
are rather cumbersome, see, e.g., \cite{4F3}.

We are able, however, to find a simple approximate expressions for
$\Omega_{mn}$ which are valid in the WKB limit of large spins $N$.
To this end, notice that the matrix elements $\Omega_{nm}$ satisfy
second-order finite difference equations. One finds them by
applying the operator $L_{23}^2$ to the both sides of
(\ref{Racah}) and taking into account (\ref{threediagonal}) 
\be
2(m+2)(m+1)\Omega_{mn}=
 \frac{f_{n+1}}{n+2}\Omega_{m,n+1}
+\frac{(2n+3)f_n}{(n+1)(n+2)}\Omega_{m,n}
+\frac{f_{n-1}}{n+1}\Omega_{m,n-1}\,,
\labeltest{Omega-rec}
\ee
where $m,n=0,\ldots,N$ and $\Omega_{m,-1}=0$.
Let
\be
\Omega_{mn}=(-1)^n \frac{\omega_m(n)}{f_n}
\labeltest{omega}
\ee
so that Eqs.~(\ref{Omega-rec}) and (\ref{reality}) are replaced by
\begin{eqnarray}
&
2(m+2)(m+1)\omega_m(n)
=
-\frac{f_n}{n+2}(\omega_m(n+1)-\omega_m(n))
 +\frac{f_n}{n+1}(\omega_m(n)-\omega_m(n-1)),
&
\nonumber
\\
&
\sum_{n=0}^N \frac1{f_n}\omega_m(n)\omega_l(n)=f_m\delta_{ml},
&
\labeltest{omega-rec}
\\
&
\omega_m(n)=\omega_n(m),
&
\nonumber
\end{eqnarray}
defining the system of orthogonal polynomials $\omega_m(n)$ in the
discrete variable $n$ with $m=0,\ldots,N$. The initial condition for the
recurrence relations in (\ref{omega-rec}) can be found from the relation
$
\Phi^{1(23)}_k(z_i)=\sum_{n=0}^N \Omega_{kn}\Phi^{(12)3}_n(z_i)
$
in the limit $z_1-z_2\to 0$ using (\ref{expand2}) and (\ref{asymphi})
as
\be
\omega_n(0)=\frac{2(-1)^N}{(N+2)(N+3)} f_n (2n+3)\,.
\labeltest{bc-omega}
\ee

There are two limiting cases when the recursion relations for $\omega_m(n)$
can be solved analytically.
In the first case, in the limit
\be
x=\frac{n}{N}=\mbox{fixed}\,,\quad m=\mbox{fixed}\,,
\quad N\to\infty
\label{Racahlim}
\ee
the recurrence relation in (\ref{omega-rec})
is reduced to the differential equation on the function
$\omega_m(n) \to \omega_m(x)$
\be
\omega''_m(x)-\frac1{x}\omega_m'(x)
+\frac{4(m+2)(m+1)}{1-x^2}\omega_m(x)=0\,.
\ee
Picking up the polynomial solution we obtain
\be
\omega_m(x)=N^2 x^2 (1-x^2) C_{m}^{3/2}(1-2x^2)\,.
\labeltest{Am}
\ee
The normalization is fixed by the second relation
in (\ref{omega-rec}) where one replaces $\sum_n\to N\int_0^1 dx$.
Finally, substituting
(\ref{Am}) into (\ref{omega}) we
get
\be
\Omega_{mn}=\frac{4(-1)^n}{N} x C_{m}^{3/2}(1-2x^2) \,,\qquad
n=xN\,,
\labeltest{Omega-as}
\ee
which is valid to $O(1/N)$ accuracy in the limit specified in
({\ref{Racahlim}).

In the second case, in the limit
\be
n\,,m=\mbox{fixed}\,, \quad N\to\infty
\ee
the recurrence relation in (\ref{omega-rec}) is reduced to the condition
\be
\frac1{n+2}(\omega_m(n+1)-\omega_m(n))=
\frac1{n+1}(\omega_m(n)-\omega_m(n-1))\,.
\ee
Its solution satysfying (\ref{bc-omega}) is given by
\be
\omega_m(n)=\frac12(-1)^N (n+1)(n+2)(m+1)(m+2) \times
\left[1+\CO(1/N^2)\right]
\ee
leading to
\be
\Omega_{mn}= N^{-2} (-1)^{N+n} (2n+3)(m+1)(m+2)\times
\left[1+\CO(1/N^2)\right]\,.
\labeltest{Omega-small}
\ee

Having defined the matrix $\Omega_{nm}$ it becomes straightforward to
calculate matrix elements of the Hamiltonians ${\cal H}_{3/2}$ and
${\cal H}_{1/2}$ in the conformal basis. To this end we write
\begin{equation}
{\cal H}_{3/2}={\cal H}_{12}+{\cal P} {\cal H}_{12} {\cal P}^{-1}
+ {\cal P}^2 {\cal H}_{12} {\cal P}^{-2}
\end{equation}
and similar for ${\cal H}_{1/2}$. Then, applying this identity
to a basis function
$\Psi^{(12)3}_n(x_i)$ and using the
properties (\ref{propt}) of the $\Omega-$matrix we find that
${\cal H}_{3/2}$ can be represented in the conformal
basis by the $(N+1)\times (N+1)$ matrix $[{\cal H}_{3/2}]_{nk}$
\begin{eqnarray}
&&
{\cal H}_{3/2} \Psi^{(12)3}_n =
\sum_{k=0}^N [{\cal H}_{3/2}]_{nk}\Psi^{(12)3}_k,
\labeltest{matrix-3/2}
\\
&&
[{\cal H}_{3/2}]_{nk}=\varepsilon(n)\delta_{nk}+[(-1)^n+(-1)^k]
\sum_{m=0}^N (-1)^m
\Omega_{nm}\varepsilon(m)\Omega_{mk},
\nonumber
\end{eqnarray}
where $\varepsilon(n)$ is the energy of the two-particle Hamiltonian
defined in (\ref{twospectrum}).
It is easy to see that this derivation relies on the
two-particle structure of the Hamiltonian only and is not sensitive to
integrability properties. In particular, the similar representation
holds for the Hamiltonian ${\cal H}_{1/2}$,
with $\varepsilon(m)$ in the sum over $m$ shifted by
the exchange interaction term $1/[(m+2)(m+1)]$.

\section{Appendix: The low-energy effective Hamiltonian for
   ${\cal H}(\epsilon)$}
\labeltest{app:b}
\setcounter{equation}{0}

The eigenproblem for the matrix (\ref{kin+pot}) takes a well known form
once we interpret the integer $k$ as a discrete momentum variable.
Denoting the corresponding
eigenvector as $\vec c=\{c_k\}$
we  construct the wave function in the configuration $x-$space as
\begin{equation}
\chi(x)=\sum_{k=-k_{\rm max}}^{k_{\rm max}} c_k\, \e^{\frac{2i\pi}3
x(k+\delta_N/2)}\,.
\labeltest{Fourrier}
\end{equation}
The restriction to $|k|\le k_{\rm max}$ serves
to remind
that (\ref{kin+pot}) presents an approximation to the Hamiltonian
${\cal H}(\epsilon)$ which is only valid for $|k|\ll \ln \eta$.
It is natural to expect that the lowest energy levels of
the matrix (\ref{kin+pot}) are not sensitive to the UV cut-off $k_{\rm max}$,
 or, equivalently,
the corresponding eigenstates $\chi(x)$ are smooth functions of $x$
at short distances $\Delta x\sim 1/\ln\eta$. To the extent that this
`decoupling' property holds true, which
we are going to verify {\it a posteriori\/}, the low-lying levels
of the Hamiltonian ${\cal H}(\epsilon)$ coincide with the lowest
eigenstates of (\ref{kin+pot}) so that the latter can be considered as
the effective low-energy Hamiltonian for the former. Having this in mind,
we temporally send the UV cutoff $k_{\rm max}$ to infinity
and assume the matrix (\ref{kin+pot}) to be of infinite size.

Using the transformation (\ref{Fourrier}) one can map the eigenproblem
for the matrix (\ref{kin+pot}) into a one-dimensional
 Schr\"odinger equation for the wave function $\chi(x)$.
It follows from (\ref{Fourrier}) that $\chi(x)$ is a
(anti)periodic function of $x$ with the period $3$ for even
(odd) values of $N$, respectively:
\begin{equation}
\chi(x+1)=(-1)^N \chi(x)\,.
\end{equation}
The two cases should, therefore, be treated separately.
Let us first consider the
case of even $N$ and  split the wave function into the sum of three terms
\begin{eqnarray}
\chi(x)&=&\chi_{+}(x)+\chi_{-}(x)+\chi_{0}(x),
\labeltest{chi}
\\
\chi_\alpha&=&\sum_{k=-\infty}^{\infty} c_{3k+\alpha} \e^{\frac{2i\pi}3
x(3k+\alpha)}\,,
\nonumber
\end{eqnarray}
where  $\alpha=\pm 1\,, 0$. Each component presents
 a (quasiperiodic) Bloch--Floquet wave function with the period $1$
and the quasimomentum $2\pi\alpha/3$:
\begin{equation}
\chi_0(x+1)=\chi_0(x)\,,\qquad \chi_\pm(x+1)=e^{\pm 2i\pi/3} \chi_\pm(x).
\labeltest{Bloch}
\end{equation}
It is straightforward to show that the eigenvalue problem for matrix
(\ref{kin+pot}) is equivalent to the Schr\"{o}dinger equation for the three
Bloch--Floquet waves $\chi_\alpha(x)$ 
propagating through a periodic array of
$\delta-$function potentials and interacting with each other:
\begin{equation}
-\frac1{2m} \partial_x^2 \chi_\alpha(x) - \frac{1}2 \sum_{k=-\infty}^\infty
\delta(x-k)
\left[3\chi_\alpha(x)-\sum_{\beta=\pm1\,,0} \chi_\beta(x)
\e^{\frac{2i\pi}{3}x(\alpha-\beta)}\right] = E \chi_\alpha(x)
\labeltest{Schrodinger}
\end{equation}
with
\begin{equation}
{\cal E}(\epsilon)-{E}_0=g E\,.
\end{equation}
This Schr\"odinger equation generalizes the famous Kronig--Penney model of
a single particle in a periodic $\delta-$function potential and its solution
follows the same procedure \cite{QM}.
Namely, the solution to (\ref{Schrodinger})
on the intervals of periodicity $n < x <n+1$ with $n$ integer are given by
the plane waves
\begin{equation}
\chi_\alpha(x)=a_\alpha(n)\, \e^{2ipx}+b_\alpha(n)\, \e^{-2ipx},
\labeltest{chi-alpha}
\end{equation}
with the coefficients $a_\alpha(n)$ and $b_\alpha(n)$ depending on $n$.
The corresponding values of the energy take the simple form
\begin{equation}
{\cal E}(\epsilon)\equiv {E}_0 +g E = {E}_0 + \frac{2gp^2}{m}
= {E}_0 + \frac{2\zeta(3)}{\ln^2(\eta\e^{\gamma_E})} p^2
\,.
\labeltest{Energy-p2}
\end{equation}
The possible values of the momentum $p$ are restricted by the quantization
conditions that one establishes by requiring $\chi_\alpha(x)$ to be
a continuous function of $x$ satisfying the periodicity condition (\ref{Bloch})
and its derivative $\partial_x \chi_\alpha(x)$ to have a discontinuity
at $x=n$ which can be calculated by integrating the both sides of
(\ref{Schrodinger})
\begin{equation}
\partial_x \chi_\alpha(x)\bigg|_{n-\delta}^{n+\delta} =
-m \left[3\chi_\alpha(n)-\sum_{\beta=\pm1\,,0} \chi_\beta(n)
\,\e^{\frac{2i\pi}{3}n(\alpha-\beta)}\right]
\end{equation}
with $\delta \to 0$.
We find that for {\it even\/} $N$ the quantized $q$ have to satisfy one
of the following three conditions
\begin{eqnarray}
\sin p &=& 0\,,
\labeltest{free-even}
\\
p\,\cot p\,(3-\tan^2p) &=& 3m\,,
\labeltest{branch-even}
\\
p\,\tan p \,\frac{3-\tan^2p}{1-\tan^2p} &=& -\frac32 m\,,
\labeltest{3rd-even}
\end{eqnarray}
which define three different branches for the dependence $p=p(m)$.
For {\it odd\/} $N$ the same conditions look like
\begin{eqnarray}
\cos p &=&0\,,
\labeltest{free-odd}
\\
p\,\tan p \,(3-\cot^2p)&=&-3m\,,
\labeltest{branch-odd}
\\
p\,\cot p \,\frac{3-\cot^2p}{1-\cot^2p}&=&\frac32 m\,.
\labeltest{3rd-odd}
\end{eqnarray}
The solutions to (\ref{free-even}) and (\ref{free-odd}) do not depend on the
perturbation $\epsilon$ and the corresponding energy levels
coincide with the levels of the unperturbed Hamiltonian. This
happens because for these values of the momentum $p$ the wave
function vanishes at integer points $\chi_\alpha(n)=0$ and, as
a consequence, the interaction term in (\ref{Schrodinger}) vanishes as well.

Since the matrix (\ref{kin+pot}) is hermitian, its eigenvalues ought to
be real. Then, it follows from (\ref{Energy-p2}) that quantized $p$ can
take either real or pure imaginary values. In the latter case,
the energy ${E}$ becomes negative and the wave
function (\ref{chi-alpha}) is exponentially decreasing with $x$, indicating
formation of a bound state. We will see that these bound
states are precisely the ones that are
  responsible for the decoupling of the pair
of lowest levels in Fig.~\ref{figure7}.

To this end, we solve the quantization conditions (\ref{branch-even}) and
(\ref{branch-odd}) for different values of the `mass' $m$.
The resulting dependence $p=p(m)$ defined by the
three quantization conditions in (\ref{branch-even}) and (\ref{branch-odd})
is shown in Fig.~\ref{mass-gap}.
The following comments are in order.

\begin{figure}
\centerline{\epsfxsize8.6cm\epsfysize7.6cm\epsffile{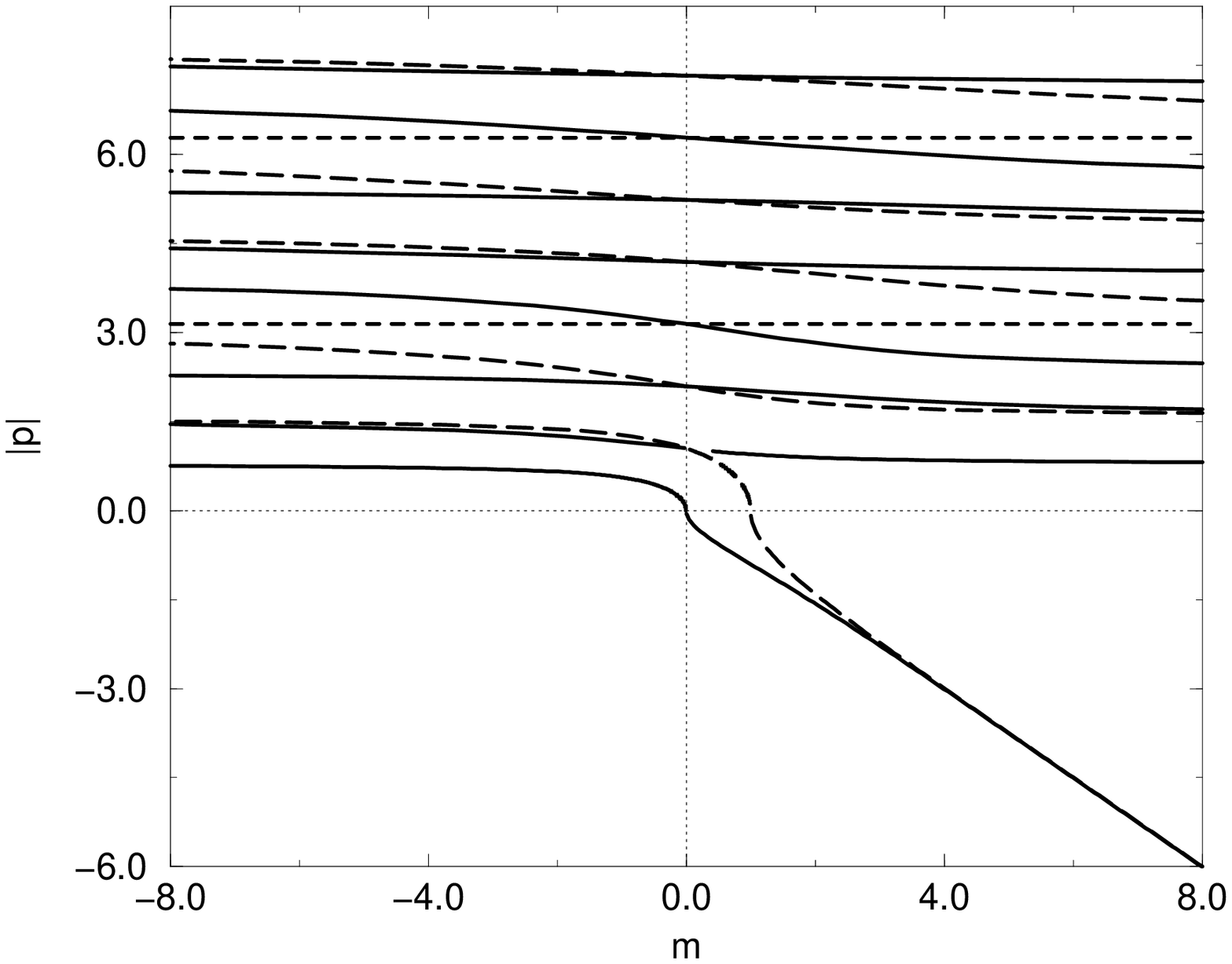}\hspace*{-0.5cm}
            \epsfxsize8.6cm\epsfysize7.6cm\epsffile{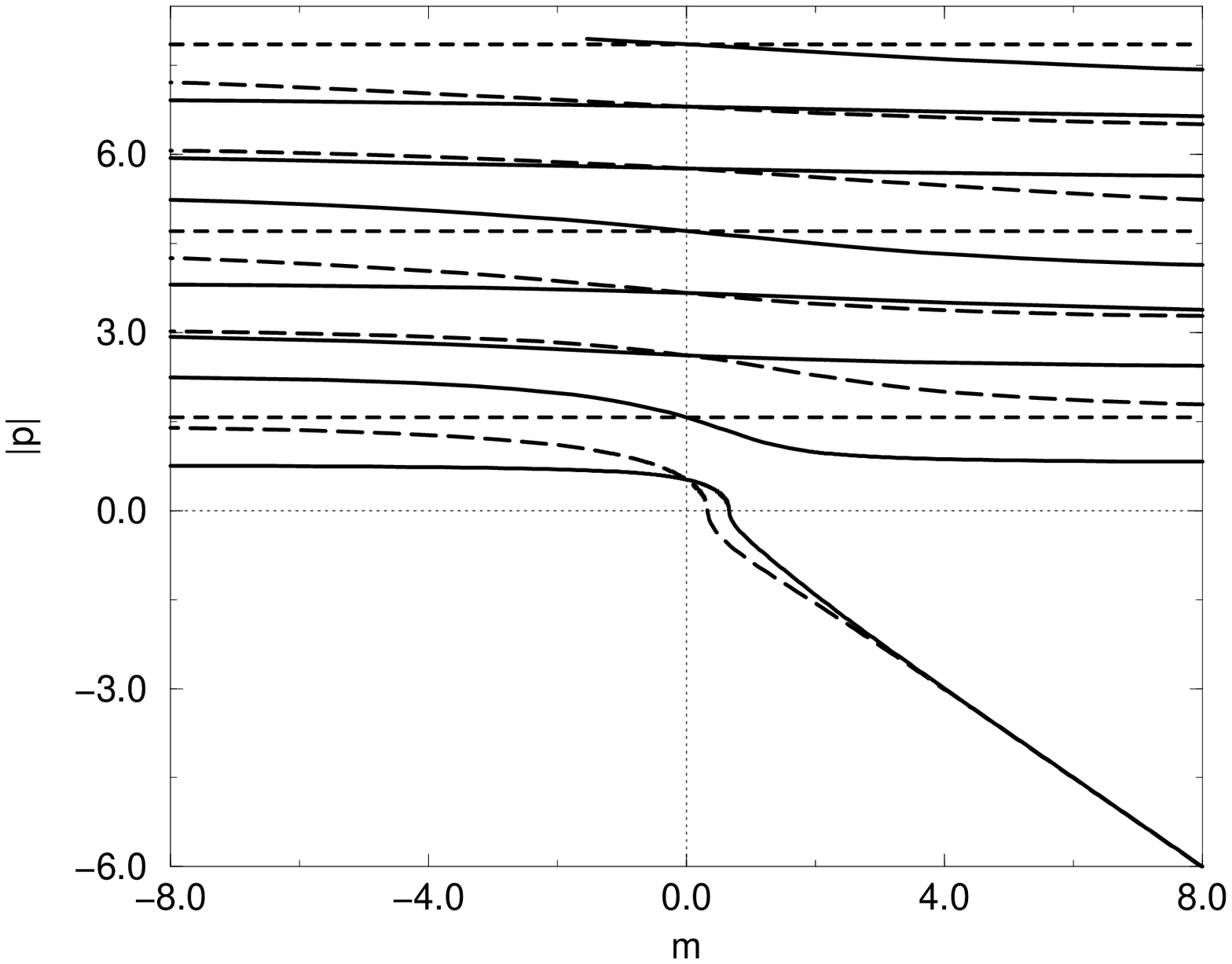}}
\caption[]{\small The flow of the quantized momenta $p=p(m)$ for even (left)
and odd (right) values of $N$. Dashed, long-dashed and solid lines
are described by three branches defined in Eqs.~(\ref{free-even}),
(\ref{branch-even}), (\ref{3rd-even})
and Eqs.~(\ref{free-odd}), (\ref{branch-odd}), (\ref{3rd-odd}), respectively.
Negative values of $|p|$ correspond to pure
imaginary momenta $p$.}
\label{mass-gap}
\end{figure}

At $m=0$ the solutions to (\ref{branch-even}) and (\ref{branch-odd}) are
given by $p_k=\pi(k/3+\delta_N/2)$ with $k$ integer and the corresponding
energies (\ref{Energy-p2}) coincide with the energy levels
(\ref{middleE}) of the Hamiltonian ${\cal H}_{3/2}$.
All solutions except the one with
$k=0$ and $N={\rm even}$ are double degenerate.

For small $m \ll 1$,
or equivalently $\epsilon\ln\eta \ll 1$, the degeneracy is removed
and each branch, $p_k=p_k^{(+)}(m)$ and $p=p_k^{(-)}(m)$, evolves
independently.  The slope of the trajectories at $m=0$ can be found from
Eqs.(\ref{free-even}) -- (\ref{3rd-odd}) as
\begin{eqnarray}
& &
\frac{d\left[p_{3k\pm1}^{(+)}\right]^2}{dm}
= -\frac34\,,\qquad
\frac{d\left[p_{3k\pm1}^{(-)}\right]^2}{dm} = -\frac14,
\nonumber
\\
& &
\frac{d\left[p_{3k}^{(+)}\right]^2}{dm} = -1\,,\qquad
\frac{d \left[p_{3k}^{(-)}\right]^2}{dm} = 0,
\labeltest{slope1}
\end{eqnarray}
with $p_{k}^{(\pm)}(m=0)=\pi(k/3+\delta_N/2)\neq 0$. For {\it even\/} $N$
and $m\to0$ we have a single non-degenerate level $p_0=0$ which evolves
as
\begin{equation}
\frac{d p_0^2}{dm} = -\frac12\,.
\labeltest{slope0}
\end{equation}
It is easy to see that Eqs.~(\ref{slope1}), (\ref{slope0}) and
(\ref{Energy-p2}) are
equivalent to (\ref{shift1}), (\ref{shift2}) and
(\ref{gap2}). Since $p_0(0)=0$, it follows from (\ref{slope0})
that $p_0$ becomes pure imaginary for an arbitrary small
$m > 0$ and the corresponding eigenstate describes a bound state with
the energy
\begin{equation}
{\cal E}_{\rm bound}(\epsilon)={E}_0-\frac{\pi^2\epsilon}{9\ln(\eta\e^{\gamma_E})}.
\end{equation}
with $\epsilon \ll 1/\ln\eta$.
For even $N$ there exists the second bound state which is formed for
nonzero mass $m\ge m_{\rm crit}$. The value $m_{\rm crit}$
corresponds to the nontrivial solution of (\ref{branch-even}) at $p=0$
\begin{equation}
m_{\rm crit}=\epsilon_{\rm crit}\frac{\pi^2}{9\zeta(3)}
\ln(\eta\e^{\gamma_E})=1\,.
\end{equation}
The similar phenomenon occurs for {\it odd\/}  $N$. In this
case, two bound states are formed for $m > 0$ and the corresponding critical
values of the masses $m_{\rm crit}$ (or equivalently $\epsilon_{\rm crit}$)
can be found from (\ref{branch-odd})
and (\ref{3rd-odd}) for $p=0$ as $m_{\rm crit}=1/3$ and $2/3$, respectively. It
is easy to see that for $m$ close to $m_{\rm crit}$ the mass gap, ${\cal
E}_{\rm bound}(\epsilon)-{E}_0$ depends linearly on the perturbation $\epsilon$
for the both bound states.

To understand what happens with the spectrum of the Hamiltonian
as $\epsilon$ varies, consider the solutions to
Eqs.~(\ref{free-even})--(\ref{3rd-odd})
in the two extreme limits: $m=-\infty$ and
$\infty$. The real quantized $p_k$ are the same in both
limits,
\begin{equation}
p_k=\frac14\pi k\,,\qquad k=1\,,2\,,...
\labeltest{p-cont}
\end{equation}
and, as a consequence, the energy
levels in the `continuum' ${g E} > 0$ are also the same and are given
by (\ref{others}). Moreover,
for large positive $m$ there are additional pure imaginary
solutions to Eqs.~(\ref{branch-even}), (\ref{3rd-even}) and
Eqs.~(\ref{branch-odd}), (\ref{3rd-odd})
\begin{equation}
ip_{\rm bound}
=-\frac34 m =
-\epsilon\frac{\pi^2}{12\zeta(3)}\ln(\eta\e^{\gamma_E}),
\labeltest{p-bound}
\end{equation}
which give rise to two bound states with the energy given by (\ref{gap1}).

Thus, the flow of the energy levels from $m=-\infty$ to
$m=\infty$ is such
that the continuum stays unchanged and
two lowest levels of the continuum `dive' into the vacuum. These
two bound states become separated from the continuum by the mass
gap, ${\cal E}_{\rm bound}(\epsilon)-{E}_0$,
whose size grows linearly with $\epsilon$ at small $m\sim m_{\rm crit}$ and
quadratically at large $m$.

Recall, now, that Eq.~(\ref{kin+pot}) presents a low-energy
approximation to the Hamiltonian ${\cal H}(\epsilon)$ and
we have to check that values of
$\bar q N$ are small on the solutions.
Since, according to (\ref{kinetic}), the kinetic energy
contribution to the Schr\"odinger equation is $\CO\left((\bar q N)^2\right)$,
the condition $\bar q N < 1$ can be expressed as
\begin{equation}
\frac{g}{6m}\int_0^3 dx\, \bigg|\partial_x\chi(x)\bigg|^2
=
\frac{g}{2m}\sum_{k=-\infty}^{\infty} k^2 |c_k|^2
=
\bigg|
\frac{2gp^2}{m}
\bigg|
< 1\,,
\labeltest{UV-cut}
\end{equation}
cf. Eq.~(\ref{Energy-p2}),
with the eigenstate $\chi(x)$ normalized as $\frac13\int_0^3
dx\,|\chi(x)|^2=1$. Since $m/g$ scales at large
$N$ as $\sim \ln^2\eta$, the restriction in (\ref{UV-cut}) imposes the
UV cut-off on the quantized momenta of the states $|p| < \ln\eta$.
It follows from (\ref{p-cont}) that for the states in the
continuum, $gE>0$, this condition is satisfied
for the $k =\CO(\ln\eta)$ lowest states only.
For the two bound states with $gE < 0$
the relations (\ref{UV-cut}) and (\ref{p-bound}) lead to the condition
\begin{equation}
g m \sim \epsilon^2 < 1\,.
\end{equation}
Thus, at large $N$, our assumption about decoupling of the low-energy levels
(smallness of $\bar q N$)  is
justified provided that $\epsilon < 1$. For $\epsilon \sim 1$
the higher-order $\bar qN$ corrections to both the kinetic energy
and the potential terms in (\ref{Schrodinger}) become significant
and eventually start to play the dominant r\^ole for $\epsilon\gg 1$.
In this case, as it was shown in
Sect.~5.3, the eigenstates of the exchange interaction $V$ provide the
appropriate basis and the Hamiltonian ${\cal H}_{3/2}$ can be
treated as a perturbation.

\end{document}